\documentclass[prb,aps,showpacs,twocolumn,preprintnumbers,amsmath,amssymb,superscriptaddress]{revtex4-2}
\usepackage[english]{babel}
\usepackage{amsmath,amssymb,amsfonts}
\usepackage{graphicx}
\usepackage[colorlinks=True,linkcolor=blue,citecolor=blue,urlcolor=blue]{hyperref}

\usepackage{booktabs}
\usepackage[dvipsnames]{xcolor}
\usepackage{comment}
\usepackage{braket}
\usepackage{bm}
\usepackage{bbm}
\usepackage{braket}
\usepackage{float}
\usepackage{multirow}
\usepackage{longtable}
\usepackage[normalem]{ulem}
\usepackage{array}
\usepackage{makecell}
\usepackage{subfigure}
\usepackage{booktabs}
\usepackage{multirow}
\usepackage{graphicx}
\graphicspath{{NewFigures/}}
\usepackage{dcolumn}
\usepackage{bm}
\usepackage{longtable}
\usepackage[dvipsnames]{xcolor}

\newcolumntype{C}{>{$}c<{$}}

\def\bk{\mathbf{k}}

\allowdisplaybreaks
\DeclareMathAlphabet{\zc}{OT1}{pzc}{m}{it}

\begin{document}
\title{Reversing non-Hermitian skin accumulation with a non-local transverse switch}

\author{Mengjie Yang}
\affiliation{Department of Physics, National University of Singapore, Singapore 117551, Singapore}

\author{Ching Hua Lee}
\email{phylch@nus.edu.sg}
\affiliation{Department of Physics, National University of Singapore, Singapore 117551, Singapore}

\date{\today}

\begin{abstract}
Asymmetrically directed couplings in non-Hermitian systems cause directional amplification that leads to boundary skin state accumulation. However, counter-intuitively, the direction of accumulation may not follow that of the directed couplings. In this work, we demonstrate new mechanisms where this accumulation can be systematically reversed without modifying the couplings at all, just by adjusting the system size or boundary conditions in a different, transverse direction. Moreover, the reversed skin dynamics can be made very robust by suppressing the amplification in the original non-reversed direction.
We motivate our approach through a series of warm-up models, culminating in a designed non-Hermitian Kagome lattice whereby wavepacket simulations demonstrate how robust reversed skin dynamics can be switched on/off in a non-local transverse manner. Our findings highlight how the non-trivial entanglement between the spectral loops of different PBC directions can be harnessed as a directional amplification switch, paving the way for new avenues of non-Hermitian sensing and lasing.
\end{abstract}

\maketitle


\section{Introduction}

The non-Hermitian skin effect (NHSE)~\cite{lee2016anomalous,yao2018edge,alvarez2018non,kunst2018biorthogonal,song2019non,lee2019anatomy,lee2019hybrid,yang2020non,li2020critical,hofmann2020reciprocal,zhu2020photonic,song2020two,liu2020helical,helbig2020generalized,zou2021observation,stegmaier2021topological,guo2021exact,lee2021many,zhang2021observation,yokomizo2021scaling,liu2021non,xue2021simple,zhang2021tidal,li2022direction,wu2022flux,gu2022transient,shang2022experimental,yang2022concentrated,arouca2020unconventional,zeng2022real,qin2023non,longhi2022self,wang2023demonstration,manna2023inner,ma2023quantum,li2023loss,meng2024exceptional,qin2023universal,zhang2023electrical,lin2023topological,tai2023zoology,jiang2023dimensional,yang2024percolation,yang2024non,nie2024multiple,xue2024topologically,xiong2024non,guo2024scale,zhang2024observation2,shen2024enhanced,li2024observation,lin2024observation,liu2024observation,yoshida2024non,gliozzi2024many,shimomura2024general,li2024emergent,yang2025beyond,qin2025many,li2025observation,shen2025non,hamanaka2025multifractality,li2025phase,rafi2025critical,shen2025observation}, characterized by extreme boundary sensitivity and directional eigenstate accumulation due to non-reciprocity and gain/loss~\cite{lee2016anomalous,alvarez2018non,lee2019anatomy,liu2020helical,zhang2021observation,yang2022concentrated,jiang2023dimensional,yang2024non,li2025phase}, has emerged as a cornerstone of contemporary condensed matter physics and photonics research. This effect has not only challenged our understanding of non-Hermitian bulk boundary-correspondences~\cite{yao2018edge,kunst2018biorthogonal,lee2019anatomy,yokomizo2019non,yang2020non,zhu2020photonic,okuma2020topological,xiao2020non} and alternative notations of criticality arising from non-locality~\cite{lee2020unraveling,li2020critical,li2021impurity,shen2022non,helbig2020generalized,pan2020non,budich2020non,yoshida2020mirror,yang2022liouvillian,zhou2022space,edvardsson2022sensitivity}, 
but has also inspired numerous experimental demonstrations for eventual technological applications, ranging from topolectrical circuits~\cite{helbig2020generalized,dong2021topolectric,zou2021observation,stegmaier2021topological,hohmann2023observation,shang2024observation,zou2025experimental} and photonic crystals~\cite{pan2018photonic,lin2024observation,liu2024observation} to quantum walks~\cite{xiao2020non,wang2021detecting,lin2022topological,wang2023demonstration,xiao2024observation}.

Recent theoretical and experimental studies have further unveiled that NHSE can surprisingly exhibit reversed accumulation, where the eigenstates accumulate opposite to the anticipated direction dictated by asymmetric coupling strengths due to peculiar interference effects. Such ``skin reversal'' phenomena have been observed both theoretically~\cite{xue2021simple,li2022direction,nie2024multiple} and experimentally~\cite{wang2023demonstration,lin2024observation}, challenging intuitive assumptions about the monotonic directional nature of NHSE. 
However, despite these advances, fundamental questions remain unresolved: is it possible to independently control or even enhance this reversal without directly modifying these asymmetric couplings? Specifically, can we harness the non-locality of non-Hermitian state pumping to modify the NHSE in a different direction? Addressing these questions is not only crucial for the fundamental physics of non-Hermitian topological phases but also for harnessing NHSE in practical applications such as directional amplification and topological lasing.

In this work, we motivate and investigate novel mechanisms of NHSE reversal driven explicitly by transverse nonlocality. Differing from prior investigations~\cite{xue2021simple,li2022direction,wang2023demonstration,nie2024multiple,lei2024activating,lin2024observation}, we systematically explore how varying the dimensions and boundary conditions in a  perpendicular (transverse) direction can distinctively alter the NHSE direction, and even cause the reversed modes to be amplified more strongly. 
Our findings emphasize significant \emph{transverse-size-dependent} behaviors and extend beyond known one-dimensional treatments of the critical NHSE~\cite{li2020critical,liu2020helical,yokomizo2021scaling,qin2023universal,rafi2022system,liu2024non}.

\section{Non-Hermitian skin effect (NHSE) reversal}

\begin{figure*}
  \centering
  \includegraphics[width=0.9\textwidth]{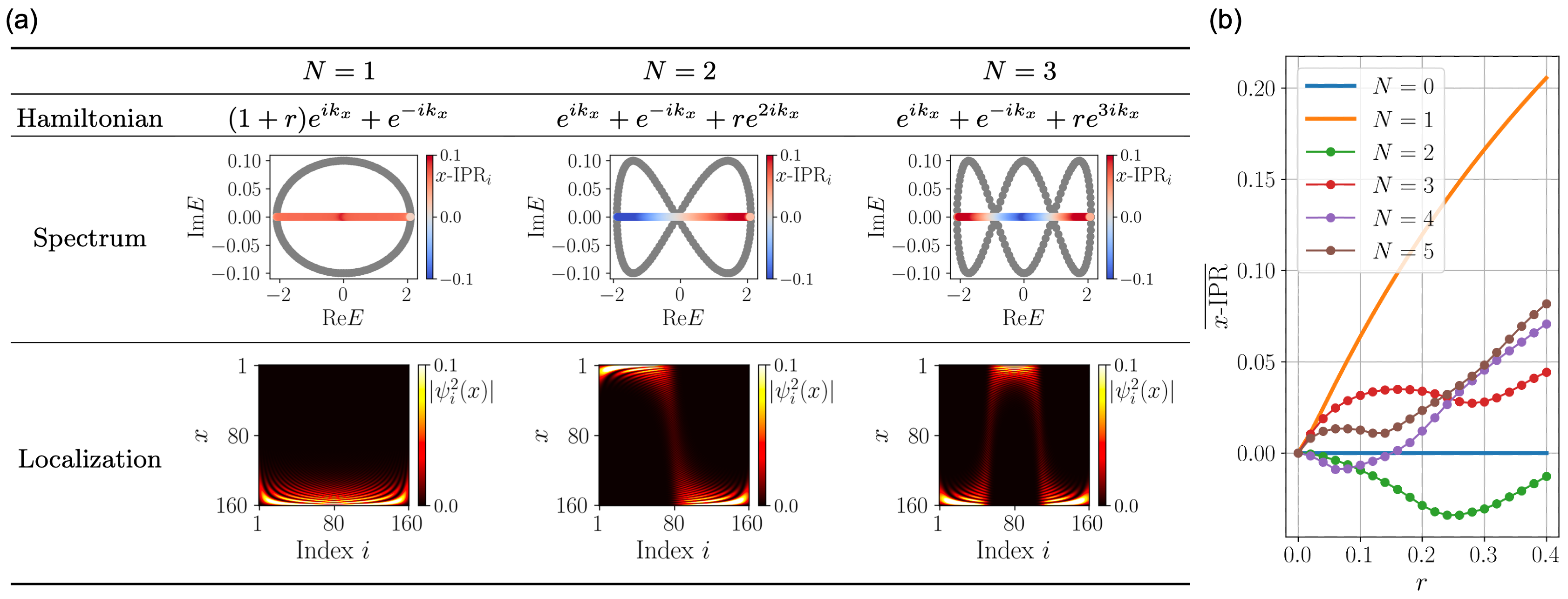}
  \caption{Spectral and spatial localization characteristics of the one-band reversal model (Eq.~\eqref{eq:1-band}) across different values of $N$. (a) The spectrum displays reversed skin states (blue dots) whose spatial localization tends leftward, with coloring determined by their respective $x\text{-IPR}_i$ values from Eq.~\eqref{eq:x-IPR}. (b) Mean $x$-IPR$_i$ ($\overline{x\text{-IPR}}$) calculated via Eq.~\eqref{eq:mean-x-IPR} plotted against non-Hermiticity parameter $r$. The $N=1$ case shows monotonically increasing $\overline{x\text{-IPR}}$ with $r$, indicating strengthening rightward skin localization, while $N\geq2$ cases exhibit substantially lower values, revealing skin accumulation away from the right boundary. The gap between $\overline{x\text{-IPR}}$ and $\overline{x\text{-IPR}}(N=1)$ quantitatively demonstrates the extent of the skin reversal. }
  \label{fig:1-band}
\end{figure*}

Given an arbitrary-dimensional system described by a real-space lattice Hamiltonian $H$ with asymmetric  hoppings, one might intuitively expect all eigenstates, and hence arbitrary time-evolved states, to accumulate at the boundary that the asymmetric hoppings point towards. 
For instance, if hopping asymmetry points only towards the $+x$ boundary, eigenstates $\psi_i(\bold r)$ satisfying $H\psi_i(\bold r)=E_i\psi_i(\bold r)$ (with $L=L_x\times L_y\times ...$ and $i=1,2,..., L$ labeling the eigenstates) would predominantly localize at the right boundary.

However, as we shall demonstrate, significant eigenstate localization may also occur on the opposite boundary—a phenomenon we refer to as ``non-Hermitian skin effect (NHSE) reversal''. This conceptually counter-intuitive phenomenon has been understood~\cite{xue2021simple,li2022direction,nie2024multiple} as arising due to destructive interference  -- as eigen-wavefunctions can be negative or complex, states that are asymmetrically pumped towards one direction may not necessarily reinforce the overall overall probability flux in that same direction. However, as we shall reveal, NHSE reversal can exhibit even more enigmatic attributes, culminating in the design of a switch for NHSE reversal that is purely controlled through the boundary conditions in a \emph{perpendicular} i.e. transverse direction.

To quantify the extent of the NHSE, whether in the expected or reversed direction, we define the normalized inverse participation ratio in the $x$-direction ($x$-IPR) for a particular eigenstate $\psi_i$:
\begin{equation}
x\text{-IPR}_i = \left(\sum_{\bold r} \frac{x-x_c^{(i)}}{L_x - x_c^{(i)}}|\psi_{i}(\bold r)|^4\right)/\left(\sum_{\bold r} |\psi_{i}(\bold r)|^2\right)^2,
 \label{eq:x-IPR}
\end{equation}
where $\bold r = (x,y,...)$ and $x_c^{(i)}=
\frac{\sum_{\bold r}\, x\,|\psi_i(\bold r)|^2}{\sum_{\bold r} |\psi_i(\bold r)|^2}$ represents the $i$-th eigenstate's center of mass coordinate. The normalization factor $\frac{1}{L_x - x_c}$ allows for consistent comparison across different system sizes. As defined, $x\text{-IPR}_i$ disappears for an eigenstate that is symmetric in the $x$-direction, and a positive (negative) value of $x\text{-IPR}_i$ indicates NHSE localization toward the right (left) boundary.

Additionally, we define the mean $x$-IPR over all eigenstates as:
\begin{equation}
\overline{x\text{-IPR}} = \frac{1}{L} \sum_{i=1}^{L} x\text{-IPR}_i, \label{eq:mean-x-IPR}
\end{equation}
which provides a global measure of the directional skin localization across the whole spectrum.
metrics in place, we now turn to our models to clearly illustrate and quantify the phenomenon of skin reversal.

\subsection{NHSE Reversal in a minimal one-band model}

To concretely illustrate one simplest manner in which NHSE reversal can occur, we consider 
the following one-band model:
\begin{equation}
H_{\text{1-band}}(k_x) = \left(e^{i k_x} + \frac{1}{e^{i k_x}}\right) + r e^{i N k_x},\label{eq:1-band}
\end{equation}
which is just the well-known Hermitian Hatano-Nelson (HN) model~\cite{hatano1996localization}, but with an additional purely rightward hopping of strength $r$ across
$N\geq2$ sites. 
We do not consider $N<2$ because if $N=0$, it reduces to the HN model with onsite potential being $r$; if $N=1$, it is the HN model with the hopping strength now biased towards the right, being $1+r$ (towards the $+x$ direction) and $1$ (towards the $-x$ direction).

One might naively expect the skin effect in this one-band model to exhibit unidirectional behavior, specifically towards the right, given that the only asymmetric hopping occurs in this direction for $r\neq 0$. While this is indeed true for the usual HN model ($N=1$ case, see Fig.~\ref{fig:1-band} where all OBC modes are red), notable NHSE reversal emerges for $N\geq2$, characterized by a substantial proportion of eigenstates (blue) localized toward the opposite boundary $x=1$.  This NHSE reversal phenomenon appears to be universal in this one-band model [Eq.~\ref{eq:1-band}], as demonstrated across multiple $N$ values in the Eq.~\eqref{eq:1-band} results presented in Fig.~\ref{fig:1-band}(a) and (b).

\begin{figure*}
  \centering
  \includegraphics[width=0.96\textwidth]{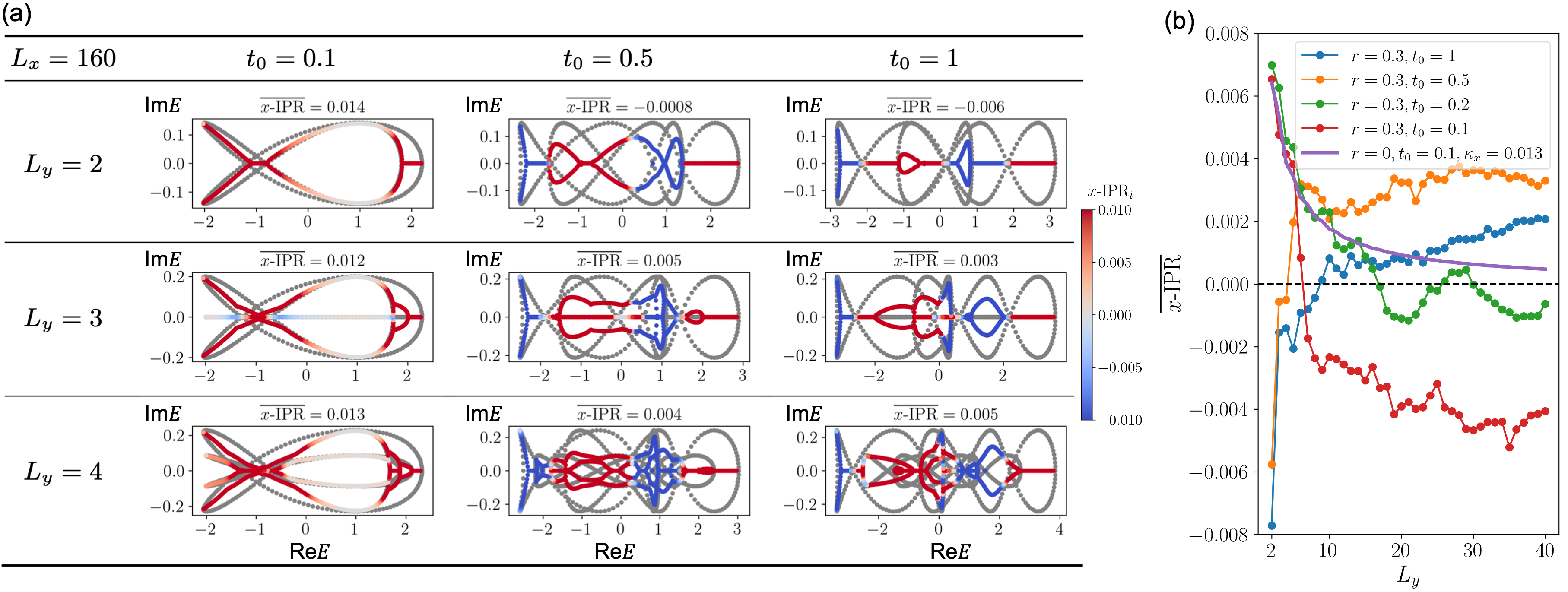}
  \caption{(a) The $x,y$-OBC spectrum for Eq.~\eqref{eq:H-size} with $N=3$, colored by $x$-IPR$_i$ values from Eq.~\eqref{eq:x-IPR}, demonstrates that the reversal of winding direction in $y$-PBC loops (gray) leads to $x$-skin reversal (blue dots). The strength of transverse hopping $t_0$ significantly influences this reversal. (b) Beyond the choice of larger or smaller $t_0$ values, an important observation is the pronounced $L_y$ dependence of the overall preference of skin localization, $\overline{x\text{-IPR}}$, (Eq.~\eqref{eq:mean-x-IPR}) across various $t_0$ values (dot-line curves). The observed $L_y$ dependence exhibits remarkably strong effects, as evidenced by comparing the variation in $\overline{x\text{-IPR}}$ between our model (Eq.~\eqref{eq:H-size}) and a comparative model with $r=0$ that incorporates a skin depth $\kappa_x$ in each $x$ chain (Eq.~\eqref{eq:Hprime-size}). Notably, the green dotted-line curve shows a significant reversal, varying from approximately 0.06 to $-0.04$, in contrast to the red curve, which ranges from about 0.06 to near zero. Both curves represent $t_0=0.1$, highlighting the strong reversal effect of $L_y$ dependence in our model. To ensure a meaningful comparison, we selected $\kappa_x=0.013$, equalizing both $\overline{x\text{-IPR}}$ values at $L_y=2$.}
  \label{fig:size-Ly}
\end{figure*}

The emergence of skin reversal can be attributed to the winding direction of the PBC loop (gray) shown in Fig.~\ref{fig:1-band}, specifically where the loop winds counter to the direction suggested by the hoppings: reversed/non-reversed OBC eigenstates, colored blue/red, occur depending on whether they are encircled clockwise/anti-clockwise by the PBC spectral loop as $k$ increases.
The correspondence between PBC spectral winding chirality and OBC skin accumulation was shown in Ref.~\cite{xue2021simple,li2022direction,ou2023non,wang2024non}, and can also be deduced from the electrostatics formalism of the NHSE~\cite{yang2022designing,xiong2024graph}.
This reversed spectral winding directly results in eigenstates accumulating towards the left boundary, despite the rightward-directed hopping. In general, for any one-band model with $H(k)= \sum_{\Delta > 0} t_\Delta e^{i\Delta k}$, where only hoppings in only one (positive) direction are permitted, the energy spectrum $E(k)=H(k)$ can still exhibit ``epicycles'' that enclose regions with reversed NHSE -- the reversed OBC skin modes contained therein can be adiabatically traced to the PBC modes represented by the epicycle circumference, whose momenta $k$ lead to effective ``destructive'' interference.

To quantify the overall extent of NHSE reversal,
we compare the $\overline{x\text{-IPR}}$ values (defined in Eq.~\eqref{eq:mean-x-IPR}) for $N\geq2$ with those obtained when $N=1$. 
The $\overline{x\text{-IPR}}$ values for $N=1$ exhibit monotonic increase with $r$, indicating unidirectional right-skin effect of increasing strength $r$. In contrast, the $\overline{x\text{-IPR}}$ values for $N\geq2$ are substantially lower than $\overline{x\text{-IPR}}(N=1)$, demonstrating a strong tendency for skin accumulation away from the right boundary. In particular, the $N=2$ case (green) exhibits an overall reversed NHSE direction when averaged over all eigenstates, across all values of $r$ shown.

\subsection{Controlling NHSE reversal with the transverse system size }

\begin{figure*}
  \centering
  \includegraphics[width=0.8\textwidth]{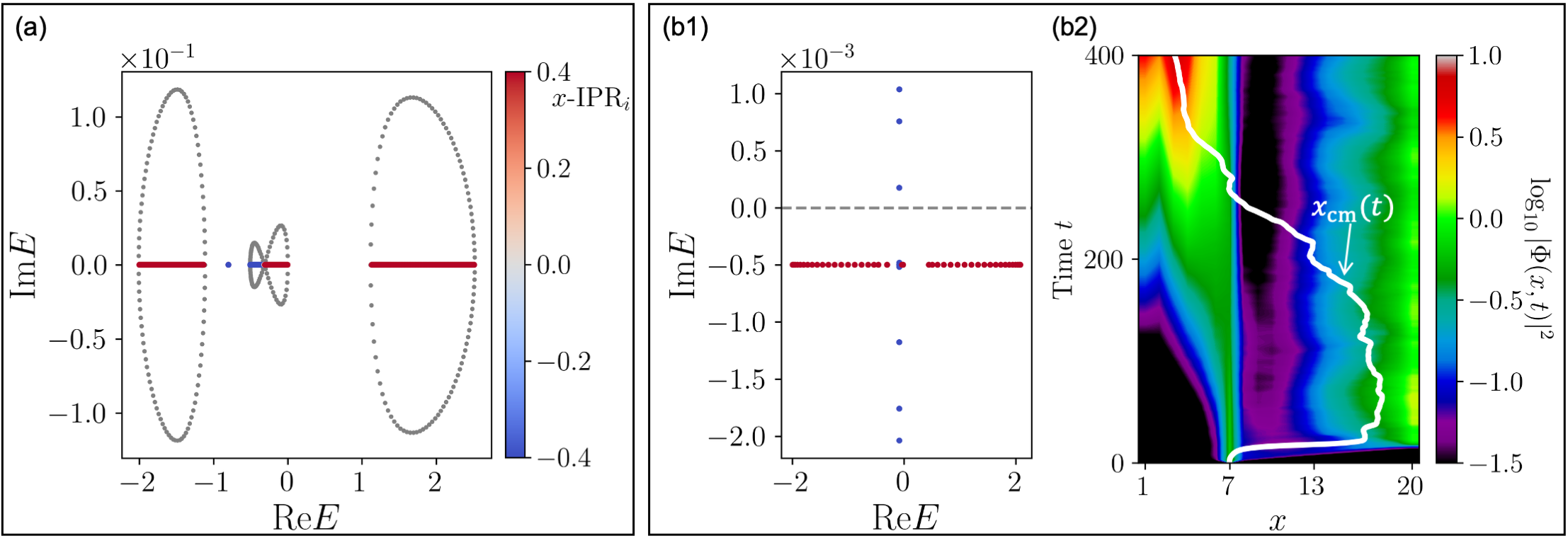}
  \caption{(a) Reversed PBC spectral winding (left lobe of central figure-8 loop) leads to NHSE reversal in the OBC eigenstates (colored blue according to its $\overline{x\text{-IPR}}$ [Eq.~\eqref{eq:mean-x-IPR}]) of $H_\text{3-band}$ [Eq.~\eqref{eq:3-band}] with $L_x=100, t_c=0.8, \gamma=0$. (b1) The reversed skin states (blue) can acquire positive Im$E$ and be hence asymptotically amplified with $L_x=20, \gamma=0.0005$, and $t_c=0.5$. (b2) Corresponding dynamical evolution $H_{\text{3-band}}\Phi(x, t)=i\dot{\Phi}(x, t)$ of a wavepacket $\Phi(x,t=0)$ initially localized in the middle, i.e., at unit cell 7. The white curve indicates the wavepacket center of mass $x_{\text{cm}}(t)=\sum_x x|\Phi(x, t)|^2/\sum_x |\Phi(x, t)|^2$. The wavepacket initially propagates to the right boundary, driven by the hopping asymmetry $u_x = 1/v_x = 1.1$. However, at around $t\approx 200$, it ultimately and inevitably reverses direction and moves towards the left boundary. }
  \label{fig:3-band}
\end{figure*}

Going beyond 1D systems, the NHSE in a particular direction can be affected by the band structure in other transverse directions. The easiest way to demonstrate this is to modify the bulk properties of the transverse subsystem slices, which can be regarded as supercells.
However, in this work, we shall only consider more enigmatic scenarios where we do \emph{not} even modify the bulk properties. Namely, we shall only adjust the (i) transverse system size and (ii) transverse boundary conditions, and show that they can already profoundly affect the NHSE reversal perpendicularly to them.

Below, we first showcase (i) by extending our minimal 1D model to 2D:
\begin{equation}
  H_{\text{size}}(\bold k)  = \left(e^{ik_x} + \frac{1}{e^{ik_x}}\right) + t_0 \left(e^{ik_y} + \frac{1}{e^{ik_y}}\right) + r e^{i N k_x}e^{ik_y}. \label{eq:H-size}
\end{equation}
This model extends the one-band model (Eq.~\eqref{eq:1-band}) to two dimensions by including nearest-neighbor hoppings of strength $t_0$ in the transverse ($y$) direction and a single rightward-directed
hopping of strength $r > 0$, which hops $N$ sites in the $x$-direction and $1$ site in the $y$-direction.

In Fig.~\ref{fig:size-Ly}(a), larger $L_y$ values are observed to give rise to more intricate spectral loops, some with reversed-NHSE OBC states (blue), consistent with the interpretation of extended $y$-degrees of freedom as supercells. 
While skin reversal occurs in this 2D extension as anticipated, a striking result is the pronounced dependence on transverse lattice size $L_y$, as shown in Fig.~\ref{fig:size-Ly}(b). Specifically, as $L_y$ increases from 2 to 40, the overall skin accumulation measure $\overline{x\text{-IPR}}$ changes drastically, demonstrating significant transverse-size-dependent skin reversal propensity.
Even in the cases of weak transverse coupling $t_0$, i.e. $t_0=0.1$, where the reversed-NHSE states (blue in Fig.~\ref{fig:size-Ly}(a)) do not appear to be prominent at small $L_y$, $\overline{x\text{-IPR}}$ still exhibits a sustained dependence on $L_y$ as the latter increases (red Fig.~\ref{fig:size-Ly}(b)). Appendix A additionally shows the NHSE reversal of the dominant eigenstate from varying $t_0$ and $L_y$.

To benchmark and quantify the extent of this NHSE reversal, we compare its $\overline{x\text{-IPR}}$ with an ``ordinary'' 2D-extended lattice at different values of $L_y$. For that, we introduce a comparative model without the uni-directional couplings (such that $r=0$): 
\begin{equation}
H'_{\text{size}}(\bold k) =\left(e^{ik_x+\kappa_x} + e^{-ik_x-\kappa_x}\right) + t_0 \left(e^{ik_y} + e^{-ik_y}\right), \label{eq:Hprime-size}
\end{equation}
which incorporates intrinsic exponential localization with non-Hermitian $x$-skin depth $\kappa_x^{-1}$. 
For a fair comparison, we set $\kappa_x=0.013$ to match initial conditions at $L_y=2$ for $t_0=0.1$. As Fig.~\ref{fig:size-Ly}(b) clearly illustrates, for the benchmark model $H'_{\text{size}}$, the $\overline{x\text{-IPR}}$ decreases rapidly with $L_y$, since its eigenstates spread out in the $y$-direction even though they remain equally localized in the $x$-direction. Yet, $\overline{x\text{-IPR}}$ for $H_{\text{size}}$ [Eq.~\ref{eq:H-size}] not only avoids generally decreasing with $L_y$, but in fact also shows dramatic reversal. 
Specifically, while at smaller $t_0$ (red, green), increasing $L_y$ causes a transition to net-reversed NHSE, the transition occurs in the opposite direction at larger $t_0$ (orange, blue). This showcases that, even in such a simple 1-band model, the transverse system size can already affect the NHSE reversal in a subtle manner.

\begin{figure*}
  \centering
  \includegraphics[width=0.98\textwidth]{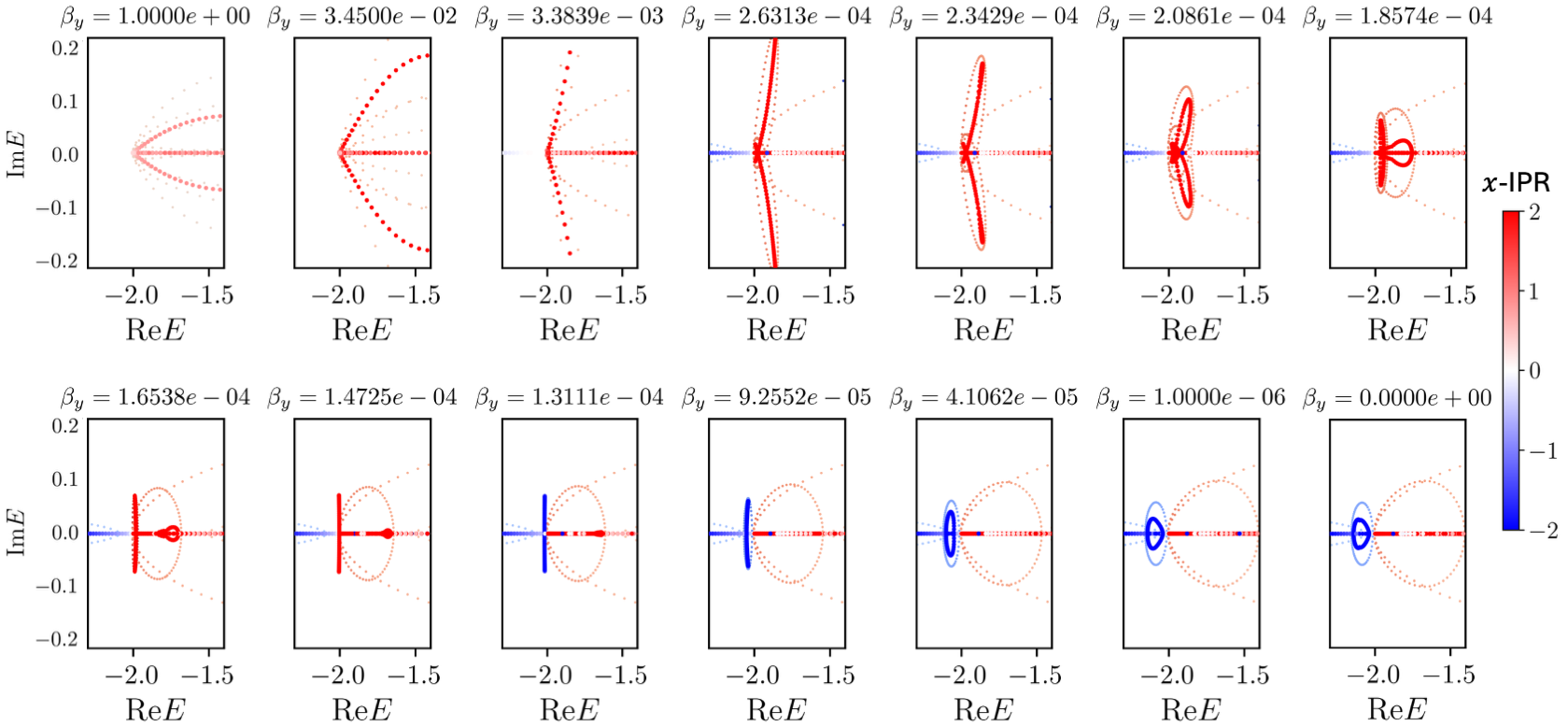}
  \caption{Spectral origin of transverse boundary-controlled NHSE reversal in the Kagome model [Eq.~\eqref{eq:kagome_model}]. Shown are its complex spectra for partial x-OBCs $\beta_x=0.2$ (lighter and thinner dots), as well as $x$-OBCs ($\beta_x=0$, darker and thicker dots), as $\beta_y$ is deformed from $1$ ($y$-PBCs) to $10^{-6}$ (almost $y$-OBCs), and to 0 ($y$-OBCs). All spectral data are colored according to their $x$-IPR [Eq.~\eqref{eq:x-IPR}]: red ($x$-IPR$>0$, right-skin), blue ($x$-IPR$<0$, left-skin). 
In general, the $\beta_x=0.2$ spectral loops encircle the $\beta_x=0$ eigenvalues, such that reversals of the $\beta_x=0.2$ spectral winding imply reversals of the $\beta_x=0$ skin accumulation. 
Initially, in the y-PBC limit $\beta_y=1.0$, the NHSE is not reversed (fully red spectrum). As $\beta_y$ decreases to $\approx 2\times  10^{-4}$, some skin states become reversed (dark blue), but only along the real line. As $\beta_y$ continues to decrease to $\approx  10^{-4}$, a complex $x$-OBC spectral branch changes from red to blue, encircled by a reversed $\beta_x=0.2$ spectral loop.
	F or smaller $\beta_y$, the reversed blue skin modes remain complex and will hence dominate in long-time dynamics,
	demonstrating that robust $x$-skin reversal can be switched on/off by toggling between $y$-OBCs/PBCs.
	 	Parameters are $L_x=120, L_y=8, u_x=1/v_x=1.1, t_c=0.5,u_y=1/v_y=3$.}
  \label{fig:zoomin-loop}
\end{figure*}

\section{Achieving robustly reversed NHSE }

\subsection{Minimal 3-band model}

Ultimately, any NHSE reversal would be more experimentally significant if the reversed skin modes experience more gain than the ``naively'' expected non-reversed skin modes. In the following, we showcase how this can be achieved by designing Hamiltonians with PBC spectral loops that twist in a way such that the reversed NHSE OBC modes have  larger Im$E$.
For this, a minimal ansatz requires 3 bands, an example which is given by the following Hamiltonian:
\begin{equation}
  H_{\text{3-band}}(k_x)  = \left(\begin{array}{ccc}
    -i\gamma & v_x+ u_x e^{-i k_x} & u_y \\
    u_x+ v_x e^{i k_x} & -i\gamma & u_y \\
    v_y & v_y & -i\gamma
  \end{array}\right),\label{eq:3-band}
\end{equation}
where $u_x,v_x,u_y,v_y$ are hopping amplitudes within this 3-component 1D model. We stipulate that $u_x>v_x$, such that each asymmetric hopping locally amplify towards the right. 
Yet, like before, it is easy to find parameters where there is an extended range of eigenstates (blue i.e. negative $x$-IPR$_i$ in Fig.~\ref{fig:3-band}(a)) that are reversed-localized towards the left edge. 

Most notably, through slight parameter adjustments by using $t_c=0.5$ in (b1) instead of $t_c=0.8$ in (a), some eigenstates experiencing NHSE reversal can be designed to possess positive imaginary eigenenergies [blue in Fig.~\ref{fig:3-band}(b1)]. 
This is significant because any dynamical time evolution would amplify only the reversed NHSE states, as the 
 amplitude of an eigenstate dynamically follows $|e^{-iE_i t}|=|e^{-i\text{Re}E_i t + \text{Im}E_i t}|=e^{\text{Im}E_i t}$. That is, 
given an initial wavepacket $\Phi(x, t=0)$ at time $t=0$ located at any cell $x$ on the lattice, after some time evolution defined by the time-dependent Schrodinger equation $H_{\text{3-band}}\Phi(x, t)=i\dot{\Phi}(x, t)$, the wavepacket will be increasingly amplified towards the left, despite the hopping asymmetry being directed to the right. 

Such NHSE reversal dynamics is demonstrated in Fig.~\ref{fig:3-band}(b2). With a wavepacket $\Phi(x,t)$ initially localized in the middle, i.e., cell 7, the wavepacket center-of-mass $x_\text{cm}(t)$ initially propagates to the right boundary, driven by the hopping asymmetry $u_x = 1/v_x = 1.1$. After remaining at the right boundary for some time, the wavepacket eventually gets dominated by the reversed-NHSE modes, and hence reverses direction and moves towards the left boundary, continuing to grow over there indefinitely (red).

\begin{figure*}
  \centering
  \includegraphics[width=0.98\textwidth]{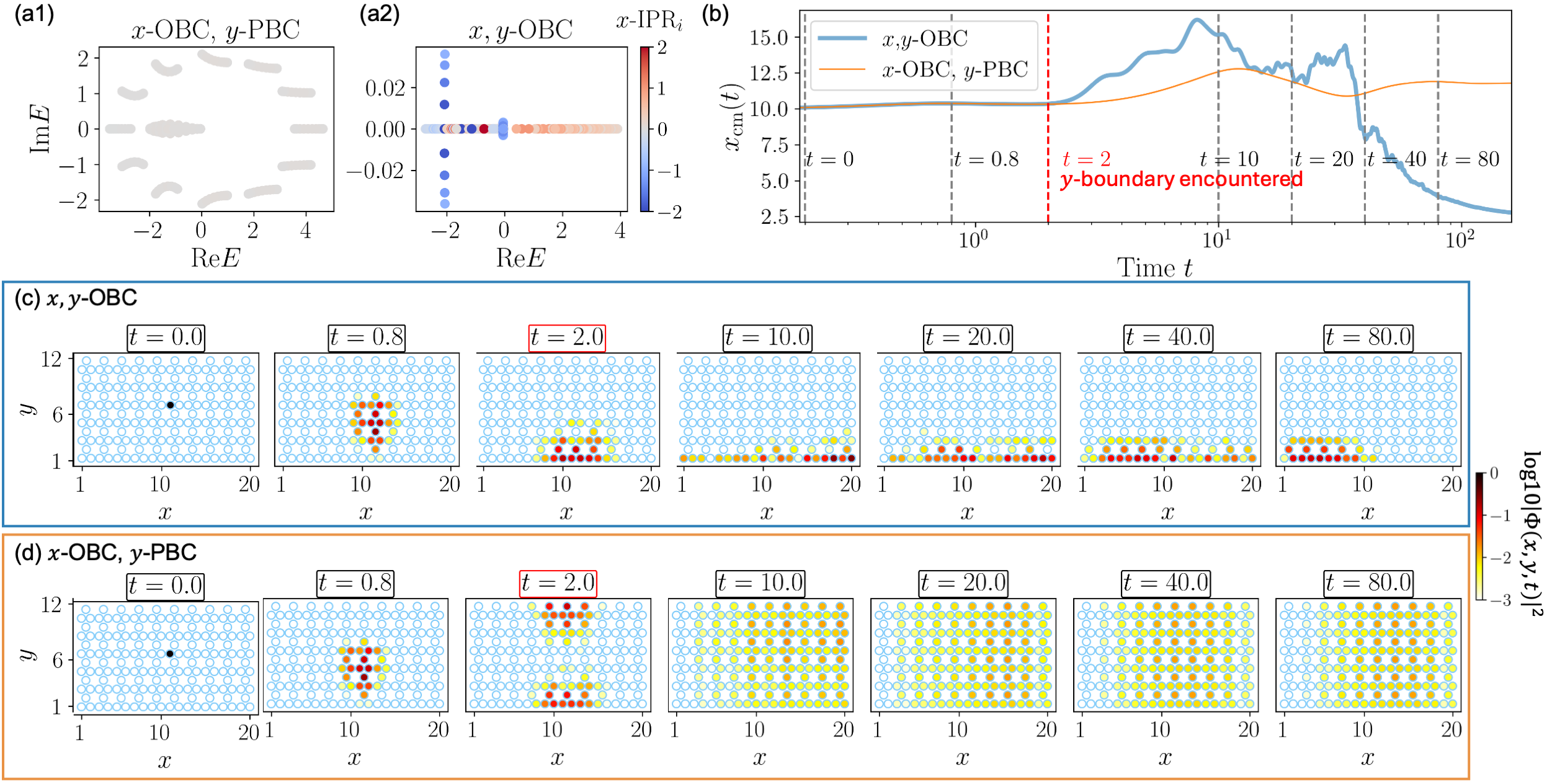}
  \caption{Dynamical consequence of switching on and off the $x$-NHSE reversal through the $y$-boundary conditions. (a1) Complex spectrum of the Kagome lattice with open $x$-boundaries but periodic $y$-boundary ($x$-OBC, $y$-PBC), given by Eq.~\eqref{eq:kagome_model} with $\beta_x=0,\beta_y=1$. (a2) Complex spectrum of the same lattice with double open boundaries ($x,y$-OBCs). Each point is colored by its $x$-IPR [Eq.~\eqref{eq:x-IPR}]: red (blue) denotes right- (left-) $x$-skin localization, and gray denotes bulk-like modes with negligible $x$-IPR. Only after the $y$-boundary is opened do pronounced left $x$-skin modes (blue) appear, indicating reversal of the original right-accumulation tendency in a \emph{perpendicular} direction. (b) Centre-of-mass trajectory $x_{\mathrm{cm}}(t)=\sum_{x,y} x|\Phi(x,y, t)|^2/\sum_{x,y} |\Phi(x,y, t)|^2$ of a wave-packet initialized at the middle of the lattice. 
	The wavepacket trajectories (blue, orange) under different $y$-boundary conditions start to diverge after the wavepacket first encounters the $y$-boundary at $t\approx 2$. For the fully open geometry (blue curve), the packet first drifts rightwards (larger $x_{\mathrm{cm}}$), 
but then subsequently reverses towards the left boundary at small $x_{\mathrm{cm}}$. With periodic $y$-direction (orange curve) the reversal never occurs. 
(c,d) Snapshots of the wavepacket probability density $|\Phi(x,y,t)|^{2}$ at various stages of the evolution, for both sets of boundary conditions. In both cases, the wavepacket intially spreads out and moves downwards. In (c), it is then blocked by the lower $y$-boundary around $t=2.0$, and evolves rightwards according to the prevailing NHSE pumping direction. However, this rightward drift is transient and the wavepacket is eventually funnelled leftward by reversed skin amplification, ultimately accumulating towards the left edge at $x=1$. 
In (d), the $y$-spreading continues unabated under $y$-PBCs, and eventually the wavepacket reaches a rather featureless steady-state without net skin accumulation. Parameters: $u_x = 1/v_x = 1.1$, $u_y = 1/v_y = 2$.}
  \label{fig:kagome-dynamics}
\end{figure*}

\section{Non-local switching of robustly reversed NHSE from transverse boundary conditions}
We have already seen how the NHSE reversal can made experimentally salient by designing the reversed modes to have larger Im$E$, and how their extent can be controlled by adjusting the transverse dimensions of the NHSE system. Below, we show that it is even possible to switch on/off the saliently reversed NHSE modes by adjusting the boundary conditions in the transverse direction. This keenly showcases that non-Hermitian bulk-boundary correspondence in one direction can crucially affect the NHSE reversal in a perpendicular direction.

For a concrete demonstration, we promote the previous 1D model Eq.~\ref{eq:3-band} to a Kagome lattice model, which is a 2D lattice with three sites per unit cell. The momentum-space Hamiltonian can be written as $H_{\text{Kagome}}(\bk)=
  V
  + e^{-i k_x}V_x^{-}
  + e^{ i k_x}V_x^{+}
  + e^{-i k_y}V_y^{+}
  + e^{ i k_y}V_y^{-}$
with
$
V=\left(\begin{matrix}
0 & v_x & u_y\\
u_x & 0   & u_y\\
v_y & v_y & 0
\end{matrix}\right),
V_x^+=\left(\begin{matrix}
0 & 0   & 0\\
v_x & 0 & 0\\
0 & 0   & 0
\end{matrix}\right),
V_x^-=\left(\begin{matrix}
0 & u_x & 0\\
0 & 0   & 0\\
0 & 0   & 0
\end{matrix}\right),
$
$
V_y^+=\left(\begin{matrix}
0 & 0   & v_y\\
0 & 0   & v_y\\
0 & 0   & 0
\end{matrix}\right),
V_y^-=\left(\begin{matrix}
0 & 0   & 0\\
0 & 0   & 0\\
u_y & u_y & 0
\end{matrix}\right).
$

We first show why switching from $y$-PBC to $y$-OBC i.e. by gradually disconnecting the $y$-boundary hoppings can profoundly affect the $x$-skin reversal. Mathematically, this can be explained through the evolution of the spectral loops as the system interpolates between $y$-PBC to $y$-OBC. We implement this by introducing boundary hopping tuning ratios $\beta_x,\beta_y$ that interpolates between x,y-PBCs (i.e $\beta_x,\beta_y=1$) and x,y-OBCs (i.e $\beta_x,\beta_y=0$), as explicitly written in real-space as $H_{\text{Kagome}}(\beta_x,\beta_y)=$
\begin{equation}
  \begin{aligned} 
     \sum_{x,x',y,y'} &[V\delta_{x,x'}\delta_{y,y'} + V_{x}^{+}\delta_{x',x+1}\delta_{y,y'} + V_{x}^{-}\delta_{x',x-1}\delta_{y,y'} \\ & + V_{y}^{+}\delta_{x,x'}\delta_{y',y+1} + V_{y}^{-}\delta_{x,x'}\delta_{y',y-1} \\ & + \beta_x\delta_{y,y'}(\delta_{x,L_x}\delta_{x',1}V_x^+ + \delta_{x,1}\delta_{x',L_x}V_x^-) \\ & + \beta_{y}\delta_{x,x'}(\delta_{y,L_{y}}\delta_{y',1}V_{y}^{+} + \delta_{y,1}\delta_{y',L_{y}}V_{y}^{-})]|x,y\rangle\langle x',y'|
  \end{aligned} 
  \label{eq:kagome_model}
  \end{equation}

\noindent Fig.~\ref{fig:zoomin-loop} shows how tuning from $y$-PBC ($\beta_y=1$) to $y$-OBC ($\beta_y=0$) leads to the appearance of complex $x,y$-OBC eigenvalues that are skin-reversed (blue, $x$-IPR$<0$). We simultaneously plot two spectra in every plot: the desired $x$-OBC ($\beta_x=0$, darker and thicker) spectrum, and its reference partial $x$-OBC ($\beta_x=0.2$, lighter and thinner) spectrum which encircles it according to the direction of $x$-skin accumulation~\footnote{We choose not to use $\beta_x=1$ because the $x$-PBC loops are too large for the $x$-OBC spectral details to be visible.}; if a $\beta_x=0.2$ spectral loop flips, the $\beta_x=0$ skin modes within it would reverse their direction of accumulation.

Initially, when $\beta_y\simeq 1$, both spectra are dominated by non-reversed right $x$-skin states (red). As $\beta_y$ is reduced, a sliver of left-skin modes (blue, negative $x$-IPR$_i$) appears, but are strictly confined to the real line. With even smaller $\beta_y\approx 2\times 10^{-4}$, these blue reversed states inflates into the complex $E$ plane, eventually forming a narrow annulus with nonzero $\mathrm{Im}\,E$. Beyond $\beta_y\lesssim 10^{-4}$, the reversed blue spectral branch becomes the dominant feature while the residual right-skin branch collapses onto the real axis. A fuller analysis (see Appendix B) reveals that the Im$E$ of these reversed states indeed dominates all other states, including those outside of the plotted spectral region. Similar conclusions can hold for many other aspect ratios $L_x/L_y$, though not all due to NHSE competition~\cite{fang2022geometry,wang2023experimental,wan2023observation,qin2024geometry,xiong2024non,yang2025tailoring} (Appendix C), and more detailed discussion of alternative spectral evolution pathways can be found in Appendix D.

In sum, the abovementioned interpolation sequence describes an on-off switch for reversed $x$-NHSE purely by tuning the transverse $y$-boundary connection: restoring $y$-periodicity suppresses skin reversal, whereas disconnecting $y$-boundary hoppings introduces a family of complex $x$-reversed skin modes (blue) that amplifies robustly compared to the real non-reversed modes (red).

Fig.~\ref{fig:kagome-dynamics} explicitly showcases the dynamical consequence of this transverse boundary-controlled skin reversal and amplification. From the full spectra plotted in Fig.~\ref{fig:kagome-dynamics}(a) [see Appendix B for their interpolation], we expect that eventually, there should be no discernible $x$-skin accumulation under $x$-OBC, $y$-PBC, but pronounced reversed left-skin eigenmodes (blue, negative $x$-IPR$_i$) when the $y$-boundary is opened. 
This is confirmed by evolving a wave packet initialized at the sample center. As summarized by the longitudinal center of mass $x_{\mathrm{cm}}(t)=\sum_{x,y} x|\Phi(x,y,t)|^2/\sum_{x,y}|\Phi(x,y,t)|^2$ motion in Fig.~\ref{fig:kagome-dynamics}(b), no significant $x$-skin accumulation occurs when the $y$ direction is periodic (orange). But with $x,y$-OBCs (blue), $x_{\mathrm{cm}}(t)$ first drifts rightward towards larger $x_{\mathrm{cm}}$, but ultimately reverses course and ends up at the left edge near $x_{\mathrm{cm}}=1$.

The detailed evolution of $x_{\mathrm{cm}}(t)$ can be understood through the state density snapshots in Figs.~\ref{fig:kagome-dynamics}(c,d), which feature time instances indicated by the dashed vertical lines in Fig.~\ref{fig:kagome-dynamics}(b). At $t=0$, the wavepacket spreads out and moves downwards and slightly rightwards. But after encountering the lower $y$-boundary at around $t=2.0$, the wavepacket under $x,y$-OBCs becomes $y$-localized and evolves according to the general Brillouin zone in $k_y$. While it may initially appear that the skin evolution is towards the right, the wavepacket subsequently reverses direction to migrate to the left boundary, confirming that the open $y$-boundary triggers $x$-skin reversal due to more robust (Im$E>0$) reversed skin modes. But for $y$-PBCs, no such reversal occurs and the wavepacket spreads out until it becaomes a rather featureless steady state.
The markedly different dynamics in Figs.~\ref{fig:kagome-dynamics}(c) and (d) underscores the pivotal role played by the transverse $y$-boundary, even though the NHSE accumulation of interest is in the $x$ direction.

\section{Conclusion}

In conclusion, we have systematically motivated and demonstrated new robust mechanisms for reversing the non-Hermitian skin effect (NHSE) by adjust the system size or boundary conditions in the perpendicular i.e. transverse direction. This runs contrary to common intuition that the skin states accumulate in the direction of the asymmetric hoppings -- the resultant NHSE pumping is in the same direction as neither the transverse mechanisms nor the longitudinal physical hoppings. Leveraging on the intrinsically non-local nature of the non-Hermitian amplification, our mechanisms can function as non-local switches for the NHSE pumping direction as well as their amplification potential.

Specifically, we demonstrated that simply switching transverse boundary conditions from periodic to open dramatically triggers pronounced skin localization opposite to the hopping asymmetry, accompanied by distinct real-to-complex eigenvalue transitions that enable robust amplification of the reversed channel. This is quantified by our $\overline{x\text{-IPR}}$  metric, as well as dynamical simulations that further validate the phenomenological manifestation of this reversal: wave packets initially propagate according to the hopping asymmetry but subsequently reverse direction and amplify towards the opposite boundary upon encountering the transverse boundaries. Our spectral winding analysis across the transverse PBC to OBC transition elucidates the fundamental topological nature of this phenomenon, underscoring the profound influence of transverse degrees of freedom on NHSE behavior.

The switching and reversal of the NHSE can be experimentally demonstrated and technologically harnessed by dynamically manipulating active elements that engender non-reciprocity. In topolectrical circuits~\cite{hofmann2019chiral,ezawa2019electric,helbig2020generalized,hofmann2020reciprocal,liu2020gain,liu2021non,stegmaier2021topological,zhang2020non,zhang2022observation,shang2022experimental,yuan2023non,stegmaier2024topological,zou2024experimental,zou2025experimental, zhu2023higher,zhang2023electrical,sahin2025topolectrical}, this can be as simple as reconfiguring an operational amplifier or adjusting a variable capacitor to reverse the direction of non-reciprocal ``hopping'', causing the voltage accumulation to flip from one end to the other. This can be similarly performed in photonic systems~\cite{pan2018photonic,xiao2020non,zhu2020photonic,song2020two,ao2020topological} where the localization of light can be switched by altering the modulation parameters that create synthetic gauge fields. This active, on-demand control over the boundary localization of energy or matter is the key to its technological potential, enabling the design of robust topological switches, highly sensitive sensors, and programmable, one-way channels for routing waves, information, or even quantum states~\cite{smith2019simulating,gou2020tunable,koh2022simulation,kirmani2022probing,frey2022realization,chen2023robust,yang2023simulating,iqbal2023creation,shen2023observation,chertkov2023characterizing,koh2023observation,koh2022stabilizing,koukoutsis2024quantum,koh2025interacting}. The demonstrated controllability and sensitivity to system geometry make our mechanism particularly promising for applications in directional amplification~\cite{abdo2013directional,li2017optical,mercier2019realization,wanjura2020topological,ramos2021topological,wanjura2021correspondence}, topological lasers~\cite{bandres2018topological,harari2018topological,zeng2020electrically,yang2022topological,ota2020active}, and asymmetric waveguides~\cite{kulishov2005nonreciprocal,shemuly2013asymmetric,sounas2017non,hu2019routing}, where precise control over directional signal propagation is essential.

As such, our insights carry significant theoretical and practical implications, challenging conventional perspectives on the NHSE and highlighting the necessity of considering transverse dimensions and boundary conditions to fully characterize non-Hermitian Brillouin zones~\cite{yao2018edge,yang2020non,jiang2023dimensional,liu2024non,li2025phase,meng2025generalized}. Recent investigations into higher-dimensional platforms~\cite{li2021imaging,gu2016holographic,liu2010oscillatory,liu2013symmetry,liu2024simulating,song2020two,zhao2025two,kawabata2020higher,zhang2022universal,zhang2021observation,wang2023experimental,schindler2021dislocation}, topological interplay~\cite{yao2018edge,kunst2018a,kunst2019non,yokomizo2019non,li2020critical,ashida2020non,okuma2020topological,bergholtz2021exceptional,longhi2019metal,mandal2020nonreciprocal,zhang2020non,longhi2019topological,weidemann2022topological,weidemann2020topological,yang2024percolation,yang2025beyond}, and novel experimental implementations~\cite{ghatak2020observation,wang2022non,weidemann2020topological,liang2022dynamic,helbig2020generalized,stegmaier2021topological,wang2023experimental,xiao2022bound,koh2022simulation,zhang2023many,shen2023proposal,koh2023measurement} promise to further harness this versatile and powerful reversal mechanism, reinforcing transverse nonlocality as an emerging cornerstone concept in the field of non-Hermitian physics.

\onecolumngrid
\appendix 

\section{Dominant eigenstate reversal from transverse system size and hopping strength}

In the main text, we discussed how non-Hermitian skin effect localization and reversal is sensitive to both the transverse hopping strength $t_0$ and the transverse system size $L_y$ of the model (Eq.~\eqref{eq:H-size} of the main text):
\begin{equation}
  H_{\text{size}}(\bold k)  = \left(e^{ik_x} + \frac{1}{e^{ik_x}}\right) + t_0 \left(e^{ik_y} + \frac{1}{e^{ik_y}}\right) + r e^{i N k_x}e^{ik_y}. \label{apendixeq:H-size}
\end{equation}

Here, to provide a more direct visualization of this phenomenon, we plot the spatial distribution of the dominant eigenstate for different parameter choices.

\begin{figure*}[h]
  \centering
  \subfigure[$L_y=2$]{\includegraphics[width=0.24\textwidth]{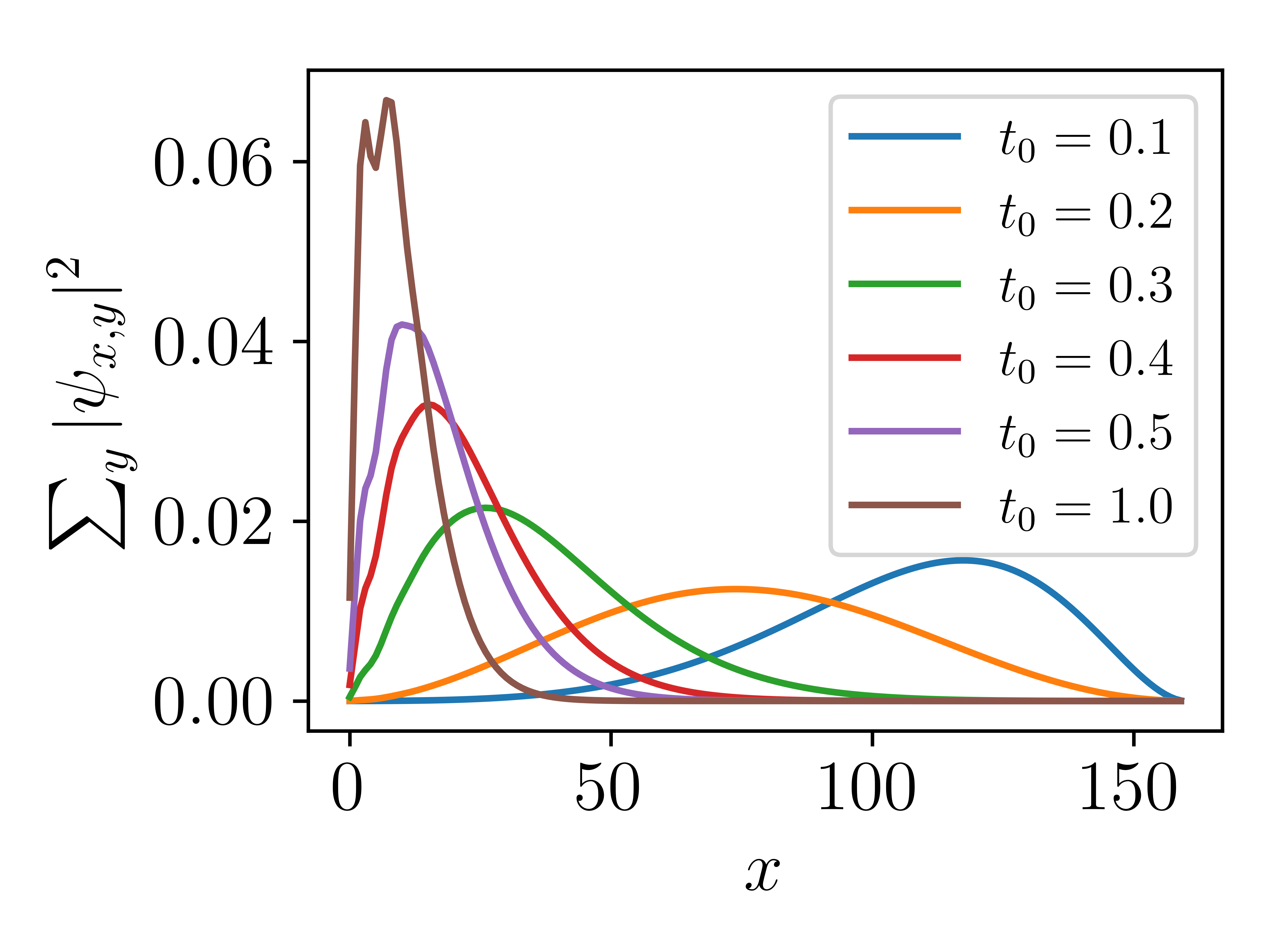}}
  \subfigure[$L_y=3$]{\includegraphics[width=0.24\textwidth]{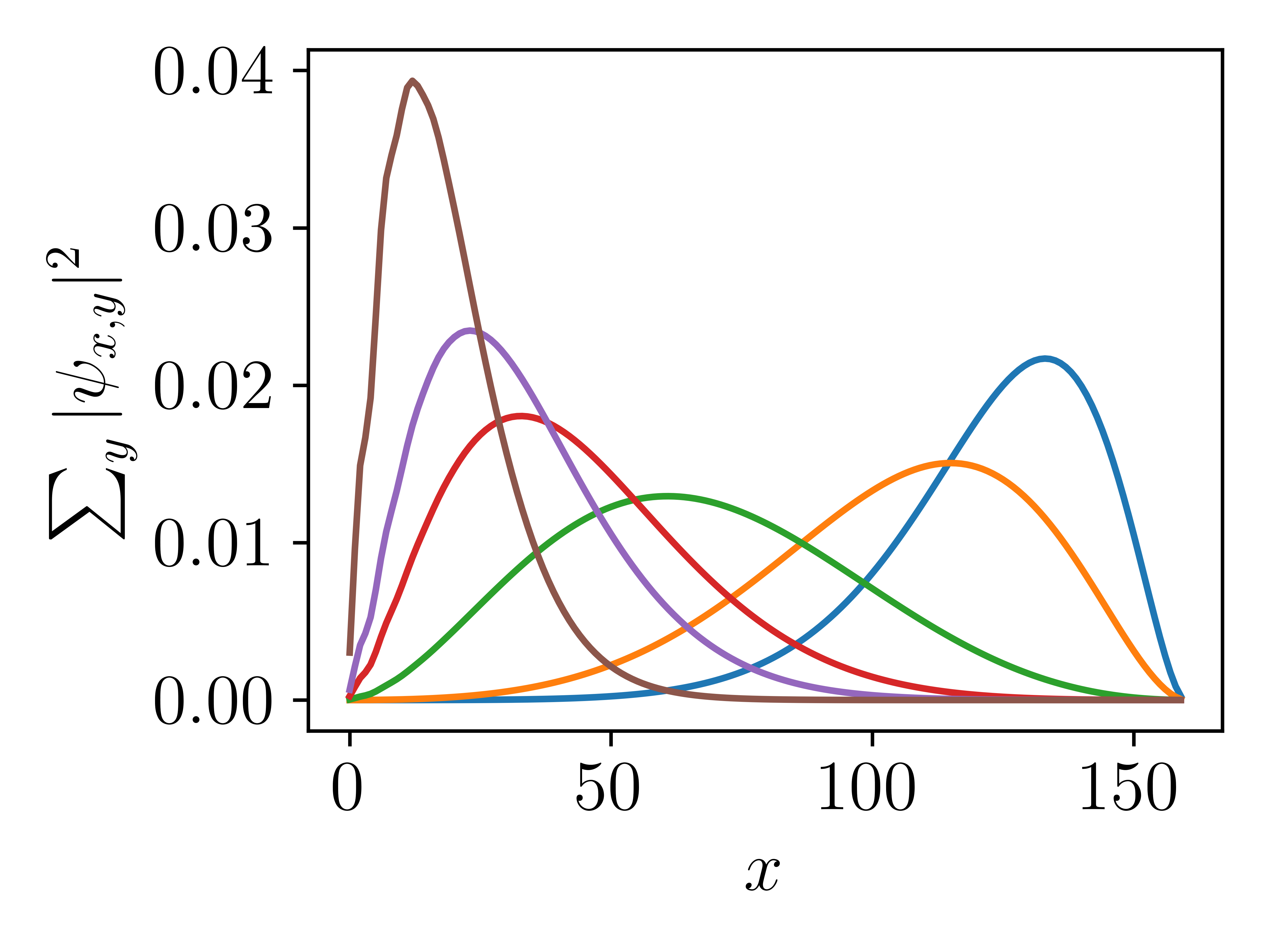}}
  \subfigure[$L_y=4$]{\includegraphics[width=0.24\textwidth]{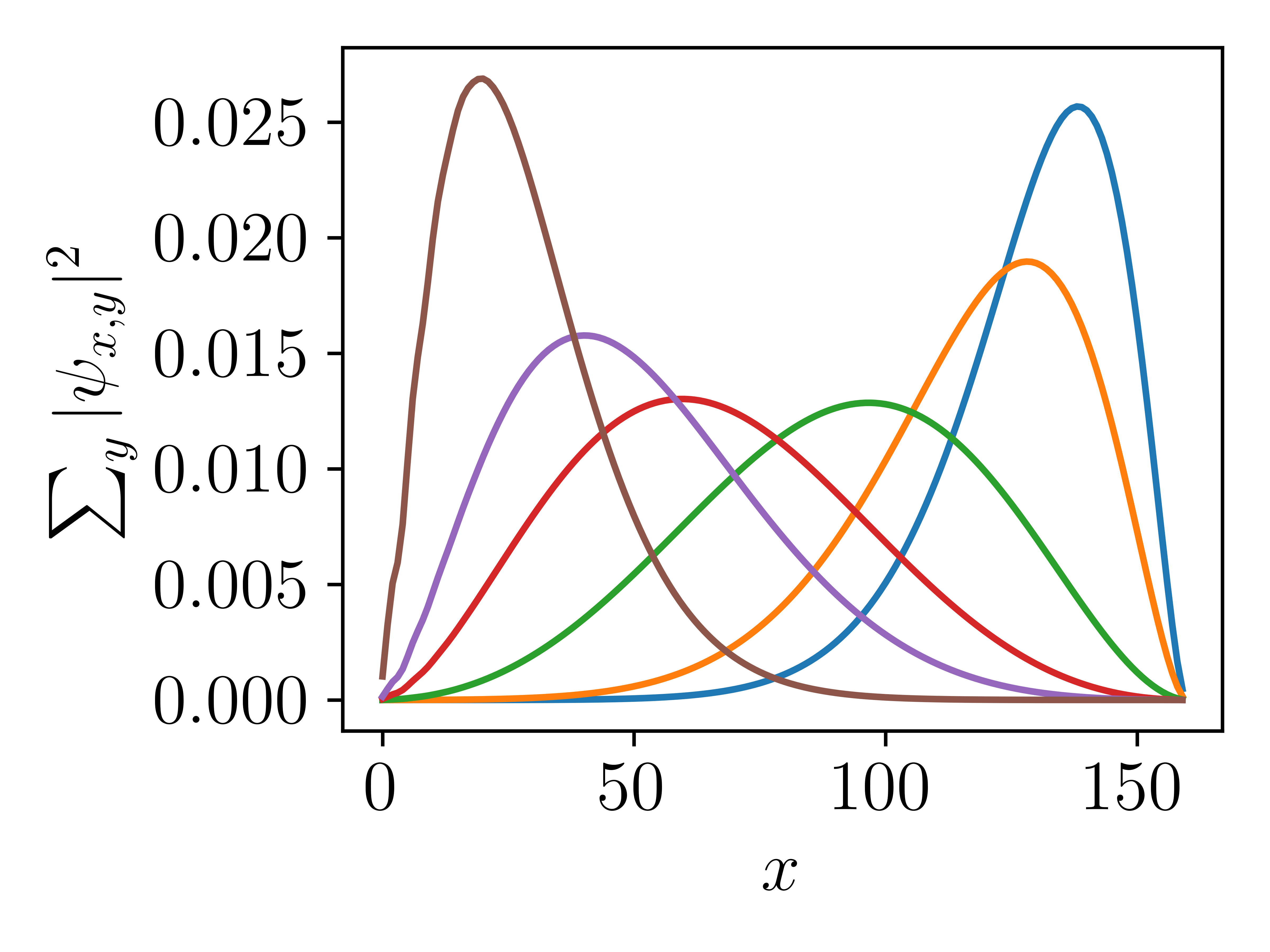}}
  \subfigure[$L_y=10$]{\includegraphics[width=0.24\textwidth]{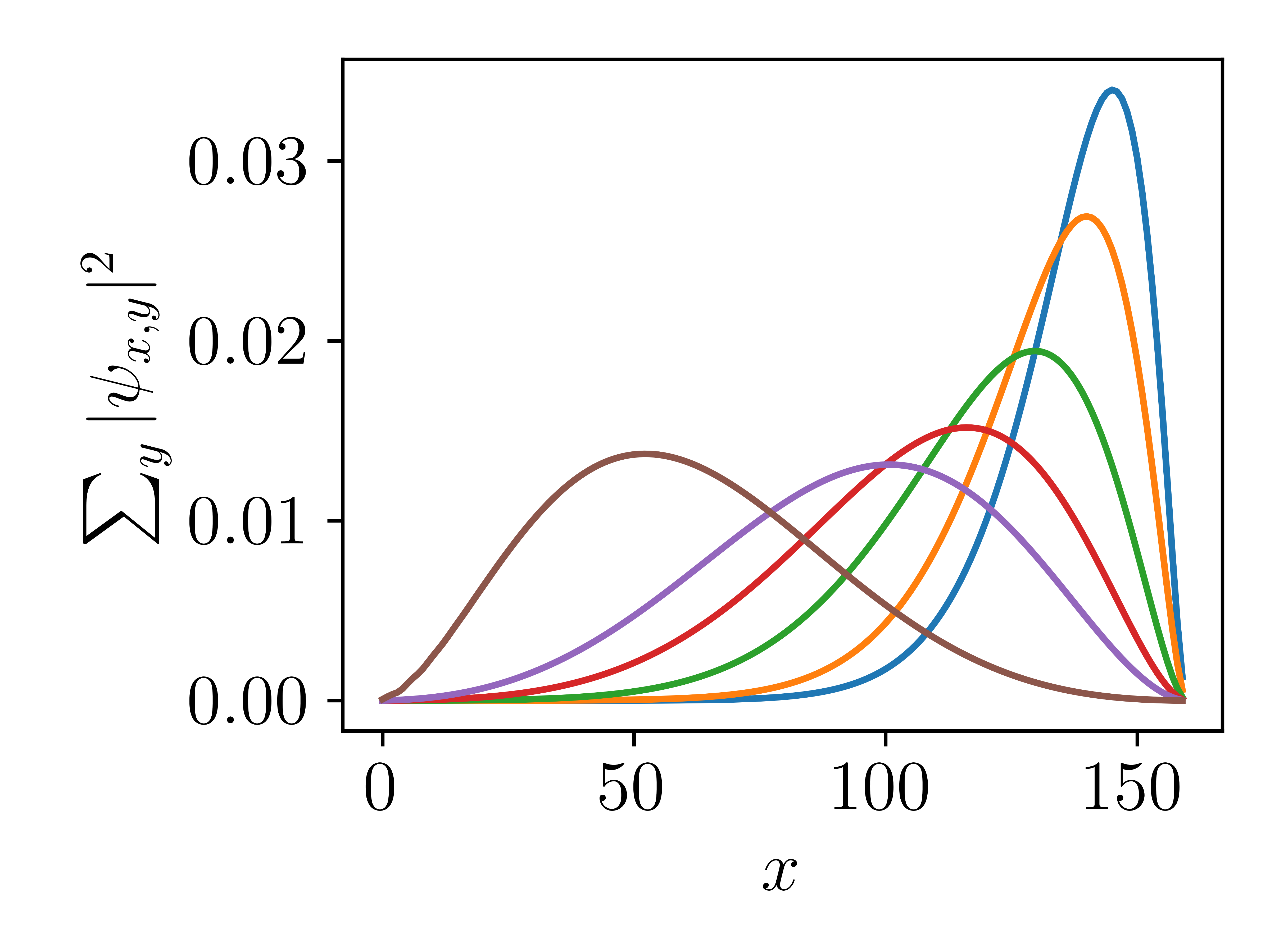}}
  \caption{Effect of varying the transverse hopping $t_0$ and system size $L_y$ on the spatial profile of the skin mode with largest imaginary eigenvalue [the model in main text Eq.~\eqref{eq:H-size} and Eq.~\eqref{apendixeq:H-size}]. The probability density summed over the $y$-direction, $\sum_y\left|\psi_{x, y}\right|^2$, is plotted as a function of position $x$. The different curves correspond to varying transverse hopping strengths, from $t_0=0.1$ (localized on the right) to $t_0=1.0$ (localized on the left). Each panel shows results for a different system width: (a) $L_y=2$, (b) $L_y=3$, (c) $L_y=4$, and (d) $L_y=10$. The plots demonstrate that increasing $t_0$ reverses the localization direction, while increasing $L_y$ weakens the degree of localization. As shown in Fig.~2b, as $L_y$ increases, the reversal is more significant. }
  \label{fig:appendix-t0-effect}
\end{figure*}

Fig.~\ref{fig:appendix-t0-effect} shows the probability density summed over the $y$-direction, $\sum_y\left|\psi_{x, y}\right|^2$, for a representative eigenstate with the largest imaginary eigenenergy. Each panel corresponds to a fixed system width ($L_y=2,3,4$), while the different curves within each panel show the effect of varying $t_0$. Two key trends are immediately apparent. First, within each panel, increasing the transverse hopping $t_0$ induces a reversal in the localization direction, shifting the mode from the right edge ($x \approx 150$) to the left edge ($x \approx 0$). Second, by comparing across the panels (e.g., the brown curve for $t_0=1.0$), we see that as $L_y$ increases, the localization becomes weaker, with the peak of the distribution becoming lower and broader.

\section{The full spectrum evolution from $y$-PBC to $y$-OBC in the Kagome lattice}

\begin{figure*}[h]
  \centering
  \includegraphics[width=0.75\textwidth]{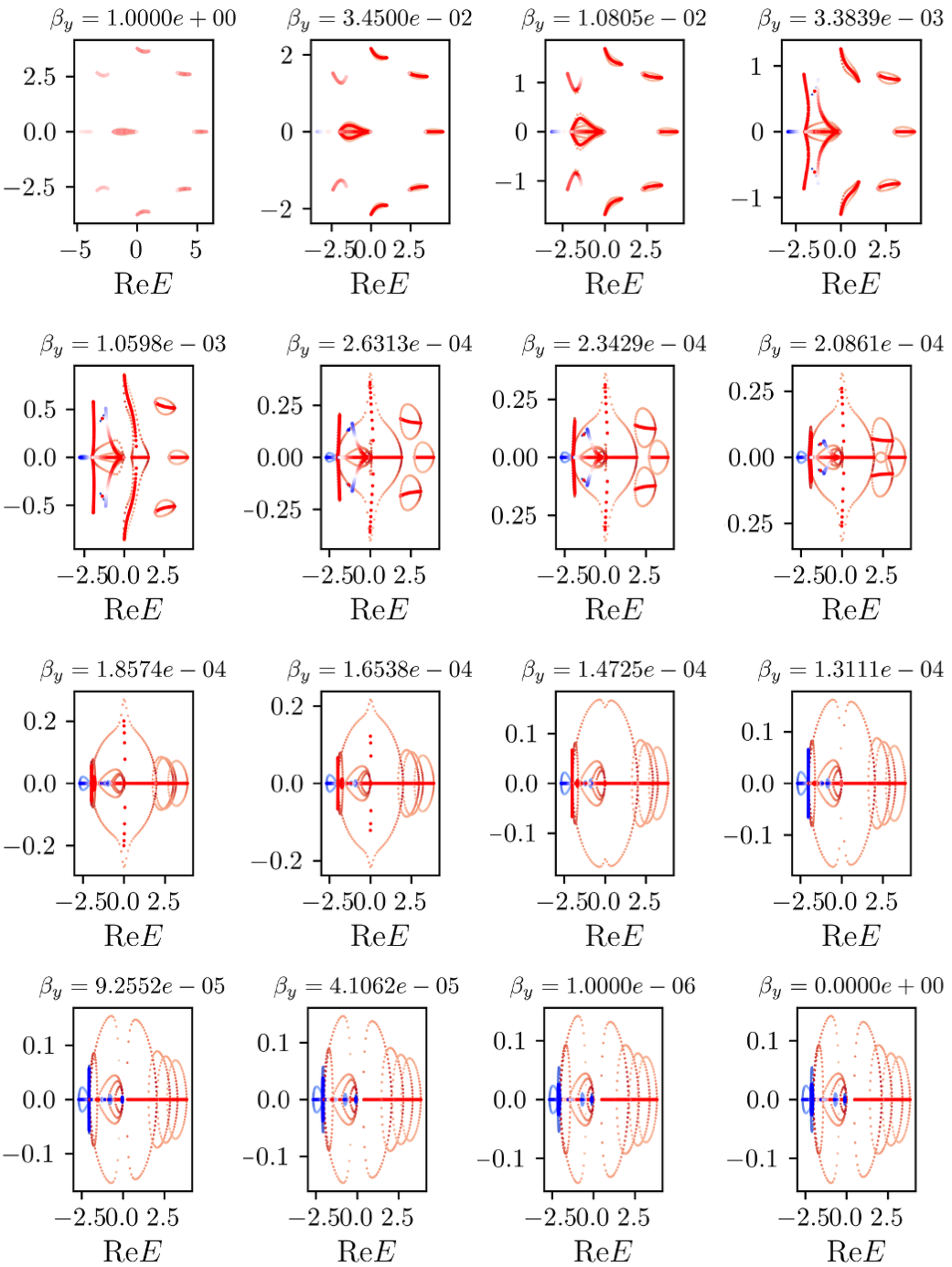}
  \caption{Full complex spectra corresponding to the zoomed panels in Fig.~\ref{fig:zoomin-loop}. Same parameters and conventions as Fig.~\ref{fig:zoomin-loop}. Each panel shows the entire spectrum for $x$-OBC ($\beta_x=0$; darker, thicker dots, colored by $x$-IPR$_i$) together with the not-fully-OBC comparison at $\beta_x=0.2$ (lighter, thinner dots), as the $y$-boundary coupling is reduced from $\beta_y=1$ to $10^{-6}$ (annotated above panels). Parameters: $L_x=120$, $L_y=8$, $u_x=1/v_x=1.1$, $t_c=0.5$, $u_y=1/v_y=3$.}
  \label{fig:appendix-full-spectrum}
\end{figure*}

In the main text, we have shown how the transverse boundary conditions can switch on/off the reversed NHSE in the Kagome lattice in Fig.~\ref{fig:zoomin-loop} with zoomed-in spectral loops. To further illustrate the full spectral evolution, we present in Fig.~\ref{fig:appendix-full-spectrum} the complete complex spectra as the transverse boundary is gradually opened by decreasing $\beta_y$ from $1$ to $10^{-6}$. The progressive emergence and growth of the reversed NHSE modes (blue) and the concurrent suppression of the original right-skin modes (red) are clearly visible, providing a comprehensive view of how the transverse boundary conditions control the NHSE reversal.

\section{Spectral winding and skin reversal under different aspect ratios in the Kagome lattice}

In the main text and preceding appendix sections, our analysis has primarily focused on systems with a fixed, large aspect ratio ($L_x \gg L_y$). However, the manifestation of the NHSE in finite two-dimensional systems can be subtly influenced by the system's geometry. Here, we investigate the effect of the aspect ratio, defined as $L_x/L_y$, on the spectral properties and the skin reversal phenomenon in the Kagome lattice.

The core of this geometric dependence lies in the quantization of the transverse momentum, $k_y$. Under PBC in the $y$-direction, the allowed values of $k_y$ are discrete: $k_y = 2\pi n_y / L_y$, where $n_y = 0, 1, \dots, L_y-1$. The full 2D PBC spectrum is effectively a superposition of $L_y$ one-dimensional spectra, each corresponding to a 1D chain with a fixed transverse momentum $k_y$. The non-Hermitian skin effect in each of these 1D chains can be different, and the overall behavior of the 2D system under open boundary conditions (OBC) arises from the competition between these modes.

\begin{figure*}[h]
  \centering
  \subfigure[]{\includegraphics[width=0.35\textwidth]{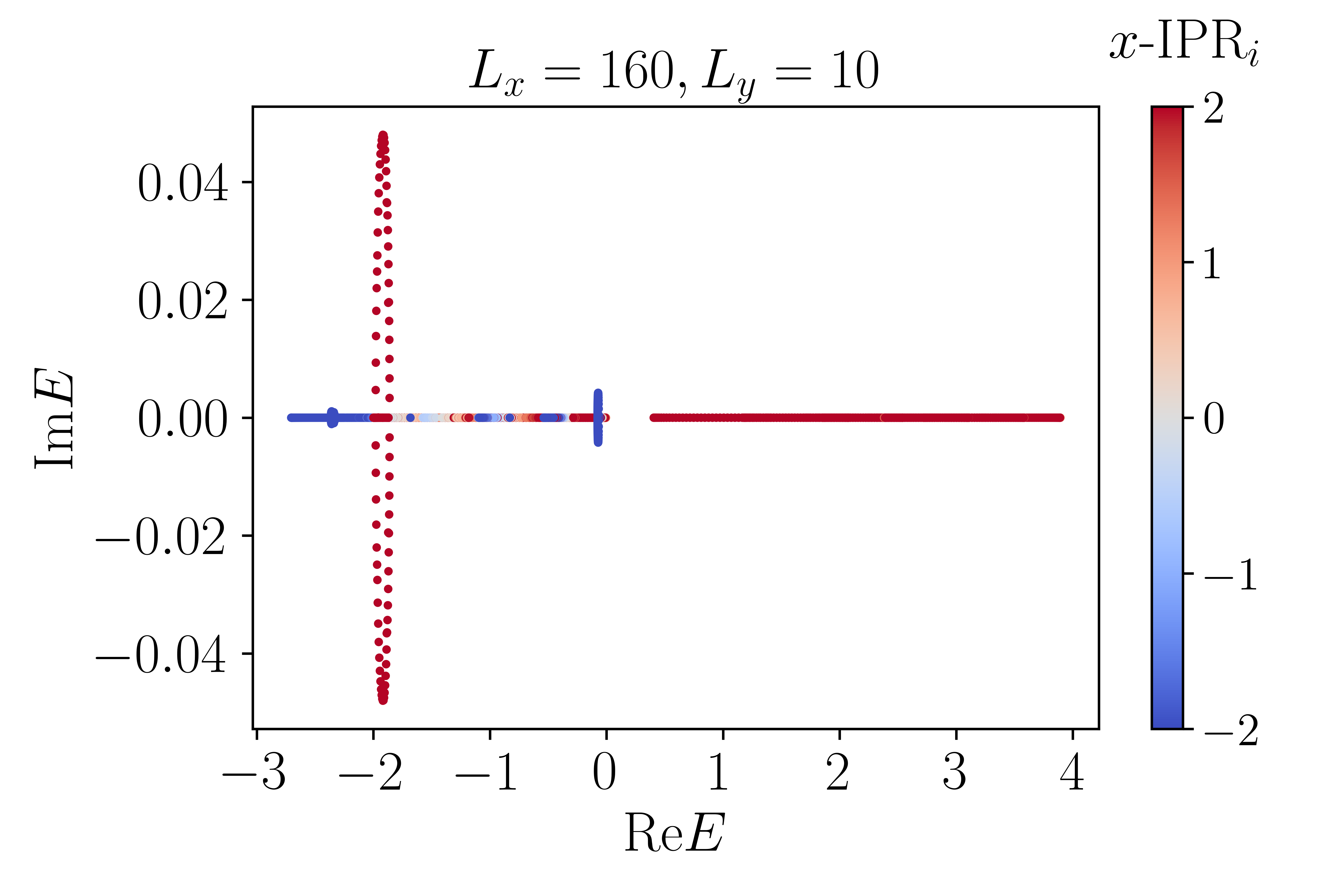}}
  \subfigure[]{\includegraphics[width=0.35\textwidth]{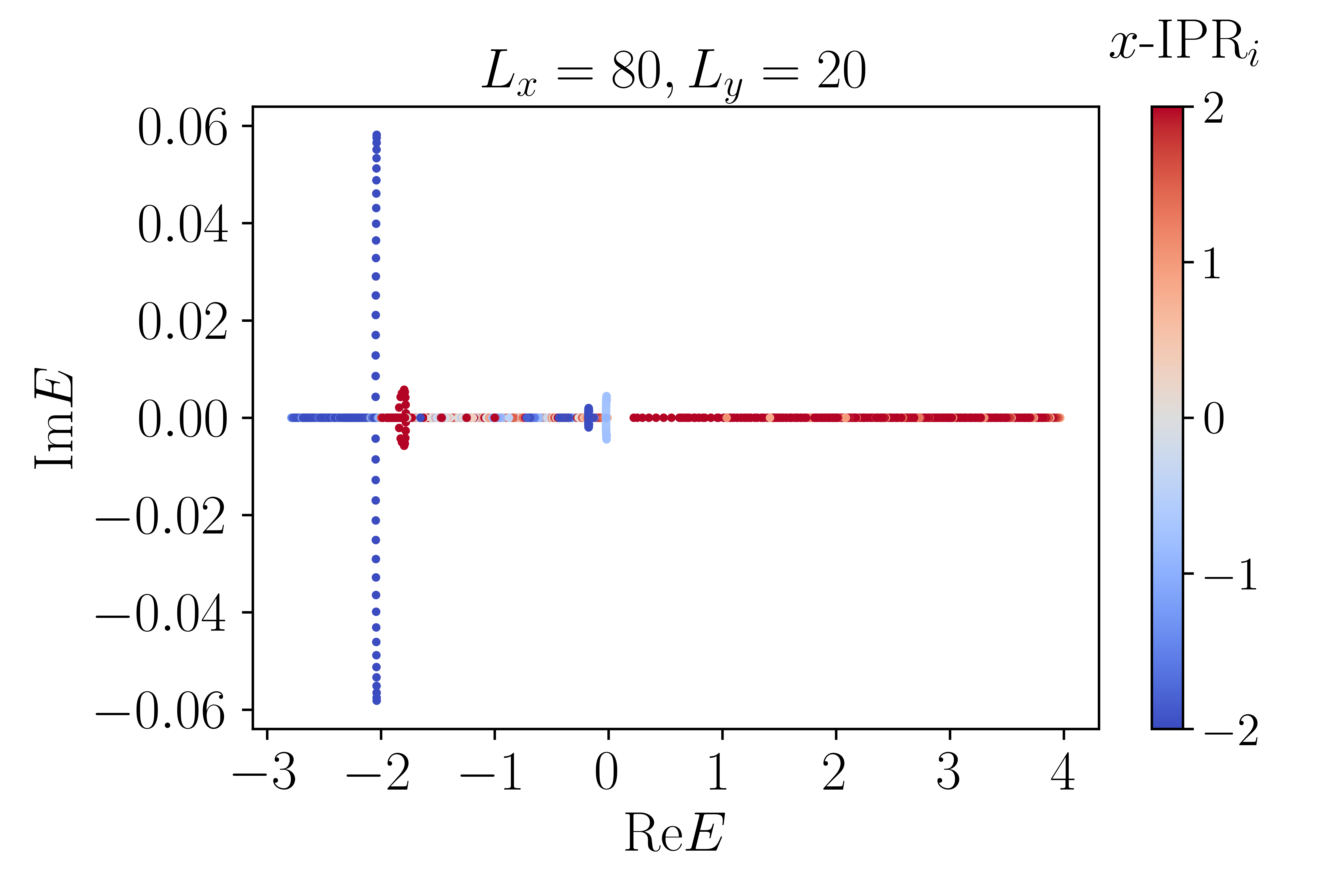}}
  \subfigure[]{\includegraphics[width=0.35\textwidth]{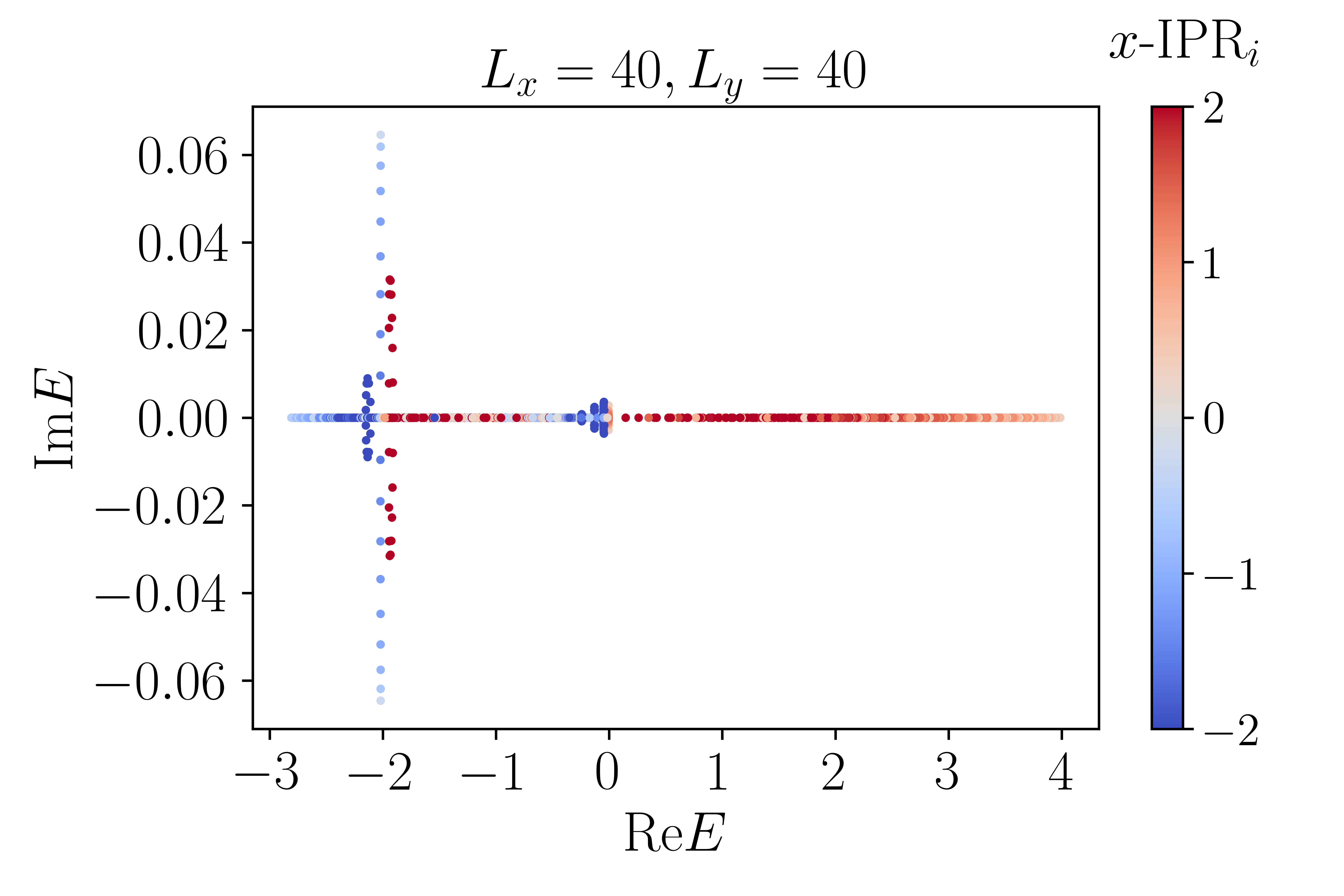}}
  \subfigure[]{\includegraphics[width=0.35\textwidth]{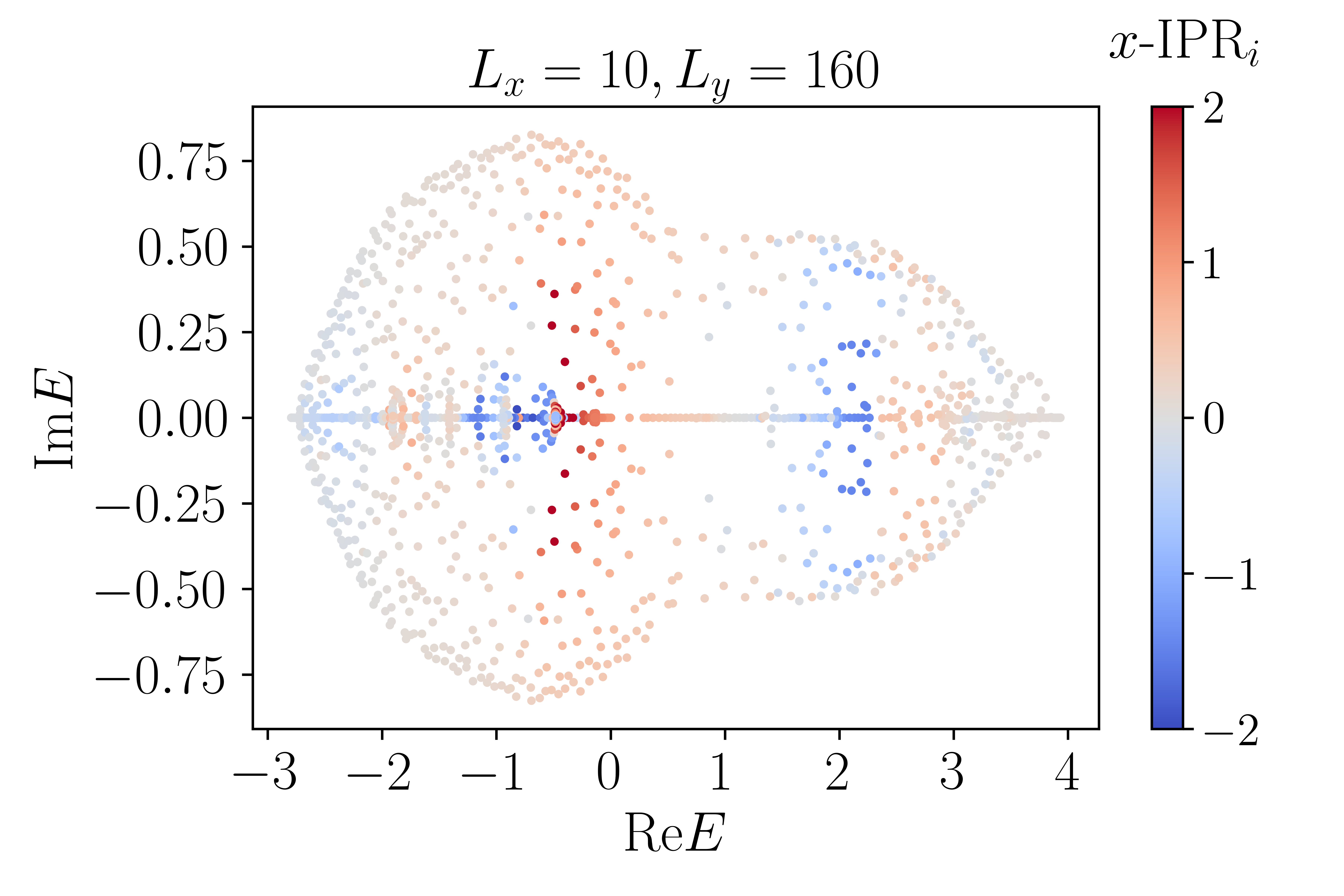}}
  \caption{Comparison of OBC spectra (dots) for differnet systems with a similar number of sites ($3\times L_x\times L_y=3\times 1600$) but different aspect ratios. (a) A long and narrow lattice ($L_x/L_y = 16$) exhibits sharply defined spectral loops corresponding to sparsely quantized $k_y$ values. (b) A lattice with an aspect ratio closer to one ($L_x/L_y = 4$) shows a much denser sampling of the 2D Brillouin zone, leading to PBC spectra that more closely fill the area of the continuous 2D dispersion. The skin reversal behavior under OBC is consequently modified by the geometry. (c) A square-like lattice ($L_x/L_y = 1$) with a dense collection of $k_y$ modes, leading to complex spectral features and a nuanced skin reversal transition. (d) A short and wide lattice ($L_x/L_y =1/16$) where the transverse quantization dominates, resulting in a different interplay of modes and skin localization. Parameters: $u_x=1/v_x=1.1$, $u_y=1/v_y=3$, $t_c=0.5$.}
  \label{fig:appendix-aspect-ratio}
\end{figure*}

To illustrate this, we compare the spectral evolution for two systems with a similar total number of sites but starkly different aspect ratios in Fig.~\ref{fig:appendix-aspect-ratio}.
In the case of a large aspect ratio $L_x/L_y$ (Fig.~\ref{fig:appendix-aspect-ratio}(a) and (b)), the system is quasi-one-dimensional. The spacing between allowed $k_y$ values is large, leading to a PBC spectrum composed of a few, well-separated loops. The skin reversal is governed by how these few transverse modes behave as parameters are tuned.

Conversely, for a system with a smaller aspect ratio ($L_x = L_y$ or $L_x < L_y$, Fig.~\ref{fig:appendix-aspect-ratio}(c) and (d)), the quantization of $k_x$ and $k_y$ is comparable. The PBC spectrum now consists of many closely-spaced loops that more densely sample the area enclosed by the 2D bulk continuum boundary. This dense collection of modes, which reflect the spectral flow resulting from the ``entanglement'' pattern between spectral loops of different PBC directions, can lead to a more complex competition when the boundaries are opened. For instance, the transition from right- to left-localized skin states may become smoother or occur at slightly different parameter values compared to the quasi-1D case.

This analysis underscores the fact that the
precise manifestation of the NHSE and the critical thresholds for its reversal in finite-sized numerical simulations depend on the system's geometry. The aspect ratio effectively controls the crossover from quasi-1D to fully 2D behavior, providing another knob to tune the system's non-Hermitian properties.

\section{Path-Dependence of the Spectral Flow and Boundary Opening Protocol}

In the main text, we established that transverse boundaries and couplings can qualitatively alter the reversal of NHSE. To provide a deeper, mechanistic explanation for this phenomenon, we analyze the protocol of opening system boundaries. The evolution from a double PBC system to a double OBC one is not unique, and the path taken reveals the underlying competition between directional skin effects.

\begin{figure*}
  \centering
  \subfigure[]{\includegraphics[width=0.23\textwidth]{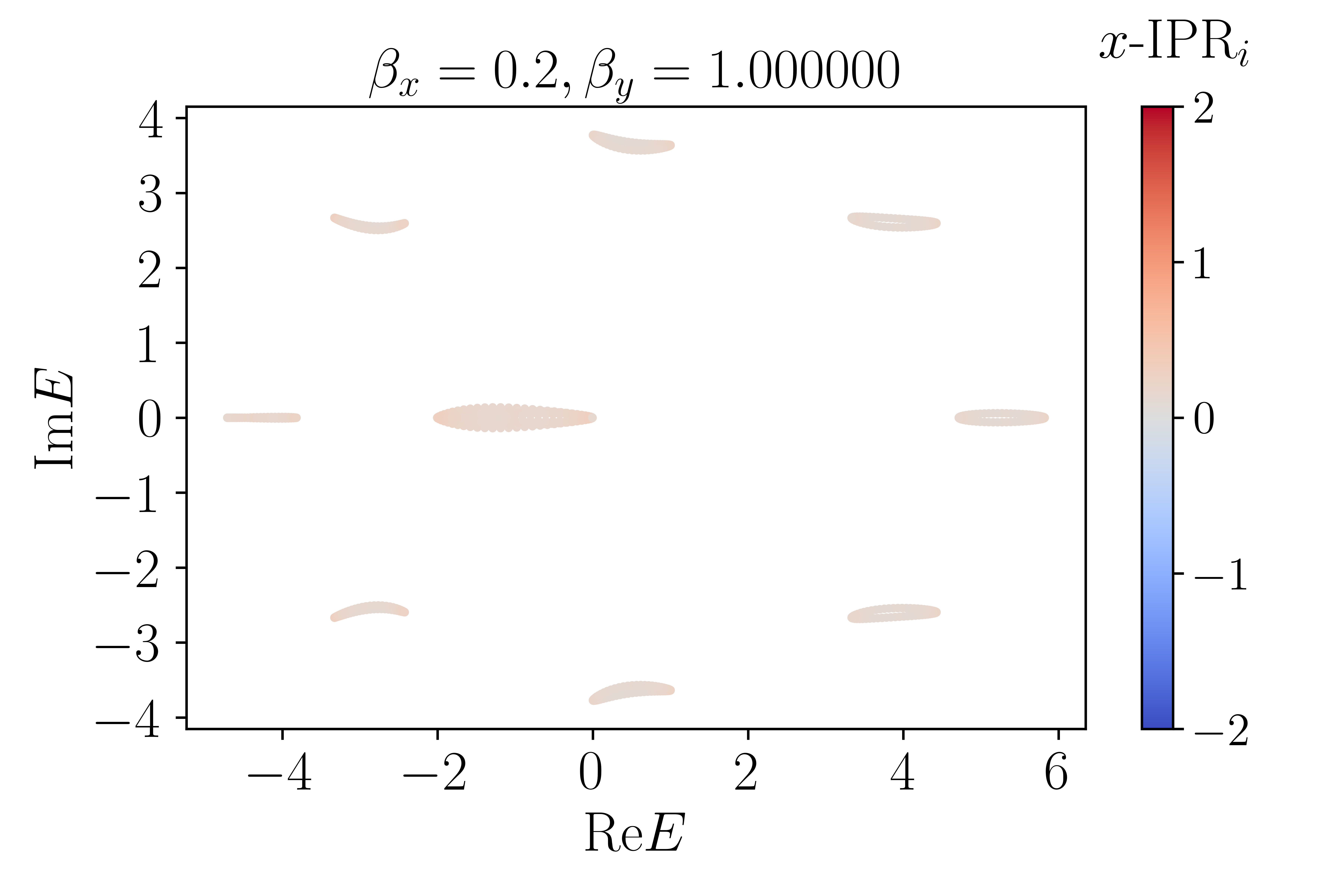}}
  \subfigure[]{\includegraphics[width=0.23\textwidth]{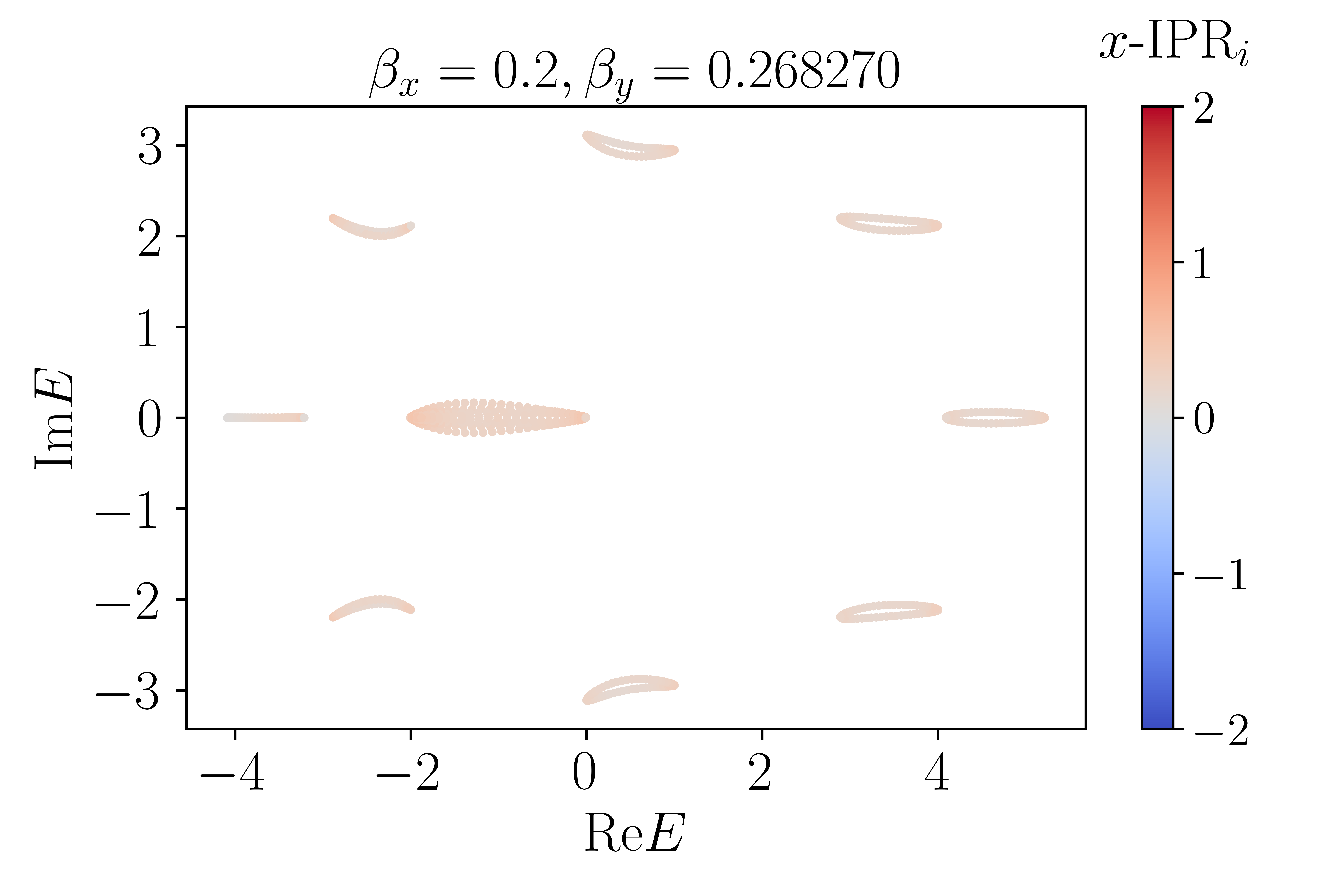}}
  \subfigure[]{\includegraphics[width=0.23\textwidth]{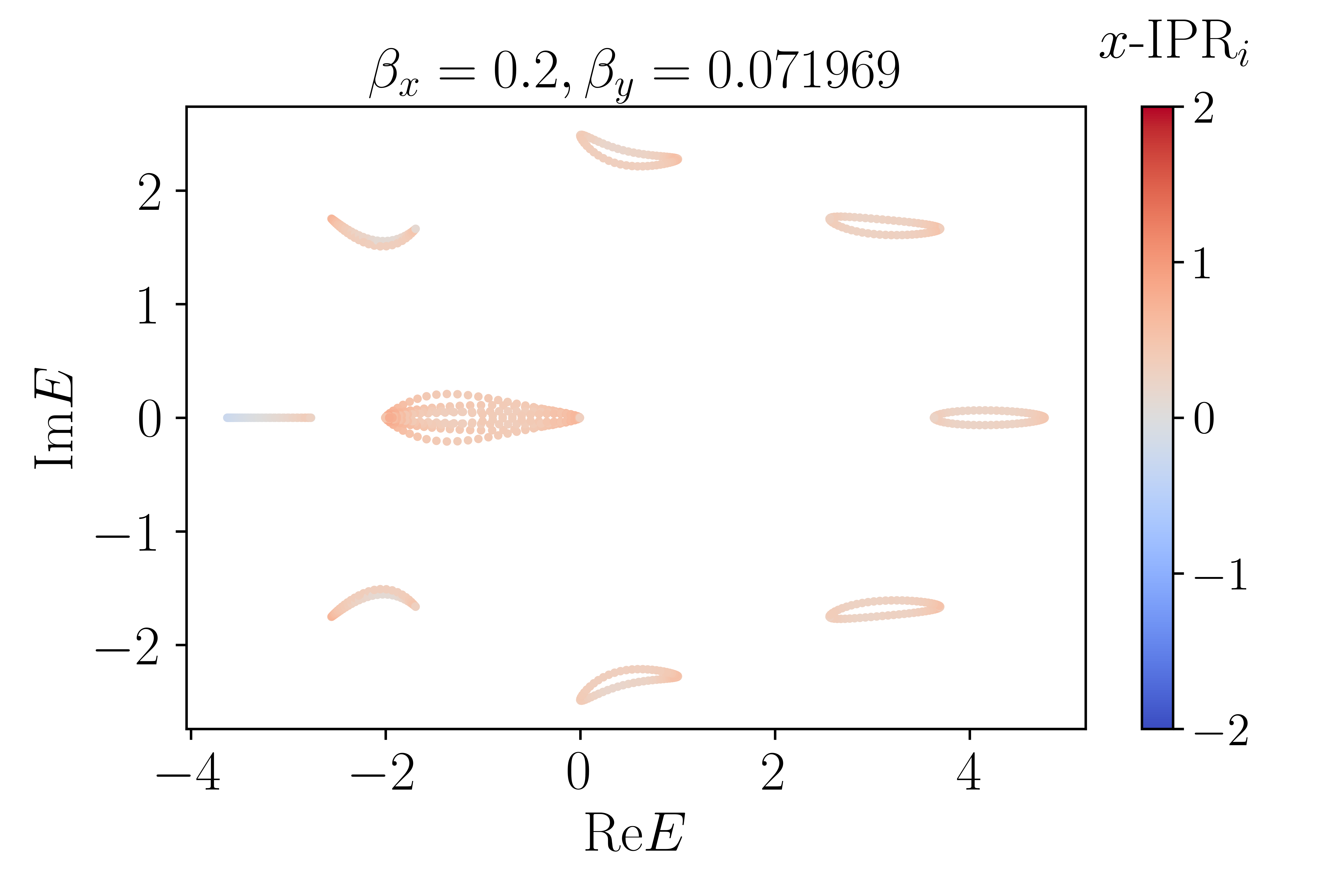}}
  \subfigure[]{\includegraphics[width=0.23\textwidth]{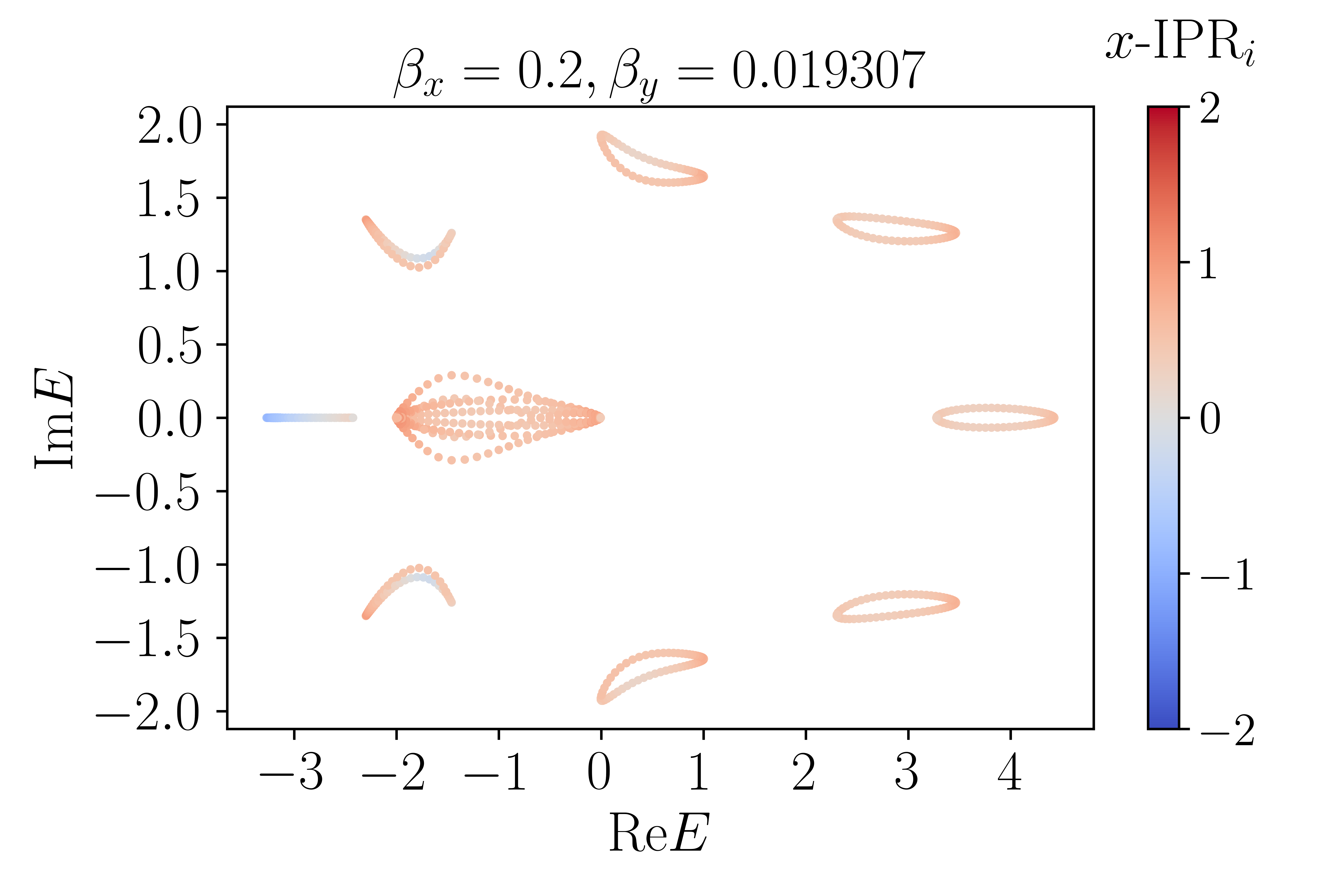}}
  \subfigure[]{\includegraphics[width=0.23\textwidth]{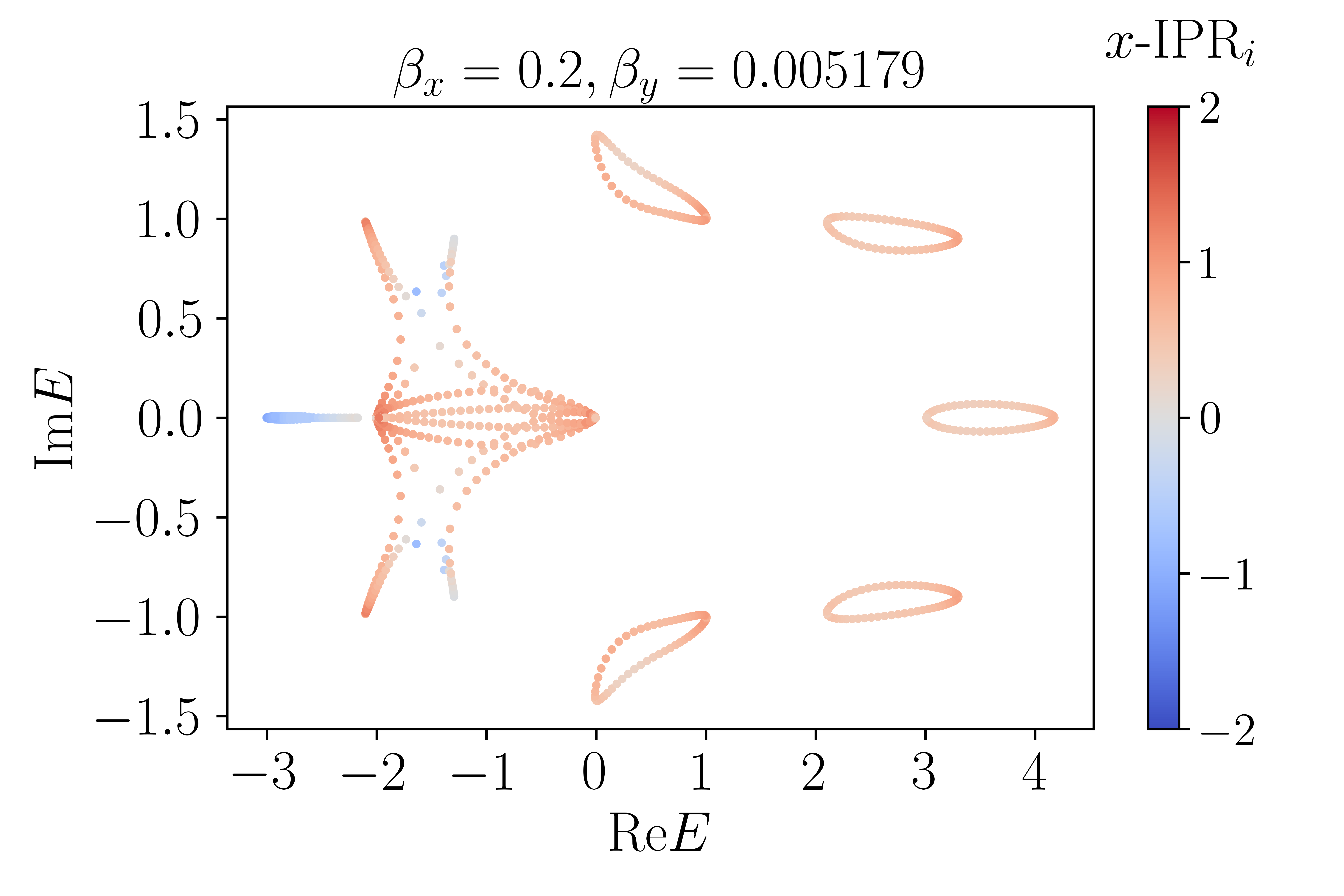}}
  \subfigure[]{\includegraphics[width=0.23\textwidth]{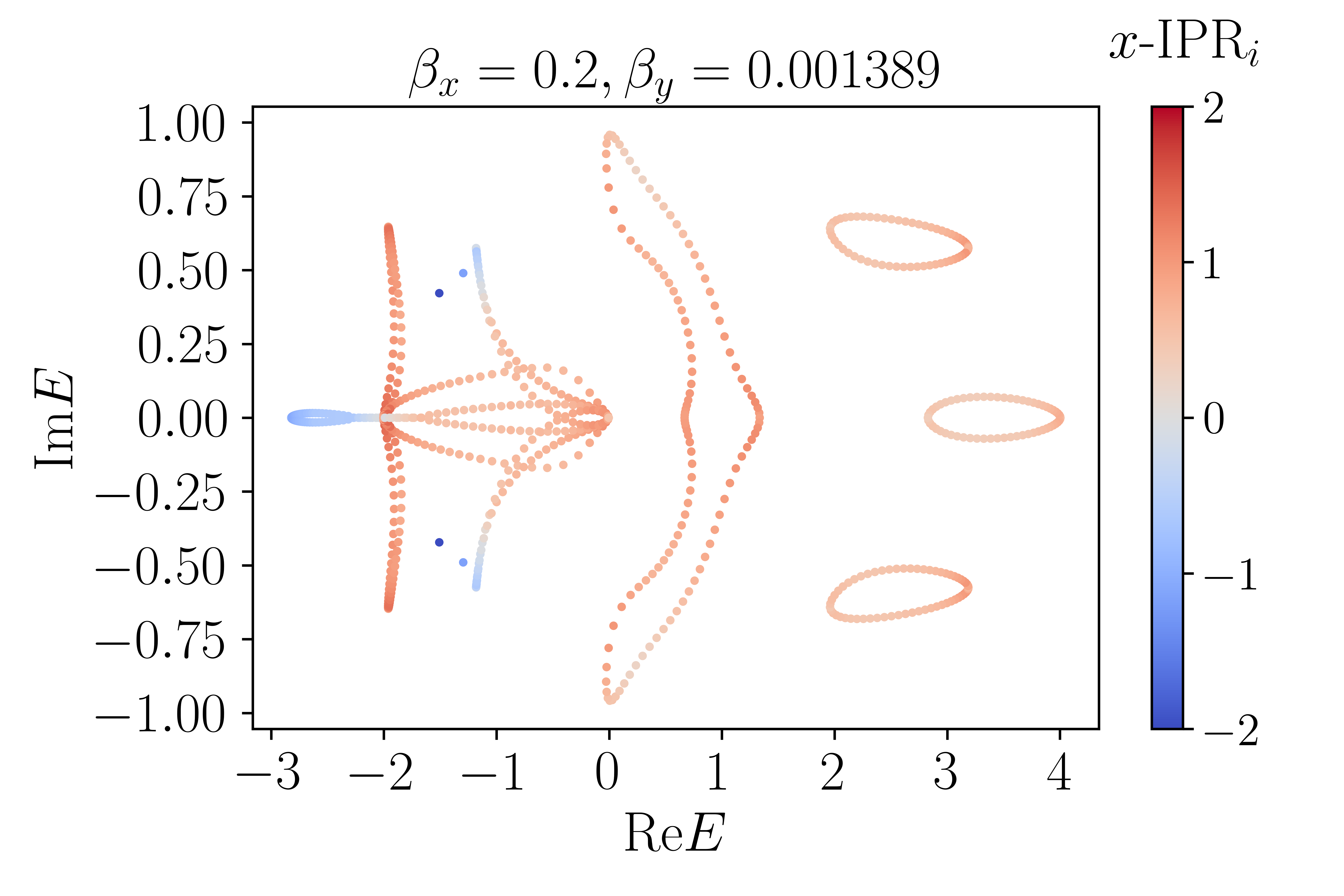}}
  \subfigure[]{\includegraphics[width=0.23\textwidth]{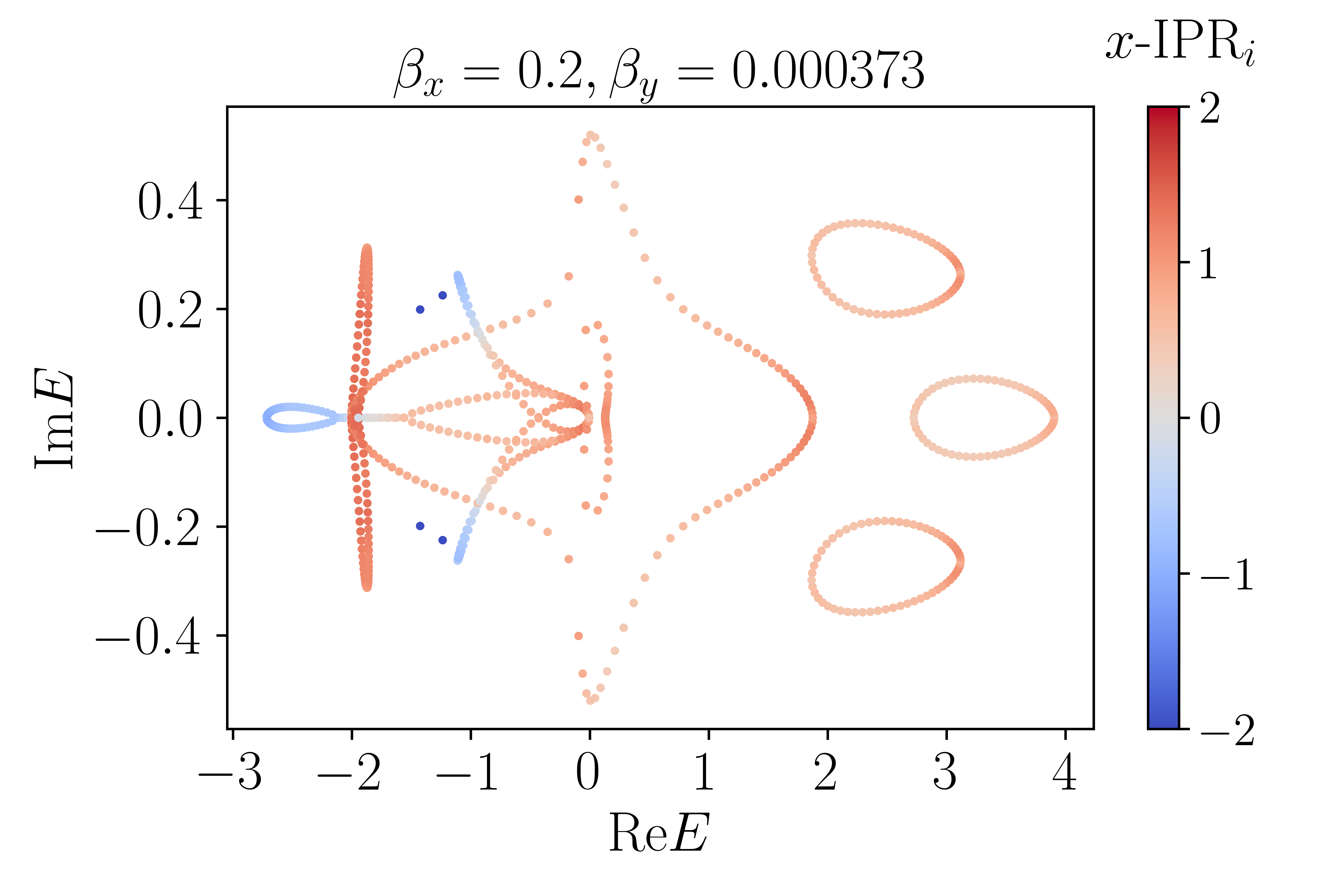}}
  \subfigure[]{\includegraphics[width=0.23\textwidth]{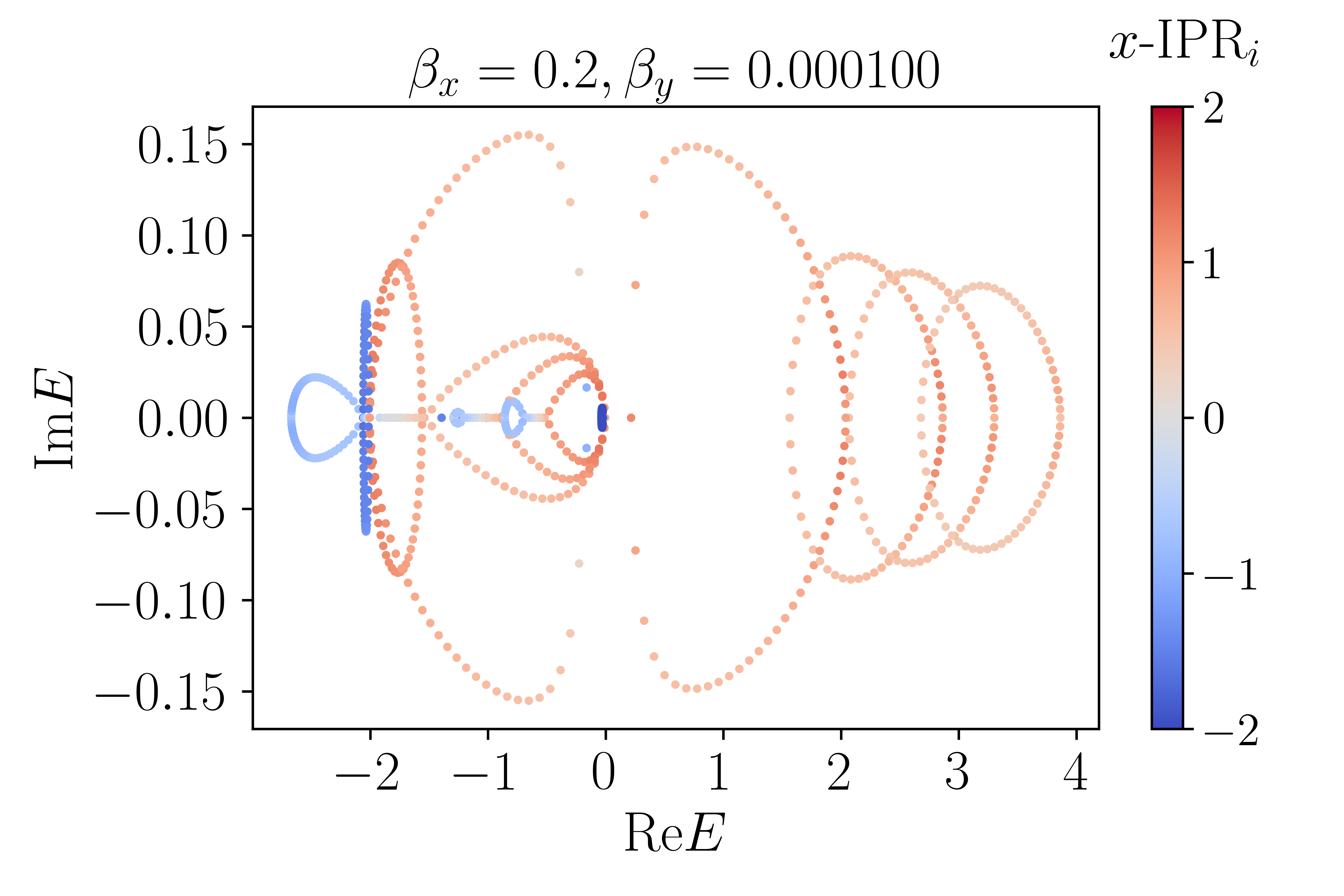}}
  \caption{Geometric Spectral Reconfiguration (Path A). Here, the longitudinal boundary is held nearly open ($\beta_x=0.2$, reflecting the quasi-1D nature of $L_x \gg L_y$), while the transverse boundary coupling $\beta_y$ is tuned from PBC to OBC. As $\beta_y$ decreases, the spectral loops corresponding to different transverse momenta $k_y$ undergo a dramatic \textbf{geometric transformation}: they deform, merge, and reconfigure the topology of the complex spectrum. The NHSE reversal is a direct consequence of this geometric evolution, as seen by the color change from red to blue in panels (d-h) which occurs concurrently with the loop fusion.
  Parameters: $L_x=120$, $L_y=8$, $u_x=1/v_x=1.1$, $t_c=0.5$, $u_y=1/v_y=3$.}
  \label{fig:appendix-path-dependence1}
\end{figure*}

\begin{figure*}
  \centering
  \subfigure[]{\includegraphics[width=0.23\textwidth]{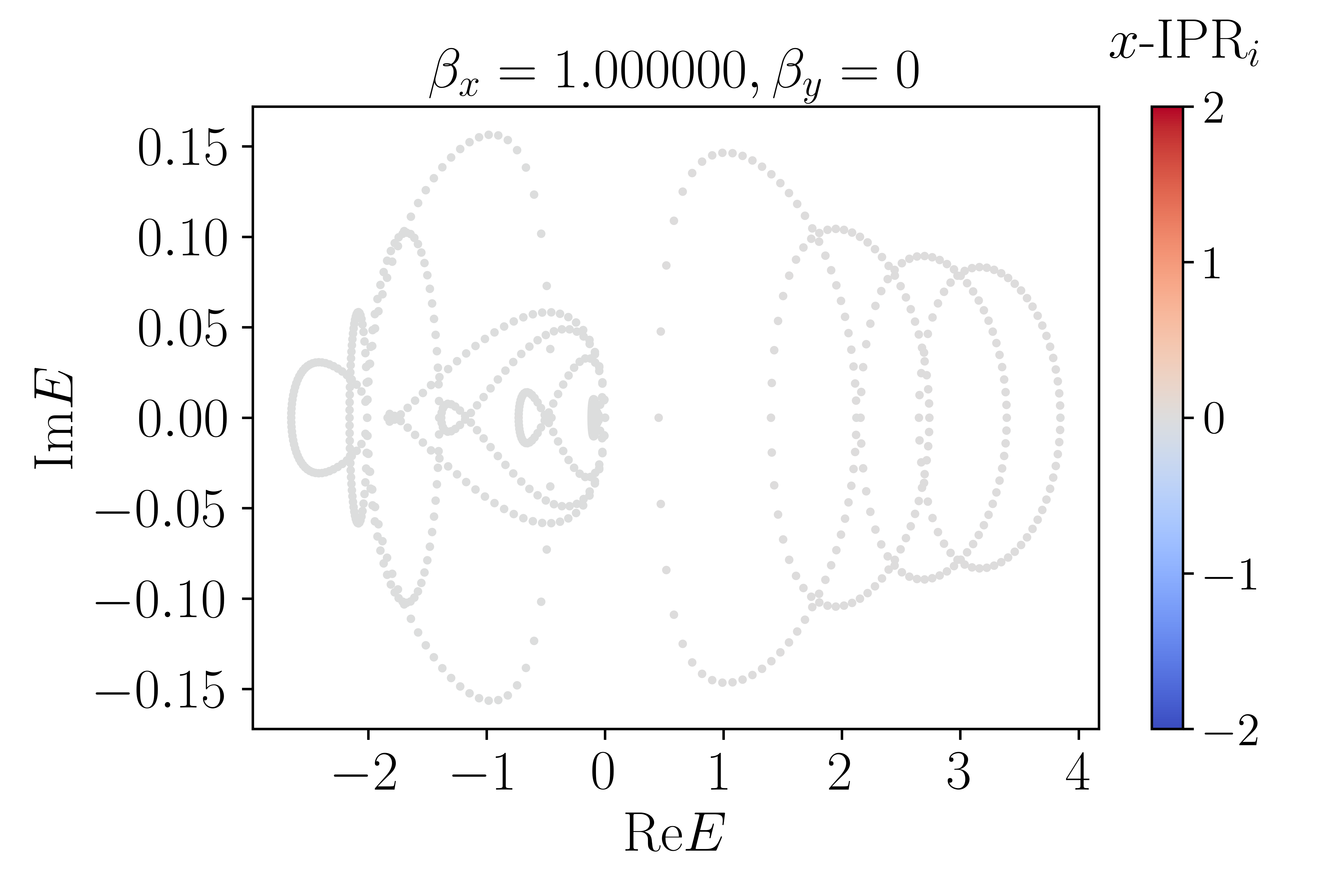}}
  \subfigure[]{\includegraphics[width=0.23\textwidth]{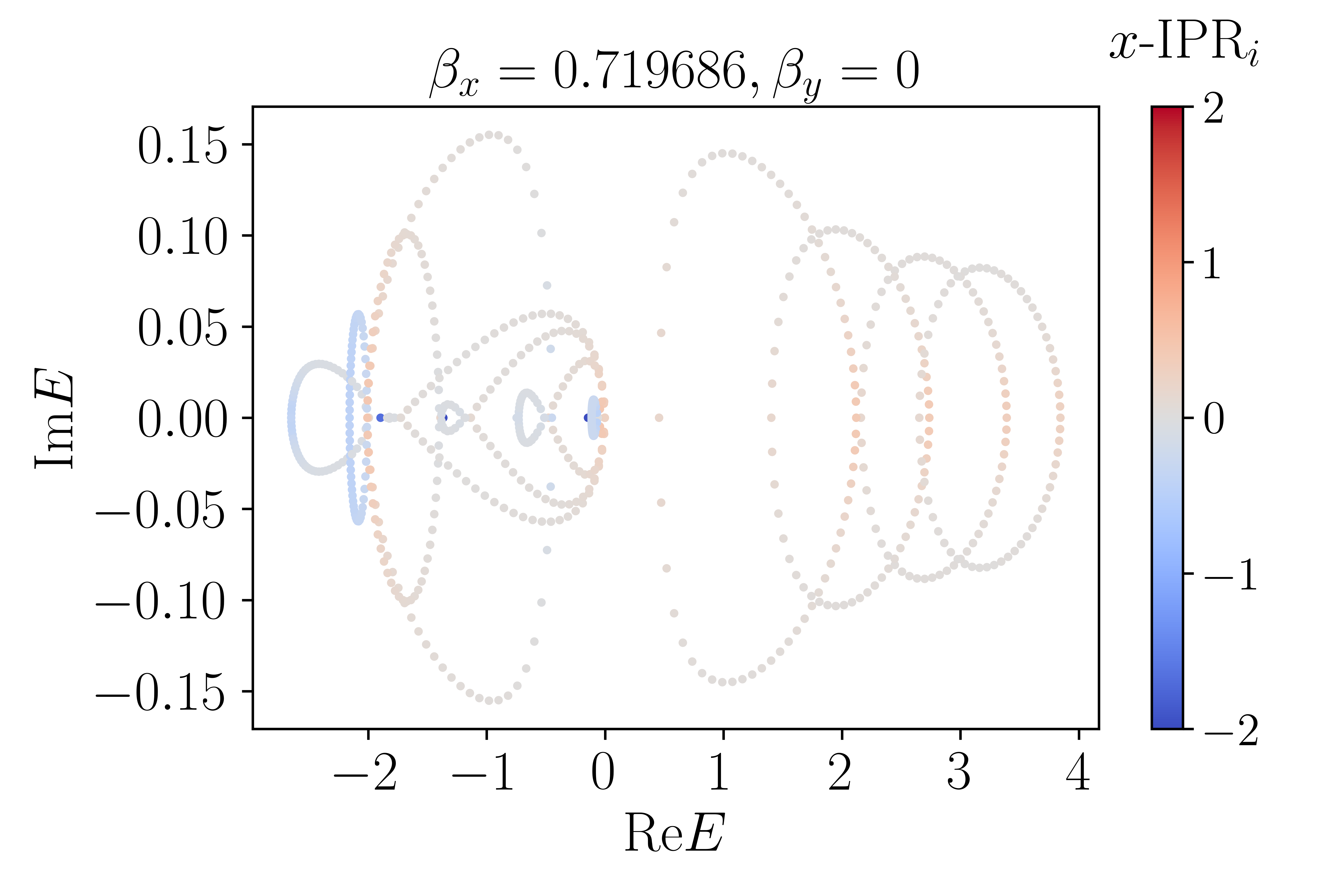}}
  \subfigure[]{\includegraphics[width=0.23\textwidth]{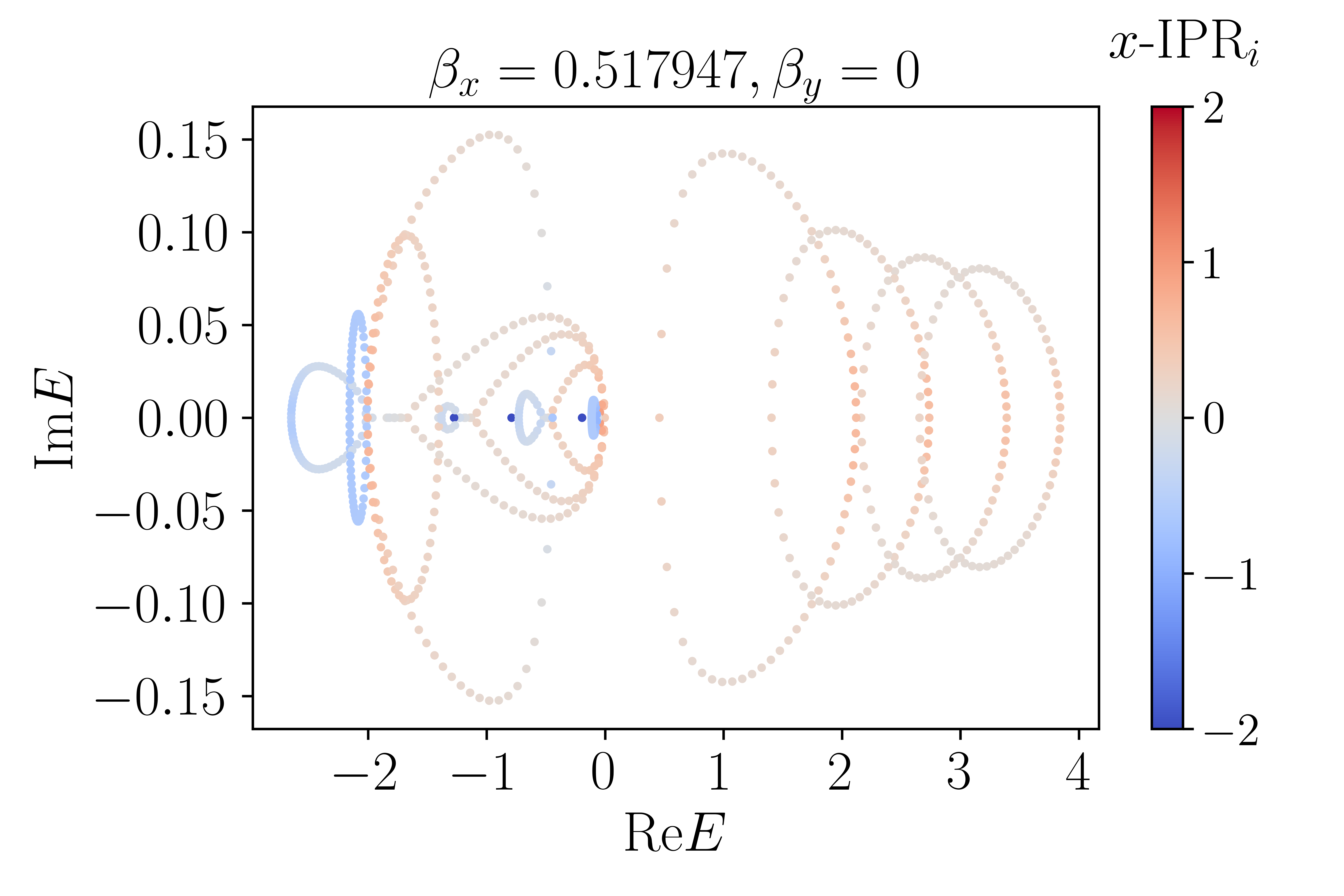}}
  \subfigure[]{\includegraphics[width=0.23\textwidth]{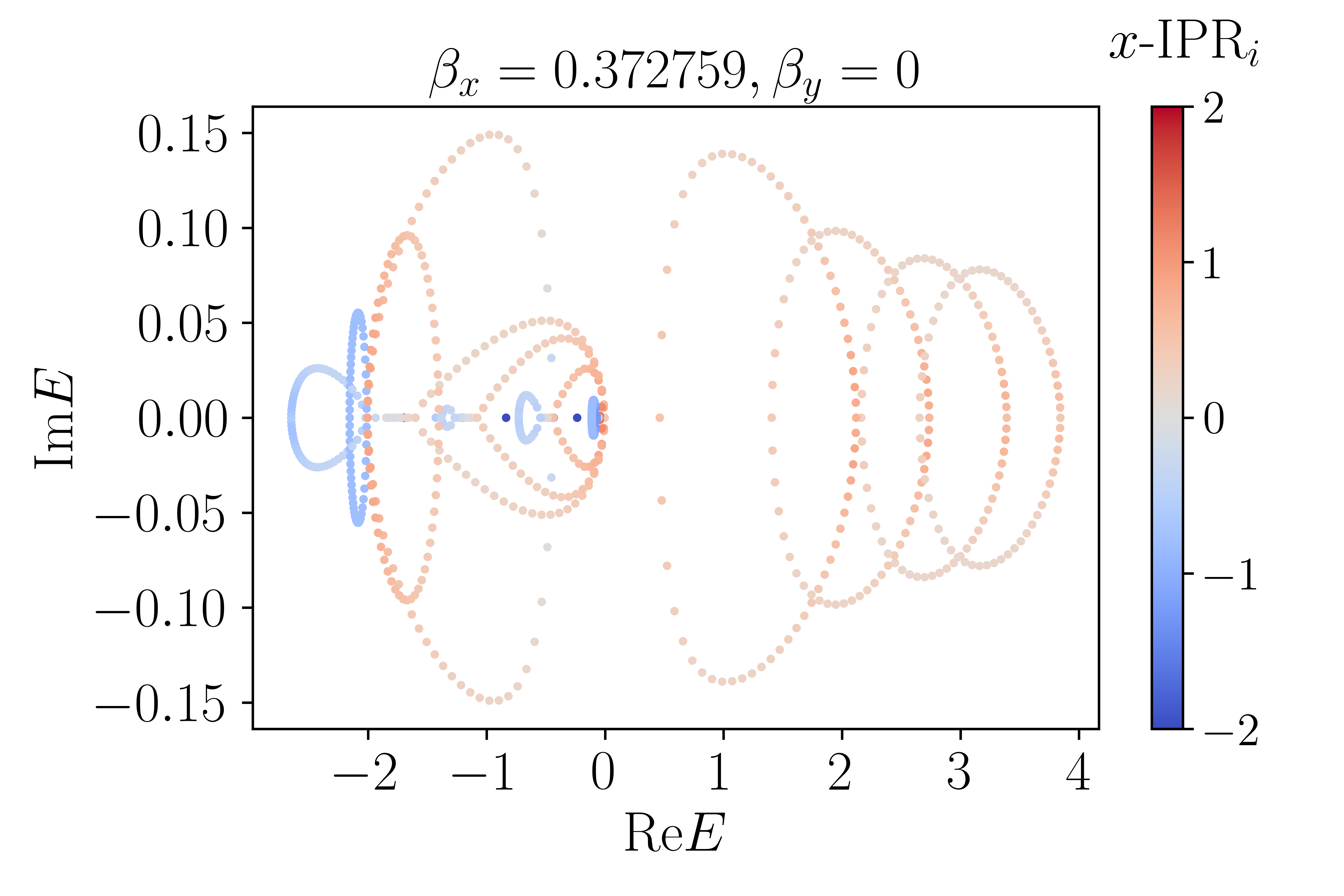}}
  \subfigure[]{\includegraphics[width=0.23\textwidth]{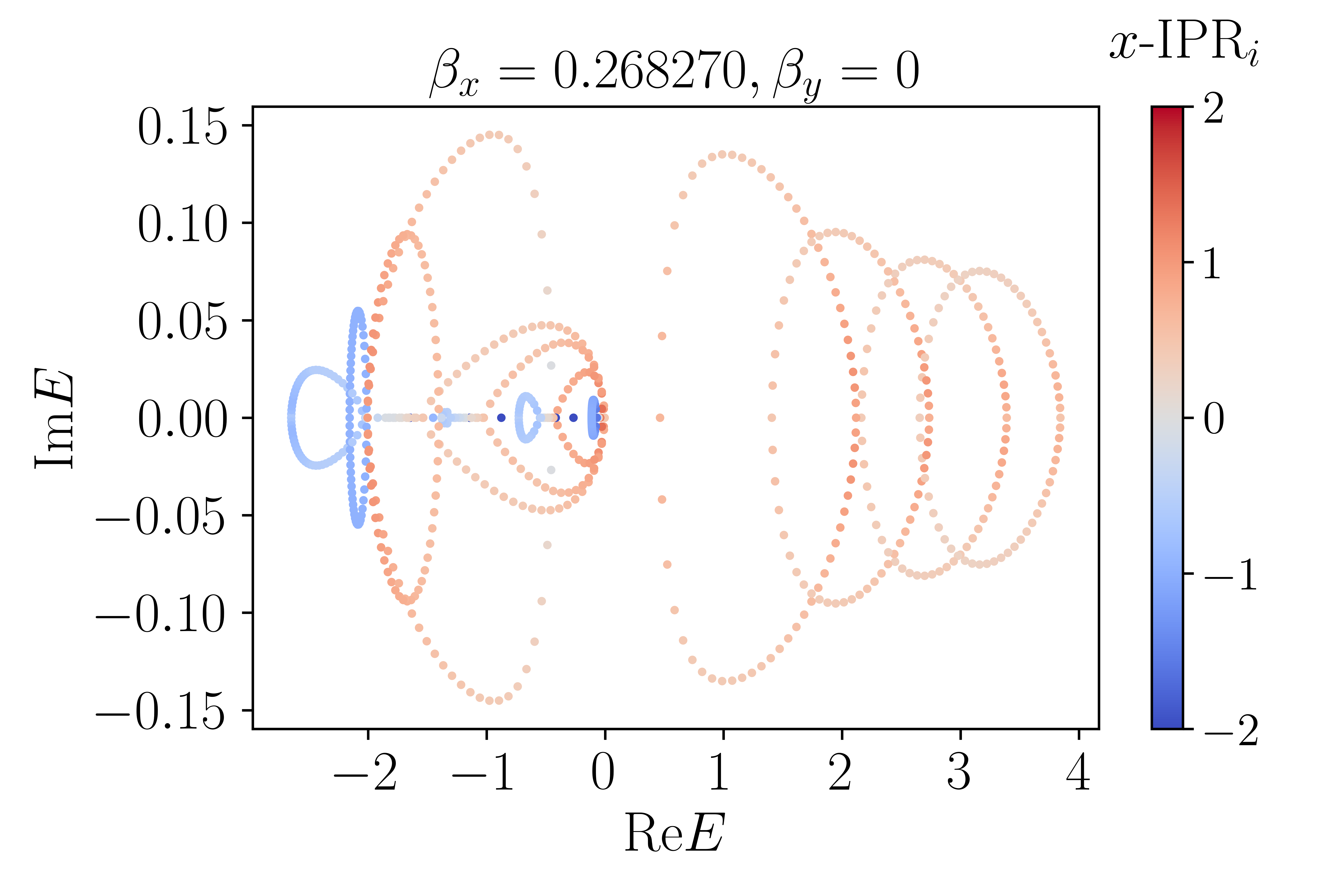}}
  \subfigure[]{\includegraphics[width=0.23\textwidth]{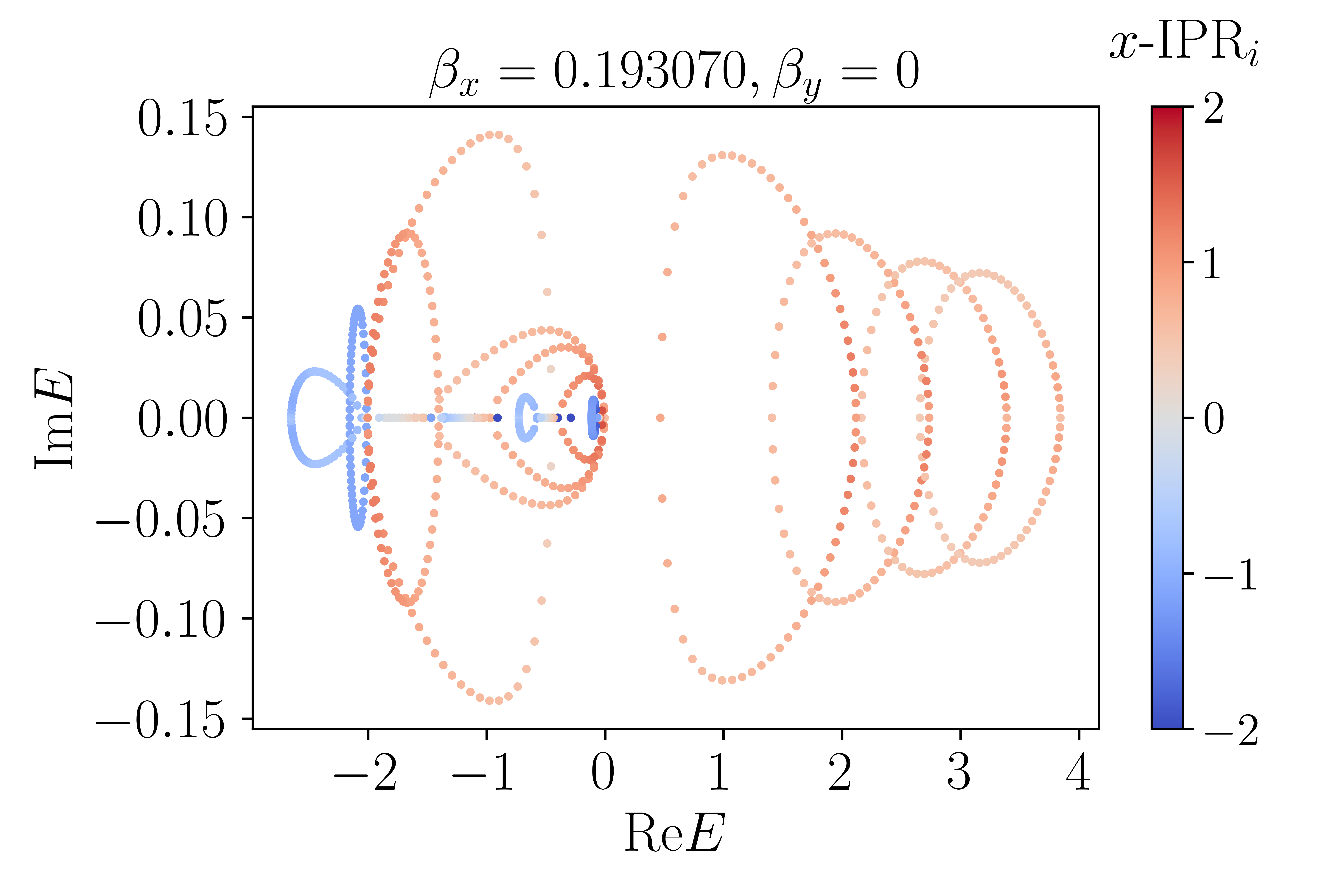}}
  \subfigure[]{\includegraphics[width=0.23\textwidth]{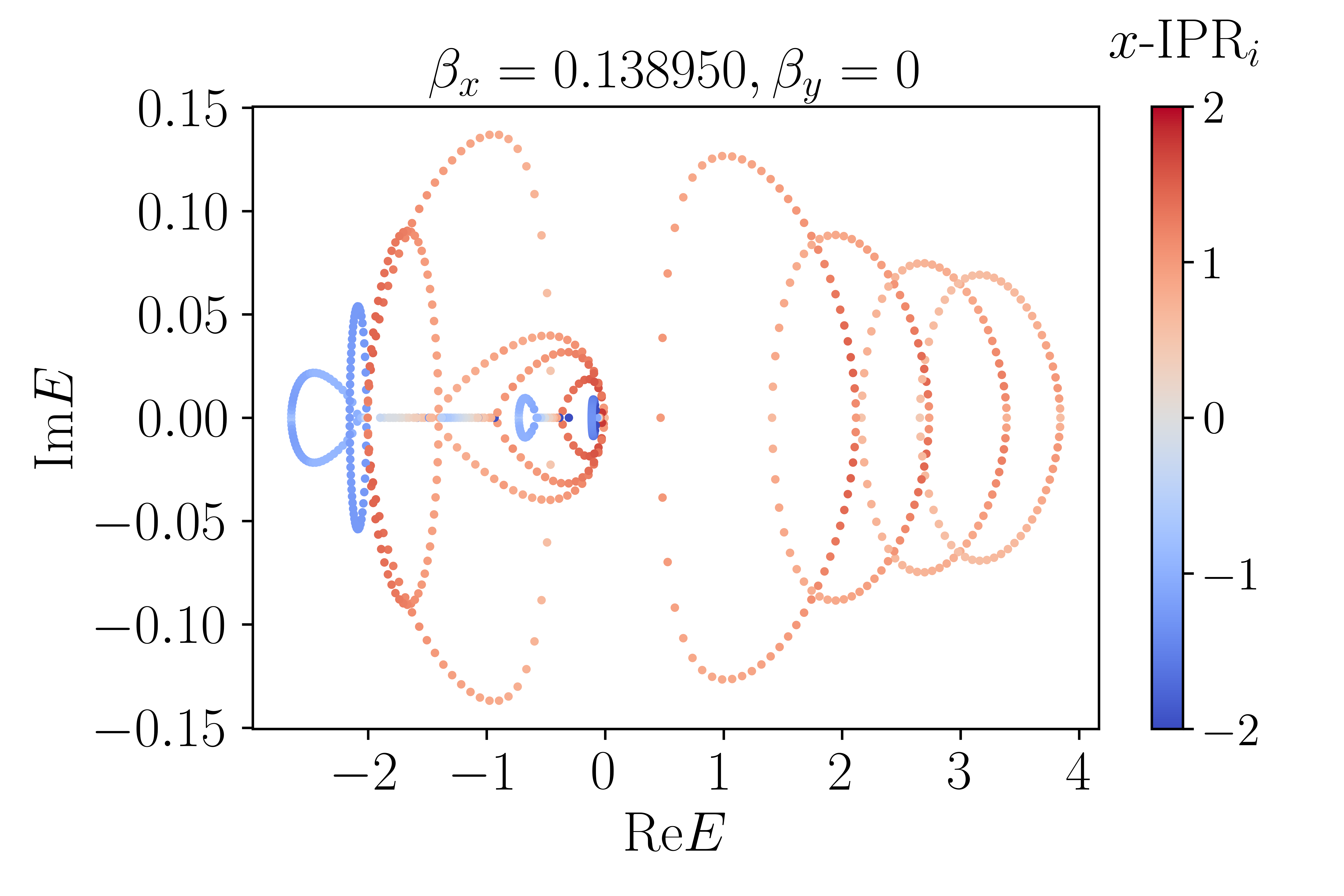}}
  \subfigure[]{\includegraphics[width=0.23\textwidth]{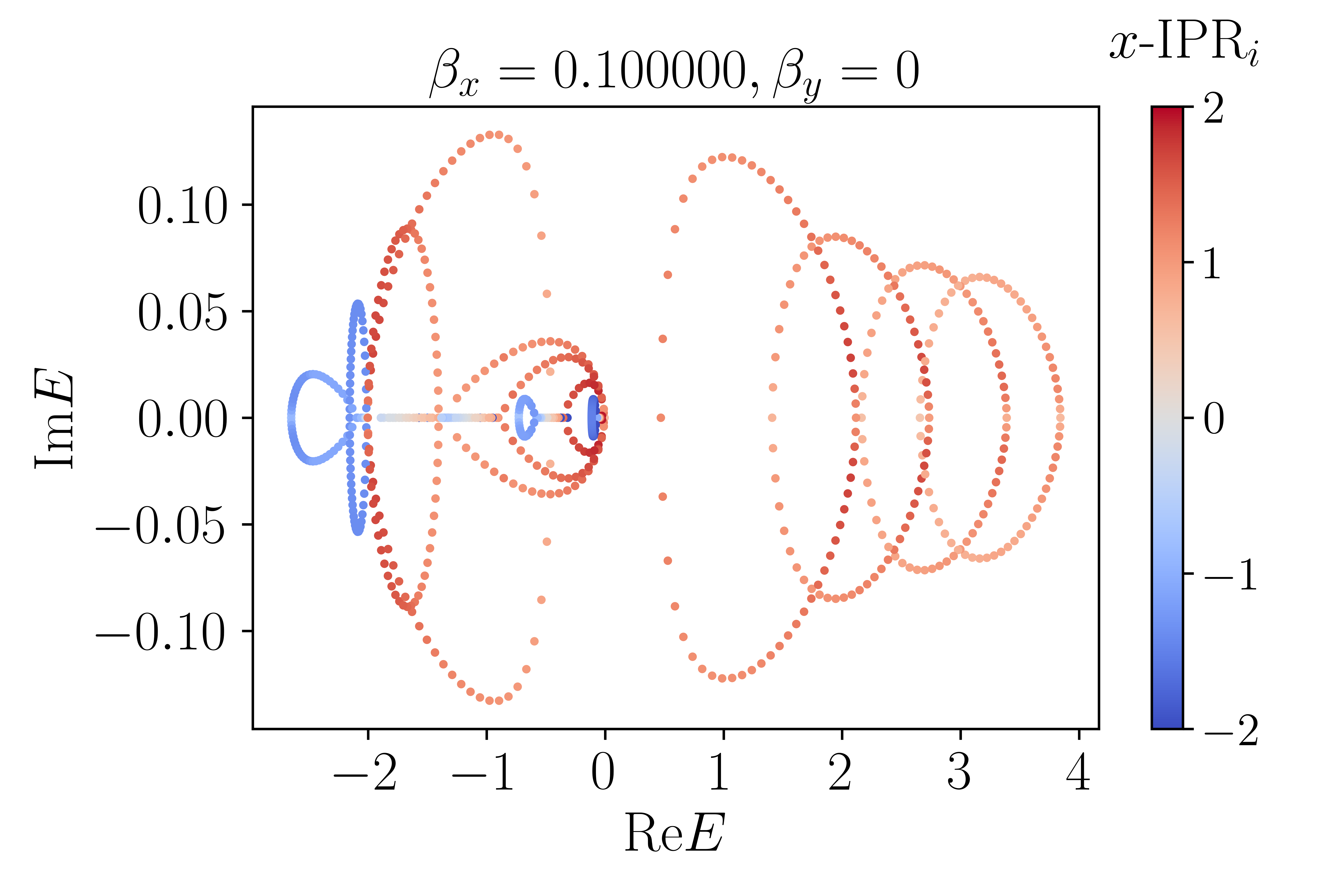}}
  \caption{No any spectrum Flow (Path B). Conversely, the transverse boundary is held fully open ($\beta_y=0$) while the longitudinal boundary $\beta_x$ is tuned. In stark contrast to Path A, the \textbf{shape of the spectral loops remains the same} throughout the entire process. The evolution is not in the geometry of the allowed energies, but in the localization properties ($x$-IPR, color) of the eigenstates. The flow visualizes the eigenstates sorting themselves from right- (red) to left-localized (blue) on a static spectral manifold. This demonstrates that once the transverse OBC is established, the fundamental spectral structure is locked in. Parameters: $L_x=120$, $L_y=8$, $u_x=1/v_x=1.1$, $t_c=0.5$, $u_y=1/v_y=3$.}
  \label{fig:appendix-path-dependence2}
\end{figure*}

This difference of two paths is visualized in Fig.~\ref{fig:appendix-path-dependence1} and Fig.~\ref{fig:appendix-path-dependence2}, where we compare two distinct paths from a periodic-like system to one with double open boundaries. While the illustrated paths do not start from the identical point nor depict the full evolution to the exact same destination, their evolutionary characteristics are profoundly different.

The first protocol, shown in Fig.~\ref{fig:appendix-path-dependence1}, is the most physically relevant for understanding the reversal phenomenon in our long-aspect-ratio ($L_x \gg L_y$) system. We start with a nearly open boundary in the long $x$-direction ($\beta_x=0.2$) and gradually open the short transverse $y$-direction. The initial state ($\beta_y=1$) consists of several distinct spectral loops, each representing a quasi-1D system indexed by a quantized transverse momentum $k_y$.

As $\beta_y$ is reduced, these loops do not simply shrink or shift; they undergo a violent geometric reconfiguration. We observe the loops deforming, expanding, and ultimately merging. This spectral fusion signifies the hybridization of the different transverse modes, which is only possible when the $y$-boundary is opened. The reversal of the skin effect—seen as the emergence of blue (left-localized) states from previously red (right-localized) loops—is a direct consequence of this topological change in the complex energy landscape. This path clearly demonstrates that opening the transverse boundary fundamentally alters the effective Hamiltonian and its spectrum.

The second protocol, Fig.~\ref{fig:appendix-path-dependence2}, provides a contrast. Here, we first open the transverse $y$-boundary fully ($\beta_y=0$) and then proceed to open the longitudinal $x$-boundary. The initial state ($\beta_x=1$) is a set of nested loops whose shape is determined by the strong localization imposed by the $y$-OBC.

As we tune $\beta_x$ from 1 to 0, the geometry of these spectral loops remains almost perfectly rigid. The allowed complex energies are effectively ``frozen'' by the dominant $y$-OBC condition.  The entire evolution is manifested in the color of the eigenstates. We are witnessing a flow of the localization property ($x$-IPR) on a static energy manifold. This path shows that once the transverse boundary is open, the fundamental structure of the problem is set, and opening the longitudinal boundary only serves to select the final localization preference of the states within that pre-existing structure.

In conclusion, this asymmetric evolution provides the answer to why the transverse boundary is so influential. The transverse boundary condition has the power to geometrically reshape the entire complex spectrum (Path A), whereas the longitudinal boundary condition primarily sorts the localization of states within a fixed spectral geometry (Path B). The skin reversal is therefore a direct outcome of the dramatic spectral metamorphosis induced by the transverse boundary conditions. The two protocols illustrate two fundamentally different mechanisms.


\begin{thebibliography}{165}%
\makeatletter
\providecommand \@ifxundefined [1]{%
 \@ifx{#1\undefined}
}%
\providecommand \@ifnum [1]{%
 \ifnum #1\expandafter \@firstoftwo
 \else \expandafter \@secondoftwo
 \fi
}%
\providecommand \@ifx [1]{%
 \ifx #1\expandafter \@firstoftwo
 \else \expandafter \@secondoftwo
 \fi
}%
\providecommand \natexlab [1]{#1}%
\providecommand \enquote  [1]{``#1''}%
\providecommand \bibnamefont  [1]{#1}%
\providecommand \bibfnamefont [1]{#1}%
\providecommand \citenamefont [1]{#1}%
\providecommand \href@noop [0]{\@secondoftwo}%
\providecommand \href [0]{\begingroup \@sanitize@url \@href}%
\providecommand \@href[1]{\@@startlink{#1}\@@href}%
\providecommand \@@href[1]{\endgroup#1\@@endlink}%
\providecommand \@sanitize@url [0]{\catcode `\\12\catcode `\$12\catcode `\&12\catcode `\#12\catcode `\^12\catcode `\_12\catcode `\%12\relax}%
\providecommand \@@startlink[1]{}%
\providecommand \@@endlink[0]{}%
\providecommand \url  [0]{\begingroup\@sanitize@url \@url }%
\providecommand \@url [1]{\endgroup\@href {#1}{\urlprefix }}%
\providecommand \urlprefix  [0]{URL }%
\providecommand \Eprint [0]{\href }%
\providecommand \doibase [0]{https://doi.org/}%
\providecommand \selectlanguage [0]{\@gobble}%
\providecommand \bibinfo  [0]{\@secondoftwo}%
\providecommand \bibfield  [0]{\@secondoftwo}%
\providecommand \translation [1]{[#1]}%
\providecommand \BibitemOpen [0]{}%
\providecommand \bibitemStop [0]{}%
\providecommand \bibitemNoStop [0]{.\EOS\space}%
\providecommand \EOS [0]{\spacefactor3000\relax}%
\providecommand \BibitemShut  [1]{\csname bibitem#1\endcsname}%
\let\auto@bib@innerbib\@empty
\bibitem [{\citenamefont {Lee}(2016)}]{lee2016anomalous}%
  \BibitemOpen
  \bibfield  {author} {\bibinfo {author} {\bibfnamefont {T.~E.}\ \bibnamefont {Lee}},\ }\bibfield  {title} {\bibinfo {title} {Anomalous edge state in a non-hermitian lattice},\ }\href@noop {} {\bibfield  {journal} {\bibinfo  {journal} {Physical review letters}\ }\textbf {\bibinfo {volume} {116}},\ \bibinfo {pages} {133903} (\bibinfo {year} {2016})}\BibitemShut {NoStop}%
\bibitem [{\citenamefont {Yao}\ and\ \citenamefont {Wang}(2018)}]{yao2018edge}%
  \BibitemOpen
  \bibfield  {author} {\bibinfo {author} {\bibfnamefont {S.}~\bibnamefont {Yao}}\ and\ \bibinfo {author} {\bibfnamefont {Z.}~\bibnamefont {Wang}},\ }\bibfield  {title} {\bibinfo {title} {Edge states and topological invariants of non-hermitian systems},\ }\href@noop {} {\bibfield  {journal} {\bibinfo  {journal} {Physical review letters}\ }\textbf {\bibinfo {volume} {121}},\ \bibinfo {pages} {086803} (\bibinfo {year} {2018})}\BibitemShut {NoStop}%
\bibitem [{\citenamefont {Alvarez}\ \emph {et~al.}(2018)\citenamefont {Alvarez}, \citenamefont {Vargas},\ and\ \citenamefont {Torres}}]{alvarez2018non}%
  \BibitemOpen
  \bibfield  {author} {\bibinfo {author} {\bibfnamefont {V.~M.}\ \bibnamefont {Alvarez}}, \bibinfo {author} {\bibfnamefont {J.~B.}\ \bibnamefont {Vargas}},\ and\ \bibinfo {author} {\bibfnamefont {L.~F.}\ \bibnamefont {Torres}},\ }\bibfield  {title} {\bibinfo {title} {Non-hermitian robust edge states in one dimension: Anomalous localization and eigenspace condensation at exceptional points},\ }\href@noop {} {\bibfield  {journal} {\bibinfo  {journal} {Physical Review B}\ }\textbf {\bibinfo {volume} {97}},\ \bibinfo {pages} {121401} (\bibinfo {year} {2018})}\BibitemShut {NoStop}%
\bibitem [{\citenamefont {Kunst}\ \emph {et~al.}(2018{\natexlab{a}})\citenamefont {Kunst}, \citenamefont {Edvardsson}, \citenamefont {Budich},\ and\ \citenamefont {Bergholtz}}]{kunst2018biorthogonal}%
  \BibitemOpen
  \bibfield  {author} {\bibinfo {author} {\bibfnamefont {F.~K.}\ \bibnamefont {Kunst}}, \bibinfo {author} {\bibfnamefont {E.}~\bibnamefont {Edvardsson}}, \bibinfo {author} {\bibfnamefont {J.~C.}\ \bibnamefont {Budich}},\ and\ \bibinfo {author} {\bibfnamefont {E.~J.}\ \bibnamefont {Bergholtz}},\ }\bibfield  {title} {\bibinfo {title} {Biorthogonal bulk-boundary correspondence in non-hermitian systems},\ }\href@noop {} {\bibfield  {journal} {\bibinfo  {journal} {Physical review letters}\ }\textbf {\bibinfo {volume} {121}},\ \bibinfo {pages} {026808} (\bibinfo {year} {2018}{\natexlab{a}})}\BibitemShut {NoStop}%
\bibitem [{\citenamefont {Song}\ \emph {et~al.}(2019)\citenamefont {Song}, \citenamefont {Yao},\ and\ \citenamefont {Wang}}]{song2019non}%
  \BibitemOpen
  \bibfield  {author} {\bibinfo {author} {\bibfnamefont {F.}~\bibnamefont {Song}}, \bibinfo {author} {\bibfnamefont {S.}~\bibnamefont {Yao}},\ and\ \bibinfo {author} {\bibfnamefont {Z.}~\bibnamefont {Wang}},\ }\bibfield  {title} {\bibinfo {title} {Non-hermitian skin effect and chiral damping in open quantum systems},\ }\href@noop {} {\bibfield  {journal} {\bibinfo  {journal} {Physical review letters}\ }\textbf {\bibinfo {volume} {123}},\ \bibinfo {pages} {170401} (\bibinfo {year} {2019})}\BibitemShut {NoStop}%
\bibitem [{\citenamefont {Lee}\ and\ \citenamefont {Thomale}(2019)}]{lee2019anatomy}%
  \BibitemOpen
  \bibfield  {author} {\bibinfo {author} {\bibfnamefont {C.~H.}\ \bibnamefont {Lee}}\ and\ \bibinfo {author} {\bibfnamefont {R.}~\bibnamefont {Thomale}},\ }\bibfield  {title} {\bibinfo {title} {Anatomy of skin modes and topology in non-hermitian systems},\ }\href@noop {} {\bibfield  {journal} {\bibinfo  {journal} {Physical Review B}\ }\textbf {\bibinfo {volume} {99}},\ \bibinfo {pages} {201103} (\bibinfo {year} {2019})}\BibitemShut {NoStop}%
\bibitem [{\citenamefont {Lee}\ \emph {et~al.}(2019)\citenamefont {Lee}, \citenamefont {Li},\ and\ \citenamefont {Gong}}]{lee2019hybrid}%
  \BibitemOpen
  \bibfield  {author} {\bibinfo {author} {\bibfnamefont {C.~H.}\ \bibnamefont {Lee}}, \bibinfo {author} {\bibfnamefont {L.}~\bibnamefont {Li}},\ and\ \bibinfo {author} {\bibfnamefont {J.}~\bibnamefont {Gong}},\ }\bibfield  {title} {\bibinfo {title} {Hybrid higher-order skin-topological modes in nonreciprocal systems},\ }\href@noop {} {\bibfield  {journal} {\bibinfo  {journal} {Physical review letters}\ }\textbf {\bibinfo {volume} {123}},\ \bibinfo {pages} {016805} (\bibinfo {year} {2019})}\BibitemShut {NoStop}%
\bibitem [{\citenamefont {Yang}\ \emph {et~al.}(2020)\citenamefont {Yang}, \citenamefont {Zhang}, \citenamefont {Fang},\ and\ \citenamefont {Hu}}]{yang2020non}%
  \BibitemOpen
  \bibfield  {author} {\bibinfo {author} {\bibfnamefont {Z.}~\bibnamefont {Yang}}, \bibinfo {author} {\bibfnamefont {K.}~\bibnamefont {Zhang}}, \bibinfo {author} {\bibfnamefont {C.}~\bibnamefont {Fang}},\ and\ \bibinfo {author} {\bibfnamefont {J.}~\bibnamefont {Hu}},\ }\bibfield  {title} {\bibinfo {title} {Non-hermitian bulk-boundary correspondence and auxiliary generalized brillouin zone theory},\ }\href@noop {} {\bibfield  {journal} {\bibinfo  {journal} {Physical Review Letters}\ }\textbf {\bibinfo {volume} {125}},\ \bibinfo {pages} {226402} (\bibinfo {year} {2020})}\BibitemShut {NoStop}%
\bibitem [{\citenamefont {Li}\ \emph {et~al.}(2020)\citenamefont {Li}, \citenamefont {Lee}, \citenamefont {Mu},\ and\ \citenamefont {Gong}}]{li2020critical}%
  \BibitemOpen
  \bibfield  {author} {\bibinfo {author} {\bibfnamefont {L.}~\bibnamefont {Li}}, \bibinfo {author} {\bibfnamefont {C.~H.}\ \bibnamefont {Lee}}, \bibinfo {author} {\bibfnamefont {S.}~\bibnamefont {Mu}},\ and\ \bibinfo {author} {\bibfnamefont {J.}~\bibnamefont {Gong}},\ }\bibfield  {title} {\bibinfo {title} {Critical non-hermitian skin effect},\ }\href@noop {} {\bibfield  {journal} {\bibinfo  {journal} {Nature communications}\ }\textbf {\bibinfo {volume} {11}},\ \bibinfo {pages} {5491} (\bibinfo {year} {2020})}\BibitemShut {NoStop}%
\bibitem [{\citenamefont {Hofmann}\ \emph {et~al.}(2020)\citenamefont {Hofmann}, \citenamefont {Helbig}, \citenamefont {Schindler}, \citenamefont {Salgo}, \citenamefont {Brzezi{\'n}ska}, \citenamefont {Greiter}, \citenamefont {Kiessling}, \citenamefont {Wolf}, \citenamefont {Vollhardt}, \citenamefont {Kaba{\v{s}}i} \emph {et~al.}}]{hofmann2020reciprocal}%
  \BibitemOpen
  \bibfield  {author} {\bibinfo {author} {\bibfnamefont {T.}~\bibnamefont {Hofmann}}, \bibinfo {author} {\bibfnamefont {T.}~\bibnamefont {Helbig}}, \bibinfo {author} {\bibfnamefont {F.}~\bibnamefont {Schindler}}, \bibinfo {author} {\bibfnamefont {N.}~\bibnamefont {Salgo}}, \bibinfo {author} {\bibfnamefont {M.}~\bibnamefont {Brzezi{\'n}ska}}, \bibinfo {author} {\bibfnamefont {M.}~\bibnamefont {Greiter}}, \bibinfo {author} {\bibfnamefont {T.}~\bibnamefont {Kiessling}}, \bibinfo {author} {\bibfnamefont {D.}~\bibnamefont {Wolf}}, \bibinfo {author} {\bibfnamefont {A.}~\bibnamefont {Vollhardt}}, \bibinfo {author} {\bibfnamefont {A.}~\bibnamefont {Kaba{\v{s}}i}}, \emph {et~al.},\ }\bibfield  {title} {\bibinfo {title} {Reciprocal skin effect and its realization in a topolectrical circuit},\ }\href@noop {} {\bibfield  {journal} {\bibinfo  {journal} {Physical Review Research}\ }\textbf {\bibinfo {volume} {2}},\ \bibinfo {pages} {023265} (\bibinfo {year} {2020})}\BibitemShut {NoStop}%
\bibitem [{\citenamefont {Zhu}\ \emph {et~al.}(2020)\citenamefont {Zhu}, \citenamefont {Wang}, \citenamefont {Gupta}, \citenamefont {Zhang}, \citenamefont {Xie}, \citenamefont {Lu},\ and\ \citenamefont {Chen}}]{zhu2020photonic}%
  \BibitemOpen
  \bibfield  {author} {\bibinfo {author} {\bibfnamefont {X.}~\bibnamefont {Zhu}}, \bibinfo {author} {\bibfnamefont {H.}~\bibnamefont {Wang}}, \bibinfo {author} {\bibfnamefont {S.~K.}\ \bibnamefont {Gupta}}, \bibinfo {author} {\bibfnamefont {H.}~\bibnamefont {Zhang}}, \bibinfo {author} {\bibfnamefont {B.}~\bibnamefont {Xie}}, \bibinfo {author} {\bibfnamefont {M.}~\bibnamefont {Lu}},\ and\ \bibinfo {author} {\bibfnamefont {Y.}~\bibnamefont {Chen}},\ }\bibfield  {title} {\bibinfo {title} {Photonic non-hermitian skin effect and non-bloch bulk-boundary correspondence},\ }\href@noop {} {\bibfield  {journal} {\bibinfo  {journal} {Physical Review Research}\ }\textbf {\bibinfo {volume} {2}},\ \bibinfo {pages} {013280} (\bibinfo {year} {2020})}\BibitemShut {NoStop}%
\bibitem [{\citenamefont {Song}\ \emph {et~al.}(2020)\citenamefont {Song}, \citenamefont {Liu}, \citenamefont {Zheng}, \citenamefont {Zhang}, \citenamefont {Wang},\ and\ \citenamefont {Lu}}]{song2020two}%
  \BibitemOpen
  \bibfield  {author} {\bibinfo {author} {\bibfnamefont {Y.}~\bibnamefont {Song}}, \bibinfo {author} {\bibfnamefont {W.}~\bibnamefont {Liu}}, \bibinfo {author} {\bibfnamefont {L.}~\bibnamefont {Zheng}}, \bibinfo {author} {\bibfnamefont {Y.}~\bibnamefont {Zhang}}, \bibinfo {author} {\bibfnamefont {B.}~\bibnamefont {Wang}},\ and\ \bibinfo {author} {\bibfnamefont {P.}~\bibnamefont {Lu}},\ }\bibfield  {title} {\bibinfo {title} {Two-dimensional non-hermitian skin effect in a synthetic photonic lattice},\ }\href@noop {} {\bibfield  {journal} {\bibinfo  {journal} {Physical Review Applied}\ }\textbf {\bibinfo {volume} {14}},\ \bibinfo {pages} {064076} (\bibinfo {year} {2020})}\BibitemShut {NoStop}%
\bibitem [{\citenamefont {Liu}\ \emph {et~al.}(2020{\natexlab{a}})\citenamefont {Liu}, \citenamefont {Zhang}, \citenamefont {Yang},\ and\ \citenamefont {Chen}}]{liu2020helical}%
  \BibitemOpen
  \bibfield  {author} {\bibinfo {author} {\bibfnamefont {C.-H.}\ \bibnamefont {Liu}}, \bibinfo {author} {\bibfnamefont {K.}~\bibnamefont {Zhang}}, \bibinfo {author} {\bibfnamefont {Z.}~\bibnamefont {Yang}},\ and\ \bibinfo {author} {\bibfnamefont {S.}~\bibnamefont {Chen}},\ }\bibfield  {title} {\bibinfo {title} {Helical damping and dynamical critical skin effect in open quantum systems},\ }\href@noop {} {\bibfield  {journal} {\bibinfo  {journal} {Physical Review Research}\ }\textbf {\bibinfo {volume} {2}},\ \bibinfo {pages} {043167} (\bibinfo {year} {2020}{\natexlab{a}})}\BibitemShut {NoStop}%
\bibitem [{\citenamefont {Helbig}\ \emph {et~al.}(2020)\citenamefont {Helbig}, \citenamefont {Hofmann}, \citenamefont {Imhof}, \citenamefont {Abdelghany}, \citenamefont {Kiessling}, \citenamefont {Molenkamp}, \citenamefont {Lee}, \citenamefont {Szameit}, \citenamefont {Greiter},\ and\ \citenamefont {Thomale}}]{helbig2020generalized}%
  \BibitemOpen
  \bibfield  {author} {\bibinfo {author} {\bibfnamefont {T.}~\bibnamefont {Helbig}}, \bibinfo {author} {\bibfnamefont {T.}~\bibnamefont {Hofmann}}, \bibinfo {author} {\bibfnamefont {S.}~\bibnamefont {Imhof}}, \bibinfo {author} {\bibfnamefont {M.}~\bibnamefont {Abdelghany}}, \bibinfo {author} {\bibfnamefont {T.}~\bibnamefont {Kiessling}}, \bibinfo {author} {\bibfnamefont {L.}~\bibnamefont {Molenkamp}}, \bibinfo {author} {\bibfnamefont {C.}~\bibnamefont {Lee}}, \bibinfo {author} {\bibfnamefont {A.}~\bibnamefont {Szameit}}, \bibinfo {author} {\bibfnamefont {M.}~\bibnamefont {Greiter}},\ and\ \bibinfo {author} {\bibfnamefont {R.}~\bibnamefont {Thomale}},\ }\bibfield  {title} {\bibinfo {title} {Generalized bulk--boundary correspondence in non-hermitian topolectrical circuits},\ }\href@noop {} {\bibfield  {journal} {\bibinfo  {journal} {Nature Physics}\ }\textbf {\bibinfo {volume} {16}},\ \bibinfo {pages} {747} (\bibinfo {year} {2020})}\BibitemShut {NoStop}%
\bibitem [{\citenamefont {Zou}\ \emph {et~al.}(2021)\citenamefont {Zou}, \citenamefont {Chen}, \citenamefont {He}, \citenamefont {Bao}, \citenamefont {Lee}, \citenamefont {Sun},\ and\ \citenamefont {Zhang}}]{zou2021observation}%
  \BibitemOpen
  \bibfield  {author} {\bibinfo {author} {\bibfnamefont {D.}~\bibnamefont {Zou}}, \bibinfo {author} {\bibfnamefont {T.}~\bibnamefont {Chen}}, \bibinfo {author} {\bibfnamefont {W.}~\bibnamefont {He}}, \bibinfo {author} {\bibfnamefont {J.}~\bibnamefont {Bao}}, \bibinfo {author} {\bibfnamefont {C.~H.}\ \bibnamefont {Lee}}, \bibinfo {author} {\bibfnamefont {H.}~\bibnamefont {Sun}},\ and\ \bibinfo {author} {\bibfnamefont {X.}~\bibnamefont {Zhang}},\ }\bibfield  {title} {\bibinfo {title} {Observation of hybrid higher-order skin-topological effect in non-hermitian topolectrical circuits},\ }\href@noop {} {\bibfield  {journal} {\bibinfo  {journal} {Nature Communications}\ }\textbf {\bibinfo {volume} {12}},\ \bibinfo {pages} {7201} (\bibinfo {year} {2021})}\BibitemShut {NoStop}%
\bibitem [{\citenamefont {Stegmaier}\ \emph {et~al.}(2021)\citenamefont {Stegmaier}, \citenamefont {Imhof}, \citenamefont {Helbig}, \citenamefont {Hofmann}, \citenamefont {Lee}, \citenamefont {Kremer}, \citenamefont {Fritzsche}, \citenamefont {Feichtner}, \citenamefont {Klembt}, \citenamefont {H{\"o}fling} \emph {et~al.}}]{stegmaier2021topological}%
  \BibitemOpen
  \bibfield  {author} {\bibinfo {author} {\bibfnamefont {A.}~\bibnamefont {Stegmaier}}, \bibinfo {author} {\bibfnamefont {S.}~\bibnamefont {Imhof}}, \bibinfo {author} {\bibfnamefont {T.}~\bibnamefont {Helbig}}, \bibinfo {author} {\bibfnamefont {T.}~\bibnamefont {Hofmann}}, \bibinfo {author} {\bibfnamefont {C.~H.}\ \bibnamefont {Lee}}, \bibinfo {author} {\bibfnamefont {M.}~\bibnamefont {Kremer}}, \bibinfo {author} {\bibfnamefont {A.}~\bibnamefont {Fritzsche}}, \bibinfo {author} {\bibfnamefont {T.}~\bibnamefont {Feichtner}}, \bibinfo {author} {\bibfnamefont {S.}~\bibnamefont {Klembt}}, \bibinfo {author} {\bibfnamefont {S.}~\bibnamefont {H{\"o}fling}}, \emph {et~al.},\ }\bibfield  {title} {\bibinfo {title} {Topological defect engineering and p t symmetry in non-hermitian electrical circuits},\ }\href@noop {} {\bibfield  {journal} {\bibinfo  {journal} {Physical Review Letters}\ }\textbf {\bibinfo {volume} {126}},\ \bibinfo {pages} {215302} (\bibinfo {year} {2021})}\BibitemShut {NoStop}%
\bibitem [{\citenamefont {Guo}\ \emph {et~al.}(2021)\citenamefont {Guo}, \citenamefont {Liu}, \citenamefont {Zhao}, \citenamefont {Liu},\ and\ \citenamefont {Chen}}]{guo2021exact}%
  \BibitemOpen
  \bibfield  {author} {\bibinfo {author} {\bibfnamefont {C.-X.}\ \bibnamefont {Guo}}, \bibinfo {author} {\bibfnamefont {C.-H.}\ \bibnamefont {Liu}}, \bibinfo {author} {\bibfnamefont {X.-M.}\ \bibnamefont {Zhao}}, \bibinfo {author} {\bibfnamefont {Y.}~\bibnamefont {Liu}},\ and\ \bibinfo {author} {\bibfnamefont {S.}~\bibnamefont {Chen}},\ }\bibfield  {title} {\bibinfo {title} {Exact solution of non-hermitian systems with generalized boundary conditions: Size-dependent boundary effect and fragility of the skin effect},\ }\href@noop {} {\bibfield  {journal} {\bibinfo  {journal} {Physical Review Letters}\ }\textbf {\bibinfo {volume} {127}},\ \bibinfo {pages} {116801} (\bibinfo {year} {2021})}\BibitemShut {NoStop}%
\bibitem [{\citenamefont {Lee}(2021)}]{lee2021many}%
  \BibitemOpen
  \bibfield  {author} {\bibinfo {author} {\bibfnamefont {C.~H.}\ \bibnamefont {Lee}},\ }\bibfield  {title} {\bibinfo {title} {Many-body topological and skin states without open boundaries},\ }\href@noop {} {\bibfield  {journal} {\bibinfo  {journal} {Physical Review B}\ }\textbf {\bibinfo {volume} {104}},\ \bibinfo {pages} {195102} (\bibinfo {year} {2021})}\BibitemShut {NoStop}%
\bibitem [{\citenamefont {Zhang}\ \emph {et~al.}(2021{\natexlab{a}})\citenamefont {Zhang}, \citenamefont {Tian}, \citenamefont {Jiang}, \citenamefont {Lu},\ and\ \citenamefont {Chen}}]{zhang2021observation}%
  \BibitemOpen
  \bibfield  {author} {\bibinfo {author} {\bibfnamefont {X.}~\bibnamefont {Zhang}}, \bibinfo {author} {\bibfnamefont {Y.}~\bibnamefont {Tian}}, \bibinfo {author} {\bibfnamefont {J.-H.}\ \bibnamefont {Jiang}}, \bibinfo {author} {\bibfnamefont {M.-H.}\ \bibnamefont {Lu}},\ and\ \bibinfo {author} {\bibfnamefont {Y.-F.}\ \bibnamefont {Chen}},\ }\bibfield  {title} {\bibinfo {title} {Observation of higher-order non-hermitian skin effect},\ }\href@noop {} {\bibfield  {journal} {\bibinfo  {journal} {Nature communications}\ }\textbf {\bibinfo {volume} {12}},\ \bibinfo {pages} {5377} (\bibinfo {year} {2021}{\natexlab{a}})}\BibitemShut {NoStop}%
\bibitem [{\citenamefont {Yokomizo}\ and\ \citenamefont {Murakami}(2021)}]{yokomizo2021scaling}%
  \BibitemOpen
  \bibfield  {author} {\bibinfo {author} {\bibfnamefont {K.}~\bibnamefont {Yokomizo}}\ and\ \bibinfo {author} {\bibfnamefont {S.}~\bibnamefont {Murakami}},\ }\bibfield  {title} {\bibinfo {title} {Scaling rule for the critical non-hermitian skin effect},\ }\href@noop {} {\bibfield  {journal} {\bibinfo  {journal} {Physical Review B}\ }\textbf {\bibinfo {volume} {104}},\ \bibinfo {pages} {165117} (\bibinfo {year} {2021})}\BibitemShut {NoStop}%
\bibitem [{\citenamefont {Liu}\ \emph {et~al.}(2021)\citenamefont {Liu}, \citenamefont {Shao}, \citenamefont {Ma}, \citenamefont {Zhang}, \citenamefont {You}, \citenamefont {Wu}, \citenamefont {Xiang}, \citenamefont {Cui},\ and\ \citenamefont {Zhang}}]{liu2021non}%
  \BibitemOpen
  \bibfield  {author} {\bibinfo {author} {\bibfnamefont {S.}~\bibnamefont {Liu}}, \bibinfo {author} {\bibfnamefont {R.}~\bibnamefont {Shao}}, \bibinfo {author} {\bibfnamefont {S.}~\bibnamefont {Ma}}, \bibinfo {author} {\bibfnamefont {L.}~\bibnamefont {Zhang}}, \bibinfo {author} {\bibfnamefont {O.}~\bibnamefont {You}}, \bibinfo {author} {\bibfnamefont {H.}~\bibnamefont {Wu}}, \bibinfo {author} {\bibfnamefont {Y.~J.}\ \bibnamefont {Xiang}}, \bibinfo {author} {\bibfnamefont {T.~J.}\ \bibnamefont {Cui}},\ and\ \bibinfo {author} {\bibfnamefont {S.}~\bibnamefont {Zhang}},\ }\bibfield  {title} {\bibinfo {title} {Non-hermitian skin effect in a non-hermitian electrical circuit},\ }\href@noop {} {\bibfield  {journal} {\bibinfo  {journal} {Research}\ } (\bibinfo {year} {2021})}\BibitemShut {NoStop}%
\bibitem [{\citenamefont {Xue}\ \emph {et~al.}(2021)\citenamefont {Xue}, \citenamefont {Li}, \citenamefont {Hu}, \citenamefont {Song},\ and\ \citenamefont {Wang}}]{xue2021simple}%
  \BibitemOpen
  \bibfield  {author} {\bibinfo {author} {\bibfnamefont {W.-T.}\ \bibnamefont {Xue}}, \bibinfo {author} {\bibfnamefont {M.-R.}\ \bibnamefont {Li}}, \bibinfo {author} {\bibfnamefont {Y.-M.}\ \bibnamefont {Hu}}, \bibinfo {author} {\bibfnamefont {F.}~\bibnamefont {Song}},\ and\ \bibinfo {author} {\bibfnamefont {Z.}~\bibnamefont {Wang}},\ }\bibfield  {title} {\bibinfo {title} {Simple formulas of directional amplification from non-bloch band theory},\ }\href@noop {} {\bibfield  {journal} {\bibinfo  {journal} {Physical Review B}\ }\textbf {\bibinfo {volume} {103}},\ \bibinfo {pages} {L241408} (\bibinfo {year} {2021})}\BibitemShut {NoStop}%
\bibitem [{\citenamefont {Zhang}\ \emph {et~al.}(2021{\natexlab{b}})\citenamefont {Zhang}, \citenamefont {Li}, \citenamefont {Liu}, \citenamefont {Tai}, \citenamefont {Thomale},\ and\ \citenamefont {Lee}}]{zhang2021tidal}%
  \BibitemOpen
  \bibfield  {author} {\bibinfo {author} {\bibfnamefont {X.}~\bibnamefont {Zhang}}, \bibinfo {author} {\bibfnamefont {G.}~\bibnamefont {Li}}, \bibinfo {author} {\bibfnamefont {Y.}~\bibnamefont {Liu}}, \bibinfo {author} {\bibfnamefont {T.}~\bibnamefont {Tai}}, \bibinfo {author} {\bibfnamefont {R.}~\bibnamefont {Thomale}},\ and\ \bibinfo {author} {\bibfnamefont {C.~H.}\ \bibnamefont {Lee}},\ }\bibfield  {title} {\bibinfo {title} {Tidal surface states as fingerprints of non-hermitian nodal knot metals},\ }\href@noop {} {\bibfield  {journal} {\bibinfo  {journal} {Communications Physics}\ }\textbf {\bibinfo {volume} {4}},\ \bibinfo {pages} {47} (\bibinfo {year} {2021}{\natexlab{b}})}\BibitemShut {NoStop}%
\bibitem [{\citenamefont {Li}\ \emph {et~al.}(2022)\citenamefont {Li}, \citenamefont {Teo}, \citenamefont {Mu},\ and\ \citenamefont {Gong}}]{li2022direction}%
  \BibitemOpen
  \bibfield  {author} {\bibinfo {author} {\bibfnamefont {L.}~\bibnamefont {Li}}, \bibinfo {author} {\bibfnamefont {W.~X.}\ \bibnamefont {Teo}}, \bibinfo {author} {\bibfnamefont {S.}~\bibnamefont {Mu}},\ and\ \bibinfo {author} {\bibfnamefont {J.}~\bibnamefont {Gong}},\ }\bibfield  {title} {\bibinfo {title} {Direction reversal of non-hermitian skin effect via coherent coupling},\ }\href@noop {} {\bibfield  {journal} {\bibinfo  {journal} {Physical Review B}\ }\textbf {\bibinfo {volume} {106}},\ \bibinfo {pages} {085427} (\bibinfo {year} {2022})}\BibitemShut {NoStop}%
\bibitem [{\citenamefont {Wu}\ \emph {et~al.}(2022)\citenamefont {Wu}, \citenamefont {Yang}, \citenamefont {Tang}, \citenamefont {Liu},\ and\ \citenamefont {Chen}}]{wu2022flux}%
  \BibitemOpen
  \bibfield  {author} {\bibinfo {author} {\bibfnamefont {C.}~\bibnamefont {Wu}}, \bibinfo {author} {\bibfnamefont {Z.}~\bibnamefont {Yang}}, \bibinfo {author} {\bibfnamefont {J.}~\bibnamefont {Tang}}, \bibinfo {author} {\bibfnamefont {N.}~\bibnamefont {Liu}},\ and\ \bibinfo {author} {\bibfnamefont {G.}~\bibnamefont {Chen}},\ }\bibfield  {title} {\bibinfo {title} {Flux-controlled skin effect and topological transition in a dissipative two-leg ladder model},\ }\href@noop {} {\bibfield  {journal} {\bibinfo  {journal} {Physical Review A}\ }\textbf {\bibinfo {volume} {106}},\ \bibinfo {pages} {062206} (\bibinfo {year} {2022})}\BibitemShut {NoStop}%
\bibitem [{\citenamefont {Gu}\ \emph {et~al.}(2022)\citenamefont {Gu}, \citenamefont {Gao}, \citenamefont {Xue}, \citenamefont {Li}, \citenamefont {Su},\ and\ \citenamefont {Zhu}}]{gu2022transient}%
  \BibitemOpen
  \bibfield  {author} {\bibinfo {author} {\bibfnamefont {Z.}~\bibnamefont {Gu}}, \bibinfo {author} {\bibfnamefont {H.}~\bibnamefont {Gao}}, \bibinfo {author} {\bibfnamefont {H.}~\bibnamefont {Xue}}, \bibinfo {author} {\bibfnamefont {J.}~\bibnamefont {Li}}, \bibinfo {author} {\bibfnamefont {Z.}~\bibnamefont {Su}},\ and\ \bibinfo {author} {\bibfnamefont {J.}~\bibnamefont {Zhu}},\ }\bibfield  {title} {\bibinfo {title} {Transient non-hermitian skin effect},\ }\href@noop {} {\bibfield  {journal} {\bibinfo  {journal} {Nature Communications}\ }\textbf {\bibinfo {volume} {13}},\ \bibinfo {pages} {7668} (\bibinfo {year} {2022})}\BibitemShut {NoStop}%
\bibitem [{\citenamefont {Shang}\ \emph {et~al.}(2022)\citenamefont {Shang}, \citenamefont {Liu}, \citenamefont {Shao}, \citenamefont {Han}, \citenamefont {Zang}, \citenamefont {Zhang}, \citenamefont {Salama}, \citenamefont {Gao}, \citenamefont {Lee}, \citenamefont {Thomale} \emph {et~al.}}]{shang2022experimental}%
  \BibitemOpen
  \bibfield  {author} {\bibinfo {author} {\bibfnamefont {C.}~\bibnamefont {Shang}}, \bibinfo {author} {\bibfnamefont {S.}~\bibnamefont {Liu}}, \bibinfo {author} {\bibfnamefont {R.}~\bibnamefont {Shao}}, \bibinfo {author} {\bibfnamefont {P.}~\bibnamefont {Han}}, \bibinfo {author} {\bibfnamefont {X.}~\bibnamefont {Zang}}, \bibinfo {author} {\bibfnamefont {X.}~\bibnamefont {Zhang}}, \bibinfo {author} {\bibfnamefont {K.~N.}\ \bibnamefont {Salama}}, \bibinfo {author} {\bibfnamefont {W.}~\bibnamefont {Gao}}, \bibinfo {author} {\bibfnamefont {C.~H.}\ \bibnamefont {Lee}}, \bibinfo {author} {\bibfnamefont {R.}~\bibnamefont {Thomale}}, \emph {et~al.},\ }\bibfield  {title} {\bibinfo {title} {Experimental identification of the second-order non-hermitian skin effect with physics-graph-informed machine learning},\ }\href@noop {} {\bibfield  {journal} {\bibinfo  {journal} {Advanced Science}\ }\textbf {\bibinfo {volume} {9}},\ \bibinfo {pages} {2202922} (\bibinfo {year} {2022})}\BibitemShut {NoStop}%
\bibitem [{\citenamefont {Yang}\ \emph {et~al.}(2022{\natexlab{a}})\citenamefont {Yang}, \citenamefont {Wang}, \citenamefont {Wu}, \citenamefont {Xiao}, \citenamefont {Yu}, \citenamefont {Yuan},\ and\ \citenamefont {Chen}}]{yang2022concentrated}%
  \BibitemOpen
  \bibfield  {author} {\bibinfo {author} {\bibfnamefont {M.}~\bibnamefont {Yang}}, \bibinfo {author} {\bibfnamefont {L.}~\bibnamefont {Wang}}, \bibinfo {author} {\bibfnamefont {X.}~\bibnamefont {Wu}}, \bibinfo {author} {\bibfnamefont {H.}~\bibnamefont {Xiao}}, \bibinfo {author} {\bibfnamefont {D.}~\bibnamefont {Yu}}, \bibinfo {author} {\bibfnamefont {L.}~\bibnamefont {Yuan}},\ and\ \bibinfo {author} {\bibfnamefont {X.}~\bibnamefont {Chen}},\ }\bibfield  {title} {\bibinfo {title} {Concentrated subradiant modes in a one-dimensional atomic array coupled with chiral waveguides},\ }\href@noop {} {\bibfield  {journal} {\bibinfo  {journal} {Physical Review A}\ }\textbf {\bibinfo {volume} {106}},\ \bibinfo {pages} {043717} (\bibinfo {year} {2022}{\natexlab{a}})}\BibitemShut {NoStop}%
\bibitem [{\citenamefont {Arouca}\ \emph {et~al.}(2020)\citenamefont {Arouca}, \citenamefont {Lee},\ and\ \citenamefont {Morais~Smith}}]{arouca2020unconventional}%
  \BibitemOpen
  \bibfield  {author} {\bibinfo {author} {\bibfnamefont {R.}~\bibnamefont {Arouca}}, \bibinfo {author} {\bibfnamefont {C.}~\bibnamefont {Lee}},\ and\ \bibinfo {author} {\bibfnamefont {C.}~\bibnamefont {Morais~Smith}},\ }\bibfield  {title} {\bibinfo {title} {Unconventional scaling at non-hermitian critical points},\ }\href@noop {} {\bibfield  {journal} {\bibinfo  {journal} {Physical Review B}\ }\textbf {\bibinfo {volume} {102}},\ \bibinfo {pages} {245145} (\bibinfo {year} {2020})}\BibitemShut {NoStop}%
\bibitem [{\citenamefont {Zeng}\ and\ \citenamefont {L{\"u}}(2022)}]{zeng2022real}%
  \BibitemOpen
  \bibfield  {author} {\bibinfo {author} {\bibfnamefont {Q.-B.}\ \bibnamefont {Zeng}}\ and\ \bibinfo {author} {\bibfnamefont {R.}~\bibnamefont {L{\"u}}},\ }\bibfield  {title} {\bibinfo {title} {Real spectra and phase transition of skin effect in nonreciprocal systems},\ }\href@noop {} {\bibfield  {journal} {\bibinfo  {journal} {Physical Review B}\ }\textbf {\bibinfo {volume} {105}},\ \bibinfo {pages} {245407} (\bibinfo {year} {2022})}\BibitemShut {NoStop}%
\bibitem [{\citenamefont {Qin}\ \emph {et~al.}(2023{\natexlab{a}})\citenamefont {Qin}, \citenamefont {Shen}, \citenamefont {Lee} \emph {et~al.}}]{qin2023non}%
  \BibitemOpen
  \bibfield  {author} {\bibinfo {author} {\bibfnamefont {F.}~\bibnamefont {Qin}}, \bibinfo {author} {\bibfnamefont {R.}~\bibnamefont {Shen}}, \bibinfo {author} {\bibfnamefont {C.~H.}\ \bibnamefont {Lee}}, \emph {et~al.},\ }\bibfield  {title} {\bibinfo {title} {Non-hermitian squeezed polarons},\ }\href@noop {} {\bibfield  {journal} {\bibinfo  {journal} {Physical Review A}\ }\textbf {\bibinfo {volume} {107}},\ \bibinfo {pages} {L010202} (\bibinfo {year} {2023}{\natexlab{a}})}\BibitemShut {NoStop}%
\bibitem [{\citenamefont {Longhi}(2022)}]{longhi2022self}%
  \BibitemOpen
  \bibfield  {author} {\bibinfo {author} {\bibfnamefont {S.}~\bibnamefont {Longhi}},\ }\bibfield  {title} {\bibinfo {title} {Self-healing of non-hermitian topological skin modes},\ }\href@noop {} {\bibfield  {journal} {\bibinfo  {journal} {Physical Review Letters}\ }\textbf {\bibinfo {volume} {128}},\ \bibinfo {pages} {157601} (\bibinfo {year} {2022})}\BibitemShut {NoStop}%
\bibitem [{\citenamefont {Wang}\ \emph {et~al.}(2023{\natexlab{a}})\citenamefont {Wang}, \citenamefont {Xiao}, \citenamefont {Lin}, \citenamefont {Wang}, \citenamefont {Lin},\ and\ \citenamefont {Xue}}]{wang2023demonstration}%
  \BibitemOpen
  \bibfield  {author} {\bibinfo {author} {\bibfnamefont {X.}~\bibnamefont {Wang}}, \bibinfo {author} {\bibfnamefont {L.}~\bibnamefont {Xiao}}, \bibinfo {author} {\bibfnamefont {Q.}~\bibnamefont {Lin}}, \bibinfo {author} {\bibfnamefont {K.}~\bibnamefont {Wang}}, \bibinfo {author} {\bibfnamefont {H.}~\bibnamefont {Lin}},\ and\ \bibinfo {author} {\bibfnamefont {P.}~\bibnamefont {Xue}},\ }\bibfield  {title} {\bibinfo {title} {Demonstration of reversed non-hermitian skin effect via quantum walks on a ladder},\ }\href@noop {} {\bibfield  {journal} {\bibinfo  {journal} {New Journal of Physics}\ }\textbf {\bibinfo {volume} {25}},\ \bibinfo {pages} {113039} (\bibinfo {year} {2023}{\natexlab{a}})}\BibitemShut {NoStop}%
\bibitem [{\citenamefont {Manna}\ and\ \citenamefont {Roy}(2023)}]{manna2023inner}%
  \BibitemOpen
  \bibfield  {author} {\bibinfo {author} {\bibfnamefont {S.}~\bibnamefont {Manna}}\ and\ \bibinfo {author} {\bibfnamefont {B.}~\bibnamefont {Roy}},\ }\bibfield  {title} {\bibinfo {title} {Inner skin effects on non-hermitian topological fractals},\ }\href@noop {} {\bibfield  {journal} {\bibinfo  {journal} {communications physics}\ }\textbf {\bibinfo {volume} {6}},\ \bibinfo {pages} {10} (\bibinfo {year} {2023})}\BibitemShut {NoStop}%
\bibitem [{\citenamefont {Ma}\ and\ \citenamefont {Hughes}(2023)}]{ma2023quantum}%
  \BibitemOpen
  \bibfield  {author} {\bibinfo {author} {\bibfnamefont {Y.}~\bibnamefont {Ma}}\ and\ \bibinfo {author} {\bibfnamefont {T.~L.}\ \bibnamefont {Hughes}},\ }\bibfield  {title} {\bibinfo {title} {Quantum skin hall effect},\ }\href@noop {} {\bibfield  {journal} {\bibinfo  {journal} {Physical Review B}\ }\textbf {\bibinfo {volume} {108}},\ \bibinfo {pages} {L100301} (\bibinfo {year} {2023})}\BibitemShut {NoStop}%
\bibitem [{\citenamefont {Li}\ \emph {et~al.}(2023)\citenamefont {Li}, \citenamefont {Lu}, \citenamefont {Zhang},\ and\ \citenamefont {Liu}}]{li2023loss}%
  \BibitemOpen
  \bibfield  {author} {\bibinfo {author} {\bibfnamefont {Y.}~\bibnamefont {Li}}, \bibinfo {author} {\bibfnamefont {C.}~\bibnamefont {Lu}}, \bibinfo {author} {\bibfnamefont {S.}~\bibnamefont {Zhang}},\ and\ \bibinfo {author} {\bibfnamefont {Y.-C.}\ \bibnamefont {Liu}},\ }\bibfield  {title} {\bibinfo {title} {Loss-induced floquet non-hermitian skin effect},\ }\href@noop {} {\bibfield  {journal} {\bibinfo  {journal} {Physical Review B}\ }\textbf {\bibinfo {volume} {108}},\ \bibinfo {pages} {L220301} (\bibinfo {year} {2023})}\BibitemShut {NoStop}%
\bibitem [{\citenamefont {Meng}\ \emph {et~al.}(2024)\citenamefont {Meng}, \citenamefont {Ang},\ and\ \citenamefont {Lee}}]{meng2024exceptional}%
  \BibitemOpen
  \bibfield  {author} {\bibinfo {author} {\bibfnamefont {H.}~\bibnamefont {Meng}}, \bibinfo {author} {\bibfnamefont {Y.~S.}\ \bibnamefont {Ang}},\ and\ \bibinfo {author} {\bibfnamefont {C.~H.}\ \bibnamefont {Lee}},\ }\bibfield  {title} {\bibinfo {title} {Exceptional points in non-hermitian systems: Applications and recent developments},\ }\href@noop {} {\bibfield  {journal} {\bibinfo  {journal} {Applied Physics Letters}\ }\textbf {\bibinfo {volume} {124}} (\bibinfo {year} {2024})}\BibitemShut {NoStop}%
\bibitem [{\citenamefont {Qin}\ \emph {et~al.}(2023{\natexlab{b}})\citenamefont {Qin}, \citenamefont {Ma}, \citenamefont {Shen}, \citenamefont {Lee} \emph {et~al.}}]{qin2023universal}%
  \BibitemOpen
  \bibfield  {author} {\bibinfo {author} {\bibfnamefont {F.}~\bibnamefont {Qin}}, \bibinfo {author} {\bibfnamefont {Y.}~\bibnamefont {Ma}}, \bibinfo {author} {\bibfnamefont {R.}~\bibnamefont {Shen}}, \bibinfo {author} {\bibfnamefont {C.~H.}\ \bibnamefont {Lee}}, \emph {et~al.},\ }\bibfield  {title} {\bibinfo {title} {Universal competitive spectral scaling from the critical non-hermitian skin effect},\ }\href@noop {} {\bibfield  {journal} {\bibinfo  {journal} {Physical Review B}\ }\textbf {\bibinfo {volume} {107}},\ \bibinfo {pages} {155430} (\bibinfo {year} {2023}{\natexlab{b}})}\BibitemShut {NoStop}%
\bibitem [{\citenamefont {Zhang}\ \emph {et~al.}(2023{\natexlab{a}})\citenamefont {Zhang}, \citenamefont {Chen}, \citenamefont {Li}, \citenamefont {Lee},\ and\ \citenamefont {Zhang}}]{zhang2023electrical}%
  \BibitemOpen
  \bibfield  {author} {\bibinfo {author} {\bibfnamefont {H.}~\bibnamefont {Zhang}}, \bibinfo {author} {\bibfnamefont {T.}~\bibnamefont {Chen}}, \bibinfo {author} {\bibfnamefont {L.}~\bibnamefont {Li}}, \bibinfo {author} {\bibfnamefont {C.~H.}\ \bibnamefont {Lee}},\ and\ \bibinfo {author} {\bibfnamefont {X.}~\bibnamefont {Zhang}},\ }\bibfield  {title} {\bibinfo {title} {Electrical circuit realization of topological switching for the non-hermitian skin effect},\ }\href@noop {} {\bibfield  {journal} {\bibinfo  {journal} {Physical Review B}\ }\textbf {\bibinfo {volume} {107}},\ \bibinfo {pages} {085426} (\bibinfo {year} {2023}{\natexlab{a}})}\BibitemShut {NoStop}%
\bibitem [{\citenamefont {Lin}\ \emph {et~al.}(2023)\citenamefont {Lin}, \citenamefont {Tai}, \citenamefont {Li},\ and\ \citenamefont {Lee}}]{lin2023topological}%
  \BibitemOpen
  \bibfield  {author} {\bibinfo {author} {\bibfnamefont {R.}~\bibnamefont {Lin}}, \bibinfo {author} {\bibfnamefont {T.}~\bibnamefont {Tai}}, \bibinfo {author} {\bibfnamefont {L.}~\bibnamefont {Li}},\ and\ \bibinfo {author} {\bibfnamefont {C.~H.}\ \bibnamefont {Lee}},\ }\bibfield  {title} {\bibinfo {title} {Topological non-hermitian skin effect},\ }\href@noop {} {\bibfield  {journal} {\bibinfo  {journal} {Frontiers of Physics}\ }\textbf {\bibinfo {volume} {18}},\ \bibinfo {pages} {53605} (\bibinfo {year} {2023})}\BibitemShut {NoStop}%
\bibitem [{\citenamefont {Tai}\ and\ \citenamefont {Lee}(2023)}]{tai2023zoology}%
  \BibitemOpen
  \bibfield  {author} {\bibinfo {author} {\bibfnamefont {T.}~\bibnamefont {Tai}}\ and\ \bibinfo {author} {\bibfnamefont {C.~H.}\ \bibnamefont {Lee}},\ }\bibfield  {title} {\bibinfo {title} {Zoology of non-hermitian spectra and their graph topology},\ }\href@noop {} {\bibfield  {journal} {\bibinfo  {journal} {Physical Review B}\ }\textbf {\bibinfo {volume} {107}},\ \bibinfo {pages} {L220301} (\bibinfo {year} {2023})}\BibitemShut {NoStop}%
\bibitem [{\citenamefont {Jiang}\ and\ \citenamefont {Lee}(2023)}]{jiang2023dimensional}%
  \BibitemOpen
  \bibfield  {author} {\bibinfo {author} {\bibfnamefont {H.}~\bibnamefont {Jiang}}\ and\ \bibinfo {author} {\bibfnamefont {C.~H.}\ \bibnamefont {Lee}},\ }\bibfield  {title} {\bibinfo {title} {Dimensional transmutation from non-hermiticity},\ }\href@noop {} {\bibfield  {journal} {\bibinfo  {journal} {Physical Review Letters}\ }\textbf {\bibinfo {volume} {131}},\ \bibinfo {pages} {076401} (\bibinfo {year} {2023})}\BibitemShut {NoStop}%
\bibitem [{\citenamefont {Yang}\ and\ \citenamefont {Lee}(2024)}]{yang2024percolation}%
  \BibitemOpen
  \bibfield  {author} {\bibinfo {author} {\bibfnamefont {M.}~\bibnamefont {Yang}}\ and\ \bibinfo {author} {\bibfnamefont {C.~H.}\ \bibnamefont {Lee}},\ }\bibfield  {title} {\bibinfo {title} {Percolation-induced pt symmetry breaking},\ }\href@noop {} {\bibfield  {journal} {\bibinfo  {journal} {Physical Review Letters}\ }\textbf {\bibinfo {volume} {133}},\ \bibinfo {pages} {136602} (\bibinfo {year} {2024})}\BibitemShut {NoStop}%
\bibitem [{\citenamefont {Yang}\ \emph {et~al.}(2024)\citenamefont {Yang}, \citenamefont {Yuan},\ and\ \citenamefont {Lee}}]{yang2024non}%
  \BibitemOpen
  \bibfield  {author} {\bibinfo {author} {\bibfnamefont {M.}~\bibnamefont {Yang}}, \bibinfo {author} {\bibfnamefont {L.}~\bibnamefont {Yuan}},\ and\ \bibinfo {author} {\bibfnamefont {C.~H.}\ \bibnamefont {Lee}},\ }\bibfield  {title} {\bibinfo {title} {Non-hermitian ultra-strong bosonic condensation through interaction-induced caging},\ }\href@noop {} {\bibfield  {journal} {\bibinfo  {journal} {arXiv preprint arXiv:2410.01258}\ } (\bibinfo {year} {2024})}\BibitemShut {NoStop}%
\bibitem [{\citenamefont {Nie}\ \emph {et~al.}(2024)\citenamefont {Nie}, \citenamefont {Cui}, \citenamefont {Wang} \emph {et~al.}}]{nie2024multiple}%
  \BibitemOpen
  \bibfield  {author} {\bibinfo {author} {\bibfnamefont {X.-F.}\ \bibnamefont {Nie}}, \bibinfo {author} {\bibfnamefont {W.-X.}\ \bibnamefont {Cui}}, \bibinfo {author} {\bibfnamefont {H.-F.}\ \bibnamefont {Wang}}, \emph {et~al.},\ }\bibfield  {title} {\bibinfo {title} {Multiple skin transitions in two-band non-hermitian systems with long-range nonreciprocal hopping},\ }\href@noop {} {\bibfield  {journal} {\bibinfo  {journal} {New Journal of Physics}\ }\textbf {\bibinfo {volume} {26}},\ \bibinfo {pages} {053025} (\bibinfo {year} {2024})}\BibitemShut {NoStop}%
\bibitem [{\citenamefont {Xue}\ and\ \citenamefont {Lee}(2024)}]{xue2024topologically}%
  \BibitemOpen
  \bibfield  {author} {\bibinfo {author} {\bibfnamefont {W.-T.}\ \bibnamefont {Xue}}\ and\ \bibinfo {author} {\bibfnamefont {C.~H.}\ \bibnamefont {Lee}},\ }\bibfield  {title} {\bibinfo {title} {Topologically protected non-hermitian super-volume-law entanglement},\ }\href@noop {} {\bibfield  {journal} {\bibinfo  {journal} {arXiv preprint arXiv:2403.03259}\ } (\bibinfo {year} {2024})}\BibitemShut {NoStop}%
\bibitem [{\citenamefont {Xiong}\ \emph {et~al.}(2024)\citenamefont {Xiong}, \citenamefont {Xing},\ and\ \citenamefont {Hu}}]{xiong2024non}%
  \BibitemOpen
  \bibfield  {author} {\bibinfo {author} {\bibfnamefont {Y.}~\bibnamefont {Xiong}}, \bibinfo {author} {\bibfnamefont {Z.-Y.}\ \bibnamefont {Xing}},\ and\ \bibinfo {author} {\bibfnamefont {H.}~\bibnamefont {Hu}},\ }\bibfield  {title} {\bibinfo {title} {Non-hermitian skin effect in arbitrary dimensions: non-bloch band theory and classification},\ }\href@noop {} {\bibfield  {journal} {\bibinfo  {journal} {arXiv preprint arXiv:2407.01296}\ } (\bibinfo {year} {2024})}\BibitemShut {NoStop}%
\bibitem [{\citenamefont {Guo}\ \emph {et~al.}(2024)\citenamefont {Guo}, \citenamefont {Su}, \citenamefont {Wang}, \citenamefont {Li}, \citenamefont {Wang}, \citenamefont {Ruan}, \citenamefont {Du}, \citenamefont {Zheng}, \citenamefont {Chen},\ and\ \citenamefont {Hu}}]{guo2024scale}%
  \BibitemOpen
  \bibfield  {author} {\bibinfo {author} {\bibfnamefont {C.-X.}\ \bibnamefont {Guo}}, \bibinfo {author} {\bibfnamefont {L.}~\bibnamefont {Su}}, \bibinfo {author} {\bibfnamefont {Y.}~\bibnamefont {Wang}}, \bibinfo {author} {\bibfnamefont {L.}~\bibnamefont {Li}}, \bibinfo {author} {\bibfnamefont {J.}~\bibnamefont {Wang}}, \bibinfo {author} {\bibfnamefont {X.}~\bibnamefont {Ruan}}, \bibinfo {author} {\bibfnamefont {Y.}~\bibnamefont {Du}}, \bibinfo {author} {\bibfnamefont {D.}~\bibnamefont {Zheng}}, \bibinfo {author} {\bibfnamefont {S.}~\bibnamefont {Chen}},\ and\ \bibinfo {author} {\bibfnamefont {H.}~\bibnamefont {Hu}},\ }\bibfield  {title} {\bibinfo {title} {Scale-tailored localization and its observation in non-hermitian electrical circuits},\ }\href@noop {} {\bibfield  {journal} {\bibinfo  {journal} {Nature Communications}\ }\textbf {\bibinfo {volume} {15}},\ \bibinfo {pages} {9120} (\bibinfo {year} {2024})}\BibitemShut {NoStop}%
\bibitem [{\citenamefont {Zhang}\ \emph {et~al.}(2024)\citenamefont {Zhang}, \citenamefont {Wu}, \citenamefont {Yan}, \citenamefont {Liu}, \citenamefont {Wang},\ and\ \citenamefont {Chen}}]{zhang2024observation2}%
  \BibitemOpen
  \bibfield  {author} {\bibinfo {author} {\bibfnamefont {X.}~\bibnamefont {Zhang}}, \bibinfo {author} {\bibfnamefont {C.}~\bibnamefont {Wu}}, \bibinfo {author} {\bibfnamefont {M.}~\bibnamefont {Yan}}, \bibinfo {author} {\bibfnamefont {N.}~\bibnamefont {Liu}}, \bibinfo {author} {\bibfnamefont {Z.}~\bibnamefont {Wang}},\ and\ \bibinfo {author} {\bibfnamefont {G.}~\bibnamefont {Chen}},\ }\bibfield  {title} {\bibinfo {title} {Observation of continuum landau modes in non-hermitian electric circuits},\ }\href@noop {} {\bibfield  {journal} {\bibinfo  {journal} {Nature Communications}\ }\textbf {\bibinfo {volume} {15}},\ \bibinfo {pages} {1798} (\bibinfo {year} {2024})}\BibitemShut {NoStop}%
\bibitem [{\citenamefont {Shen}\ \emph {et~al.}(2024)\citenamefont {Shen}, \citenamefont {Qin}, \citenamefont {Desaules}, \citenamefont {Papi{\'c}},\ and\ \citenamefont {Lee}}]{shen2024enhanced}%
  \BibitemOpen
  \bibfield  {author} {\bibinfo {author} {\bibfnamefont {R.}~\bibnamefont {Shen}}, \bibinfo {author} {\bibfnamefont {F.}~\bibnamefont {Qin}}, \bibinfo {author} {\bibfnamefont {J.-Y.}\ \bibnamefont {Desaules}}, \bibinfo {author} {\bibfnamefont {Z.}~\bibnamefont {Papi{\'c}}},\ and\ \bibinfo {author} {\bibfnamefont {C.~H.}\ \bibnamefont {Lee}},\ }\bibfield  {title} {\bibinfo {title} {Enhanced many-body quantum scars from the non-hermitian fock skin effect},\ }\href@noop {} {\bibfield  {journal} {\bibinfo  {journal} {Physical Review Letters}\ }\textbf {\bibinfo {volume} {133}},\ \bibinfo {pages} {216601} (\bibinfo {year} {2024})}\BibitemShut {NoStop}%
\bibitem [{\citenamefont {Li}\ \emph {et~al.}(2024{\natexlab{a}})\citenamefont {Li}, \citenamefont {Wang}, \citenamefont {Wang}, \citenamefont {Lin}, \citenamefont {Ma},\ and\ \citenamefont {Jiang}}]{li2024observation}%
  \BibitemOpen
  \bibfield  {author} {\bibinfo {author} {\bibfnamefont {Z.}~\bibnamefont {Li}}, \bibinfo {author} {\bibfnamefont {L.-W.}\ \bibnamefont {Wang}}, \bibinfo {author} {\bibfnamefont {X.}~\bibnamefont {Wang}}, \bibinfo {author} {\bibfnamefont {Z.-K.}\ \bibnamefont {Lin}}, \bibinfo {author} {\bibfnamefont {G.}~\bibnamefont {Ma}},\ and\ \bibinfo {author} {\bibfnamefont {J.-H.}\ \bibnamefont {Jiang}},\ }\bibfield  {title} {\bibinfo {title} {Observation of dynamic non-hermitian skin effects},\ }\href@noop {} {\bibfield  {journal} {\bibinfo  {journal} {Nature Communications}\ }\textbf {\bibinfo {volume} {15}},\ \bibinfo {pages} {6544} (\bibinfo {year} {2024}{\natexlab{a}})}\BibitemShut {NoStop}%
\bibitem [{\citenamefont {Lin}\ \emph {et~al.}(2024)\citenamefont {Lin}, \citenamefont {Song}, \citenamefont {Wang}, \citenamefont {Xin}, \citenamefont {Sun}, \citenamefont {Wu}, \citenamefont {Huang}, \citenamefont {Zhu}, \citenamefont {Jiang},\ and\ \citenamefont {Li}}]{lin2024observation}%
  \BibitemOpen
  \bibfield  {author} {\bibinfo {author} {\bibfnamefont {Z.}~\bibnamefont {Lin}}, \bibinfo {author} {\bibfnamefont {W.}~\bibnamefont {Song}}, \bibinfo {author} {\bibfnamefont {L.-W.}\ \bibnamefont {Wang}}, \bibinfo {author} {\bibfnamefont {H.}~\bibnamefont {Xin}}, \bibinfo {author} {\bibfnamefont {J.}~\bibnamefont {Sun}}, \bibinfo {author} {\bibfnamefont {S.}~\bibnamefont {Wu}}, \bibinfo {author} {\bibfnamefont {C.}~\bibnamefont {Huang}}, \bibinfo {author} {\bibfnamefont {S.}~\bibnamefont {Zhu}}, \bibinfo {author} {\bibfnamefont {J.-H.}\ \bibnamefont {Jiang}},\ and\ \bibinfo {author} {\bibfnamefont {T.}~\bibnamefont {Li}},\ }\bibfield  {title} {\bibinfo {title} {Observation of topological transition in floquet non-hermitian skin effects in silicon photonics},\ }\href@noop {} {\bibfield  {journal} {\bibinfo  {journal} {Physical Review Letters}\ }\textbf {\bibinfo {volume} {133}},\ \bibinfo {pages} {073803} (\bibinfo {year} {2024})}\BibitemShut {NoStop}%
\bibitem [{\citenamefont {Liu}\ \emph {et~al.}(2024{\natexlab{a}})\citenamefont {Liu}, \citenamefont {Cao}, \citenamefont {Qi}, \citenamefont {Huang}, \citenamefont {Gao}, \citenamefont {Peng}, \citenamefont {Li},\ and\ \citenamefont {Zhu}}]{liu2024observation}%
  \BibitemOpen
  \bibfield  {author} {\bibinfo {author} {\bibfnamefont {Y.-K.}\ \bibnamefont {Liu}}, \bibinfo {author} {\bibfnamefont {P.-C.}\ \bibnamefont {Cao}}, \bibinfo {author} {\bibfnamefont {M.}~\bibnamefont {Qi}}, \bibinfo {author} {\bibfnamefont {Q.-K.-L.}\ \bibnamefont {Huang}}, \bibinfo {author} {\bibfnamefont {F.}~\bibnamefont {Gao}}, \bibinfo {author} {\bibfnamefont {Y.-G.}\ \bibnamefont {Peng}}, \bibinfo {author} {\bibfnamefont {Y.}~\bibnamefont {Li}},\ and\ \bibinfo {author} {\bibfnamefont {X.-F.}\ \bibnamefont {Zhu}},\ }\bibfield  {title} {\bibinfo {title} {Observation of non-hermitian skin effect in thermal diffusion},\ }\href@noop {} {\bibfield  {journal} {\bibinfo  {journal} {Science Bulletin}\ }\textbf {\bibinfo {volume} {69}},\ \bibinfo {pages} {1228} (\bibinfo {year} {2024}{\natexlab{a}})}\BibitemShut {NoStop}%
\bibitem [{\citenamefont {Yoshida}\ \emph {et~al.}(2024)\citenamefont {Yoshida}, \citenamefont {Zhang}, \citenamefont {Neupert},\ and\ \citenamefont {Kawakami}}]{yoshida2024non}%
  \BibitemOpen
  \bibfield  {author} {\bibinfo {author} {\bibfnamefont {T.}~\bibnamefont {Yoshida}}, \bibinfo {author} {\bibfnamefont {S.-B.}\ \bibnamefont {Zhang}}, \bibinfo {author} {\bibfnamefont {T.}~\bibnamefont {Neupert}},\ and\ \bibinfo {author} {\bibfnamefont {N.}~\bibnamefont {Kawakami}},\ }\bibfield  {title} {\bibinfo {title} {Non-hermitian mott skin effect},\ }\href@noop {} {\bibfield  {journal} {\bibinfo  {journal} {Physical Review Letters}\ }\textbf {\bibinfo {volume} {133}},\ \bibinfo {pages} {076502} (\bibinfo {year} {2024})}\BibitemShut {NoStop}%
\bibitem [{\citenamefont {Gliozzi}\ \emph {et~al.}(2024)\citenamefont {Gliozzi}, \citenamefont {De~Tomasi},\ and\ \citenamefont {Hughes}}]{gliozzi2024many}%
  \BibitemOpen
  \bibfield  {author} {\bibinfo {author} {\bibfnamefont {J.}~\bibnamefont {Gliozzi}}, \bibinfo {author} {\bibfnamefont {G.}~\bibnamefont {De~Tomasi}},\ and\ \bibinfo {author} {\bibfnamefont {T.~L.}\ \bibnamefont {Hughes}},\ }\bibfield  {title} {\bibinfo {title} {Many-body non-hermitian skin effect for multipoles},\ }\href@noop {} {\bibfield  {journal} {\bibinfo  {journal} {Physical review letters}\ }\textbf {\bibinfo {volume} {133}},\ \bibinfo {pages} {136503} (\bibinfo {year} {2024})}\BibitemShut {NoStop}%
\bibitem [{\citenamefont {Shimomura}\ and\ \citenamefont {Sato}(2024)}]{shimomura2024general}%
  \BibitemOpen
  \bibfield  {author} {\bibinfo {author} {\bibfnamefont {K.}~\bibnamefont {Shimomura}}\ and\ \bibinfo {author} {\bibfnamefont {M.}~\bibnamefont {Sato}},\ }\bibfield  {title} {\bibinfo {title} {General criterion for non-hermitian skin effects and application: Fock space skin effects in many-body systems},\ }\href@noop {} {\bibfield  {journal} {\bibinfo  {journal} {Physical Review Letters}\ }\textbf {\bibinfo {volume} {133}},\ \bibinfo {pages} {136502} (\bibinfo {year} {2024})}\BibitemShut {NoStop}%
\bibitem [{\citenamefont {Li}\ \emph {et~al.}(2024{\natexlab{b}})\citenamefont {Li}, \citenamefont {Yu},\ and\ \citenamefont {Li}}]{li2024emergent}%
  \BibitemOpen
  \bibfield  {author} {\bibinfo {author} {\bibfnamefont {S.-Z.}\ \bibnamefont {Li}}, \bibinfo {author} {\bibfnamefont {X.-J.}\ \bibnamefont {Yu}},\ and\ \bibinfo {author} {\bibfnamefont {Z.}~\bibnamefont {Li}},\ }\bibfield  {title} {\bibinfo {title} {Emergent entanglement phase transitions in non-hermitian aubry-andr{\'e}-harper chains},\ }\href@noop {} {\bibfield  {journal} {\bibinfo  {journal} {Physical Review B}\ }\textbf {\bibinfo {volume} {109}},\ \bibinfo {pages} {024306} (\bibinfo {year} {2024}{\natexlab{b}})}\BibitemShut {NoStop}%
\bibitem [{\citenamefont {Yang}\ and\ \citenamefont {Lee}(2025)}]{yang2025beyond}%
  \BibitemOpen
  \bibfield  {author} {\bibinfo {author} {\bibfnamefont {M.}~\bibnamefont {Yang}}\ and\ \bibinfo {author} {\bibfnamefont {C.~H.}\ \bibnamefont {Lee}},\ }\bibfield  {title} {\bibinfo {title} {Beyond symmetry protection: Robust feedback-enforced edge states in non-hermitian stacked quantum spin hall systems},\ }\href@noop {} {\bibfield  {journal} {\bibinfo  {journal} {arXiv preprint arXiv:2507.17295}\ } (\bibinfo {year} {2025})}\BibitemShut {NoStop}%
\bibitem [{\citenamefont {Qin}\ \emph {et~al.}(2025)\citenamefont {Qin}, \citenamefont {Ang}, \citenamefont {Lee},\ and\ \citenamefont {Li}}]{qin2025many}%
  \BibitemOpen
  \bibfield  {author} {\bibinfo {author} {\bibfnamefont {Y.}~\bibnamefont {Qin}}, \bibinfo {author} {\bibfnamefont {Y.~S.}\ \bibnamefont {Ang}}, \bibinfo {author} {\bibfnamefont {C.~H.}\ \bibnamefont {Lee}},\ and\ \bibinfo {author} {\bibfnamefont {L.}~\bibnamefont {Li}},\ }\bibfield  {title} {\bibinfo {title} {Many-body critical non-hermitian skin effect},\ }\href@noop {} {\bibfield  {journal} {\bibinfo  {journal} {arXiv preprint arXiv:2506.01383}\ } (\bibinfo {year} {2025})}\BibitemShut {NoStop}%
\bibitem [{\citenamefont {Li}\ \emph {et~al.}(2025{\natexlab{a}})\citenamefont {Li}, \citenamefont {Zhang}, \citenamefont {Kou}, \citenamefont {Xiao}, \citenamefont {Jia}, \citenamefont {Li},\ and\ \citenamefont {Mei}}]{li2025observation}%
  \BibitemOpen
  \bibfield  {author} {\bibinfo {author} {\bibfnamefont {Y.}~\bibnamefont {Li}}, \bibinfo {author} {\bibfnamefont {J.-H.}\ \bibnamefont {Zhang}}, \bibinfo {author} {\bibfnamefont {Y.}~\bibnamefont {Kou}}, \bibinfo {author} {\bibfnamefont {L.}~\bibnamefont {Xiao}}, \bibinfo {author} {\bibfnamefont {S.}~\bibnamefont {Jia}}, \bibinfo {author} {\bibfnamefont {L.}~\bibnamefont {Li}},\ and\ \bibinfo {author} {\bibfnamefont {F.}~\bibnamefont {Mei}},\ }\bibfield  {title} {\bibinfo {title} {Observation of gauge field induced non-hermitian helical spin skin effects},\ }\href@noop {} {\bibfield  {journal} {\bibinfo  {journal} {arXiv preprint arXiv:2504.18063}\ } (\bibinfo {year} {2025}{\natexlab{a}})}\BibitemShut {NoStop}%
\bibitem [{\citenamefont {Shen}\ \emph {et~al.}(2025{\natexlab{a}})\citenamefont {Shen}, \citenamefont {Chan},\ and\ \citenamefont {Lee}}]{shen2025non}%
  \BibitemOpen
  \bibfield  {author} {\bibinfo {author} {\bibfnamefont {R.}~\bibnamefont {Shen}}, \bibinfo {author} {\bibfnamefont {W.~J.}\ \bibnamefont {Chan}},\ and\ \bibinfo {author} {\bibfnamefont {C.~H.}\ \bibnamefont {Lee}},\ }\bibfield  {title} {\bibinfo {title} {Non-hermitian skin effect along hyperbolic geodesics},\ }\href@noop {} {\bibfield  {journal} {\bibinfo  {journal} {Physical Review B}\ }\textbf {\bibinfo {volume} {111}},\ \bibinfo {pages} {045420} (\bibinfo {year} {2025}{\natexlab{a}})}\BibitemShut {NoStop}%
\bibitem [{\citenamefont {Hamanaka}\ and\ \citenamefont {Kawabata}(2025)}]{hamanaka2025multifractality}%
  \BibitemOpen
  \bibfield  {author} {\bibinfo {author} {\bibfnamefont {S.}~\bibnamefont {Hamanaka}}\ and\ \bibinfo {author} {\bibfnamefont {K.}~\bibnamefont {Kawabata}},\ }\bibfield  {title} {\bibinfo {title} {Multifractality of the many-body non-hermitian skin effect},\ }\href@noop {} {\bibfield  {journal} {\bibinfo  {journal} {Physical Review B}\ }\textbf {\bibinfo {volume} {111}},\ \bibinfo {pages} {035144} (\bibinfo {year} {2025})}\BibitemShut {NoStop}%
\bibitem [{\citenamefont {Li}\ \emph {et~al.}(2025{\natexlab{b}})\citenamefont {Li}, \citenamefont {Jiang},\ and\ \citenamefont {Lee}}]{li2025phase}%
  \BibitemOpen
  \bibfield  {author} {\bibinfo {author} {\bibfnamefont {Q.}~\bibnamefont {Li}}, \bibinfo {author} {\bibfnamefont {H.}~\bibnamefont {Jiang}},\ and\ \bibinfo {author} {\bibfnamefont {C.~H.}\ \bibnamefont {Lee}},\ }\bibfield  {title} {\bibinfo {title} {Phase-space generalized brillouin zone for spatially inhomogeneous non-hermitian systems},\ }\href@noop {} {\bibfield  {journal} {\bibinfo  {journal} {arXiv preprint arXiv:2501.09785}\ } (\bibinfo {year} {2025}{\natexlab{b}})}\BibitemShut {NoStop}%
\bibitem [{\citenamefont {Rafi-Ul-Islam}\ \emph {et~al.}(2025)\citenamefont {Rafi-Ul-Islam}, \citenamefont {Siu}, \citenamefont {Razo},\ and\ \citenamefont {Jalil}}]{rafi2025critical}%
  \BibitemOpen
  \bibfield  {author} {\bibinfo {author} {\bibfnamefont {S.}~\bibnamefont {Rafi-Ul-Islam}}, \bibinfo {author} {\bibfnamefont {Z.~B.}\ \bibnamefont {Siu}}, \bibinfo {author} {\bibfnamefont {M.~S.~H.}\ \bibnamefont {Razo}},\ and\ \bibinfo {author} {\bibfnamefont {M.~B.}\ \bibnamefont {Jalil}},\ }\bibfield  {title} {\bibinfo {title} {Critical non-hermitian skin effect in a cross-coupled hermitian chain},\ }\href@noop {} {\bibfield  {journal} {\bibinfo  {journal} {Physical Review B}\ }\textbf {\bibinfo {volume} {111}},\ \bibinfo {pages} {115415} (\bibinfo {year} {2025})}\BibitemShut {NoStop}%
\bibitem [{\citenamefont {Shen}\ \emph {et~al.}(2025{\natexlab{b}})\citenamefont {Shen}, \citenamefont {Chen}, \citenamefont {Yang},\ and\ \citenamefont {Lee}}]{shen2025observation}%
  \BibitemOpen
  \bibfield  {author} {\bibinfo {author} {\bibfnamefont {R.}~\bibnamefont {Shen}}, \bibinfo {author} {\bibfnamefont {T.}~\bibnamefont {Chen}}, \bibinfo {author} {\bibfnamefont {B.}~\bibnamefont {Yang}},\ and\ \bibinfo {author} {\bibfnamefont {C.~H.}\ \bibnamefont {Lee}},\ }\bibfield  {title} {\bibinfo {title} {Observation of the non-hermitian skin effect and fermi skin on a digital quantum computer},\ }\href@noop {} {\bibfield  {journal} {\bibinfo  {journal} {Nature Communications}\ }\textbf {\bibinfo {volume} {16}},\ \bibinfo {pages} {1340} (\bibinfo {year} {2025}{\natexlab{b}})}\BibitemShut {NoStop}%
\bibitem [{\citenamefont {Yokomizo}\ and\ \citenamefont {Murakami}(2019)}]{yokomizo2019non}%
  \BibitemOpen
  \bibfield  {author} {\bibinfo {author} {\bibfnamefont {K.}~\bibnamefont {Yokomizo}}\ and\ \bibinfo {author} {\bibfnamefont {S.}~\bibnamefont {Murakami}},\ }\bibfield  {title} {\bibinfo {title} {Non-bloch band theory of non-hermitian systems},\ }\href@noop {} {\bibfield  {journal} {\bibinfo  {journal} {Physical review letters}\ }\textbf {\bibinfo {volume} {123}},\ \bibinfo {pages} {066404} (\bibinfo {year} {2019})}\BibitemShut {NoStop}%
\bibitem [{\citenamefont {Okuma}\ \emph {et~al.}(2020)\citenamefont {Okuma}, \citenamefont {Kawabata}, \citenamefont {Shiozaki},\ and\ \citenamefont {Sato}}]{okuma2020topological}%
  \BibitemOpen
  \bibfield  {author} {\bibinfo {author} {\bibfnamefont {N.}~\bibnamefont {Okuma}}, \bibinfo {author} {\bibfnamefont {K.}~\bibnamefont {Kawabata}}, \bibinfo {author} {\bibfnamefont {K.}~\bibnamefont {Shiozaki}},\ and\ \bibinfo {author} {\bibfnamefont {M.}~\bibnamefont {Sato}},\ }\bibfield  {title} {\bibinfo {title} {Topological origin of non-hermitian skin effects},\ }\href@noop {} {\bibfield  {journal} {\bibinfo  {journal} {Physical review letters}\ }\textbf {\bibinfo {volume} {124}},\ \bibinfo {pages} {086801} (\bibinfo {year} {2020})}\BibitemShut {NoStop}%
\bibitem [{\citenamefont {Xiao}\ \emph {et~al.}(2020)\citenamefont {Xiao}, \citenamefont {Deng}, \citenamefont {Wang}, \citenamefont {Zhu}, \citenamefont {Wang}, \citenamefont {Yi},\ and\ \citenamefont {Xue}}]{xiao2020non}%
  \BibitemOpen
  \bibfield  {author} {\bibinfo {author} {\bibfnamefont {L.}~\bibnamefont {Xiao}}, \bibinfo {author} {\bibfnamefont {T.}~\bibnamefont {Deng}}, \bibinfo {author} {\bibfnamefont {K.}~\bibnamefont {Wang}}, \bibinfo {author} {\bibfnamefont {G.}~\bibnamefont {Zhu}}, \bibinfo {author} {\bibfnamefont {Z.}~\bibnamefont {Wang}}, \bibinfo {author} {\bibfnamefont {W.}~\bibnamefont {Yi}},\ and\ \bibinfo {author} {\bibfnamefont {P.}~\bibnamefont {Xue}},\ }\bibfield  {title} {\bibinfo {title} {Non-hermitian bulk--boundary correspondence in quantum dynamics},\ }\href@noop {} {\bibfield  {journal} {\bibinfo  {journal} {Nature Physics}\ }\textbf {\bibinfo {volume} {16}},\ \bibinfo {pages} {761} (\bibinfo {year} {2020})}\BibitemShut {NoStop}%
\bibitem [{\citenamefont {Lee}\ \emph {et~al.}(2020)\citenamefont {Lee}, \citenamefont {Li}, \citenamefont {Thomale},\ and\ \citenamefont {Gong}}]{lee2020unraveling}%
  \BibitemOpen
  \bibfield  {author} {\bibinfo {author} {\bibfnamefont {C.~H.}\ \bibnamefont {Lee}}, \bibinfo {author} {\bibfnamefont {L.}~\bibnamefont {Li}}, \bibinfo {author} {\bibfnamefont {R.}~\bibnamefont {Thomale}},\ and\ \bibinfo {author} {\bibfnamefont {J.}~\bibnamefont {Gong}},\ }\bibfield  {title} {\bibinfo {title} {Unraveling non-hermitian pumping: emergent spectral singularities and anomalous responses},\ }\href@noop {} {\bibfield  {journal} {\bibinfo  {journal} {Physical Review B}\ }\textbf {\bibinfo {volume} {102}},\ \bibinfo {pages} {085151} (\bibinfo {year} {2020})}\BibitemShut {NoStop}%
\bibitem [{\citenamefont {Li}\ \emph {et~al.}(2021{\natexlab{a}})\citenamefont {Li}, \citenamefont {Lee},\ and\ \citenamefont {Gong}}]{li2021impurity}%
  \BibitemOpen
  \bibfield  {author} {\bibinfo {author} {\bibfnamefont {L.}~\bibnamefont {Li}}, \bibinfo {author} {\bibfnamefont {C.~H.}\ \bibnamefont {Lee}},\ and\ \bibinfo {author} {\bibfnamefont {J.}~\bibnamefont {Gong}},\ }\bibfield  {title} {\bibinfo {title} {Impurity induced scale-free localization},\ }\href@noop {} {\bibfield  {journal} {\bibinfo  {journal} {Communications Physics}\ }\textbf {\bibinfo {volume} {4}},\ \bibinfo {pages} {42} (\bibinfo {year} {2021}{\natexlab{a}})}\BibitemShut {NoStop}%
\bibitem [{\citenamefont {Shen}\ and\ \citenamefont {Lee}(2022)}]{shen2022non}%
  \BibitemOpen
  \bibfield  {author} {\bibinfo {author} {\bibfnamefont {R.}~\bibnamefont {Shen}}\ and\ \bibinfo {author} {\bibfnamefont {C.~H.}\ \bibnamefont {Lee}},\ }\bibfield  {title} {\bibinfo {title} {Non-hermitian skin clusters from strong interactions},\ }\href@noop {} {\bibfield  {journal} {\bibinfo  {journal} {Communications Physics}\ }\textbf {\bibinfo {volume} {5}},\ \bibinfo {pages} {238} (\bibinfo {year} {2022})}\BibitemShut {NoStop}%
\bibitem [{\citenamefont {Pan}\ \emph {et~al.}(2020)\citenamefont {Pan}, \citenamefont {Chen}, \citenamefont {Chen},\ and\ \citenamefont {Zhai}}]{pan2020non}%
  \BibitemOpen
  \bibfield  {author} {\bibinfo {author} {\bibfnamefont {L.}~\bibnamefont {Pan}}, \bibinfo {author} {\bibfnamefont {X.}~\bibnamefont {Chen}}, \bibinfo {author} {\bibfnamefont {Y.}~\bibnamefont {Chen}},\ and\ \bibinfo {author} {\bibfnamefont {H.}~\bibnamefont {Zhai}},\ }\bibfield  {title} {\bibinfo {title} {Non-hermitian linear response theory},\ }\href@noop {} {\bibfield  {journal} {\bibinfo  {journal} {Nature Physics}\ }\textbf {\bibinfo {volume} {16}},\ \bibinfo {pages} {767} (\bibinfo {year} {2020})}\BibitemShut {NoStop}%
\bibitem [{\citenamefont {Budich}\ and\ \citenamefont {Bergholtz}(2020)}]{budich2020non}%
  \BibitemOpen
  \bibfield  {author} {\bibinfo {author} {\bibfnamefont {J.~C.}\ \bibnamefont {Budich}}\ and\ \bibinfo {author} {\bibfnamefont {E.~J.}\ \bibnamefont {Bergholtz}},\ }\bibfield  {title} {\bibinfo {title} {Non-hermitian topological sensors},\ }\href@noop {} {\bibfield  {journal} {\bibinfo  {journal} {Physical Review Letters}\ }\textbf {\bibinfo {volume} {125}},\ \bibinfo {pages} {180403} (\bibinfo {year} {2020})}\BibitemShut {NoStop}%
\bibitem [{\citenamefont {Yoshida}\ \emph {et~al.}(2020)\citenamefont {Yoshida}, \citenamefont {Mizoguchi},\ and\ \citenamefont {Hatsugai}}]{yoshida2020mirror}%
  \BibitemOpen
  \bibfield  {author} {\bibinfo {author} {\bibfnamefont {T.}~\bibnamefont {Yoshida}}, \bibinfo {author} {\bibfnamefont {T.}~\bibnamefont {Mizoguchi}},\ and\ \bibinfo {author} {\bibfnamefont {Y.}~\bibnamefont {Hatsugai}},\ }\bibfield  {title} {\bibinfo {title} {Mirror skin effect and its electric circuit simulation},\ }\href@noop {} {\bibfield  {journal} {\bibinfo  {journal} {Physical Review Research}\ }\textbf {\bibinfo {volume} {2}},\ \bibinfo {pages} {022062} (\bibinfo {year} {2020})}\BibitemShut {NoStop}%
\bibitem [{\citenamefont {Yang}\ \emph {et~al.}(2022{\natexlab{b}})\citenamefont {Yang}, \citenamefont {Jiang},\ and\ \citenamefont {Bergholtz}}]{yang2022liouvillian}%
  \BibitemOpen
  \bibfield  {author} {\bibinfo {author} {\bibfnamefont {F.}~\bibnamefont {Yang}}, \bibinfo {author} {\bibfnamefont {Q.-D.}\ \bibnamefont {Jiang}},\ and\ \bibinfo {author} {\bibfnamefont {E.~J.}\ \bibnamefont {Bergholtz}},\ }\bibfield  {title} {\bibinfo {title} {Liouvillian skin effect in an exactly solvable model},\ }\href@noop {} {\bibfield  {journal} {\bibinfo  {journal} {Physical Review Research}\ }\textbf {\bibinfo {volume} {4}},\ \bibinfo {pages} {023160} (\bibinfo {year} {2022}{\natexlab{b}})}\BibitemShut {NoStop}%
\bibitem [{\citenamefont {Zhou}\ \emph {et~al.}(2022)\citenamefont {Zhou}, \citenamefont {Zhou}, \citenamefont {Zhang},\ and\ \citenamefont {Zhai}}]{zhou2022space}%
  \BibitemOpen
  \bibfield  {author} {\bibinfo {author} {\bibfnamefont {T.-G.}\ \bibnamefont {Zhou}}, \bibinfo {author} {\bibfnamefont {Y.-N.}\ \bibnamefont {Zhou}}, \bibinfo {author} {\bibfnamefont {P.}~\bibnamefont {Zhang}},\ and\ \bibinfo {author} {\bibfnamefont {H.}~\bibnamefont {Zhai}},\ }\bibfield  {title} {\bibinfo {title} {Space-time duality between quantum chaos and non-hermitian boundary effect},\ }\href@noop {} {\bibfield  {journal} {\bibinfo  {journal} {Physical Review Research}\ }\textbf {\bibinfo {volume} {4}},\ \bibinfo {pages} {L022039} (\bibinfo {year} {2022})}\BibitemShut {NoStop}%
\bibitem [{\citenamefont {Edvardsson}\ and\ \citenamefont {Ardonne}(2022)}]{edvardsson2022sensitivity}%
  \BibitemOpen
  \bibfield  {author} {\bibinfo {author} {\bibfnamefont {E.}~\bibnamefont {Edvardsson}}\ and\ \bibinfo {author} {\bibfnamefont {E.}~\bibnamefont {Ardonne}},\ }\bibfield  {title} {\bibinfo {title} {Sensitivity of non-hermitian systems},\ }\href@noop {} {\bibfield  {journal} {\bibinfo  {journal} {Physical Review B}\ }\textbf {\bibinfo {volume} {106}},\ \bibinfo {pages} {115107} (\bibinfo {year} {2022})}\BibitemShut {NoStop}%
\bibitem [{\citenamefont {Dong}\ \emph {et~al.}(2021)\citenamefont {Dong}, \citenamefont {Juri{\v{c}}i{\'c}},\ and\ \citenamefont {Roy}}]{dong2021topolectric}%
  \BibitemOpen
  \bibfield  {author} {\bibinfo {author} {\bibfnamefont {J.}~\bibnamefont {Dong}}, \bibinfo {author} {\bibfnamefont {V.}~\bibnamefont {Juri{\v{c}}i{\'c}}},\ and\ \bibinfo {author} {\bibfnamefont {B.}~\bibnamefont {Roy}},\ }\bibfield  {title} {\bibinfo {title} {Topolectric circuits: Theory and construction},\ }\href@noop {} {\bibfield  {journal} {\bibinfo  {journal} {Physical Review Research}\ }\textbf {\bibinfo {volume} {3}},\ \bibinfo {pages} {023056} (\bibinfo {year} {2021})}\BibitemShut {NoStop}%
\bibitem [{\citenamefont {Hohmann}\ \emph {et~al.}(2023)\citenamefont {Hohmann}, \citenamefont {Hofmann}, \citenamefont {Helbig}, \citenamefont {Imhof}, \citenamefont {Brand}, \citenamefont {Upreti}, \citenamefont {Stegmaier}, \citenamefont {Fritzsche}, \citenamefont {M{\"u}ller}, \citenamefont {Schwingenschl{\"o}gl} \emph {et~al.}}]{hohmann2023observation}%
  \BibitemOpen
  \bibfield  {author} {\bibinfo {author} {\bibfnamefont {H.}~\bibnamefont {Hohmann}}, \bibinfo {author} {\bibfnamefont {T.}~\bibnamefont {Hofmann}}, \bibinfo {author} {\bibfnamefont {T.}~\bibnamefont {Helbig}}, \bibinfo {author} {\bibfnamefont {S.}~\bibnamefont {Imhof}}, \bibinfo {author} {\bibfnamefont {H.}~\bibnamefont {Brand}}, \bibinfo {author} {\bibfnamefont {L.~K.}\ \bibnamefont {Upreti}}, \bibinfo {author} {\bibfnamefont {A.}~\bibnamefont {Stegmaier}}, \bibinfo {author} {\bibfnamefont {A.}~\bibnamefont {Fritzsche}}, \bibinfo {author} {\bibfnamefont {T.}~\bibnamefont {M{\"u}ller}}, \bibinfo {author} {\bibfnamefont {U.}~\bibnamefont {Schwingenschl{\"o}gl}}, \emph {et~al.},\ }\bibfield  {title} {\bibinfo {title} {Observation of cnoidal wave localization in nonlinear topolectric circuits},\ }\href@noop {} {\bibfield  {journal} {\bibinfo  {journal} {Physical Review Research}\ }\textbf {\bibinfo {volume} {5}},\ \bibinfo {pages} {L012041} (\bibinfo {year} {2023})}\BibitemShut {NoStop}%
\bibitem [{\citenamefont {Shang}\ \emph {et~al.}(2024)\citenamefont {Shang}, \citenamefont {Liu}, \citenamefont {Jiang}, \citenamefont {Shao}, \citenamefont {Zang}, \citenamefont {Lee}, \citenamefont {Thomale}, \citenamefont {Manchon}, \citenamefont {Cui},\ and\ \citenamefont {Schwingenschl{\"o}gl}}]{shang2024observation}%
  \BibitemOpen
  \bibfield  {author} {\bibinfo {author} {\bibfnamefont {C.}~\bibnamefont {Shang}}, \bibinfo {author} {\bibfnamefont {S.}~\bibnamefont {Liu}}, \bibinfo {author} {\bibfnamefont {C.}~\bibnamefont {Jiang}}, \bibinfo {author} {\bibfnamefont {R.}~\bibnamefont {Shao}}, \bibinfo {author} {\bibfnamefont {X.}~\bibnamefont {Zang}}, \bibinfo {author} {\bibfnamefont {C.~H.}\ \bibnamefont {Lee}}, \bibinfo {author} {\bibfnamefont {R.}~\bibnamefont {Thomale}}, \bibinfo {author} {\bibfnamefont {A.}~\bibnamefont {Manchon}}, \bibinfo {author} {\bibfnamefont {T.~J.}\ \bibnamefont {Cui}},\ and\ \bibinfo {author} {\bibfnamefont {U.}~\bibnamefont {Schwingenschl{\"o}gl}},\ }\bibfield  {title} {\bibinfo {title} {Observation of a higher-order end topological insulator in a real projective lattice},\ }\href@noop {} {\bibfield  {journal} {\bibinfo  {journal} {Advanced Science}\ }\textbf {\bibinfo {volume} {11}},\ \bibinfo {pages} {2303222} (\bibinfo {year} {2024})}\BibitemShut {NoStop}%
\bibitem [{\citenamefont {Zou}\ \emph {et~al.}(2025)\citenamefont {Zou}, \citenamefont {Chen}, \citenamefont {Lee},\ and\ \citenamefont {Zhang}}]{zou2025experimental}%
  \BibitemOpen
  \bibfield  {author} {\bibinfo {author} {\bibfnamefont {D.}~\bibnamefont {Zou}}, \bibinfo {author} {\bibfnamefont {T.}~\bibnamefont {Chen}}, \bibinfo {author} {\bibfnamefont {C.~H.}\ \bibnamefont {Lee}},\ and\ \bibinfo {author} {\bibfnamefont {X.}~\bibnamefont {Zhang}},\ }\bibfield  {title} {\bibinfo {title} {Experimental simulation of negative entanglement entropy scaling with electrical circuits},\ }\href@noop {} {\bibfield  {journal} {\bibinfo  {journal} {Physical Review B}\ }\textbf {\bibinfo {volume} {111}},\ \bibinfo {pages} {214119} (\bibinfo {year} {2025})}\BibitemShut {NoStop}%
\bibitem [{\citenamefont {Pan}\ \emph {et~al.}(2018)\citenamefont {Pan}, \citenamefont {Zhao}, \citenamefont {Miao}, \citenamefont {Longhi},\ and\ \citenamefont {Feng}}]{pan2018photonic}%
  \BibitemOpen
  \bibfield  {author} {\bibinfo {author} {\bibfnamefont {M.}~\bibnamefont {Pan}}, \bibinfo {author} {\bibfnamefont {H.}~\bibnamefont {Zhao}}, \bibinfo {author} {\bibfnamefont {P.}~\bibnamefont {Miao}}, \bibinfo {author} {\bibfnamefont {S.}~\bibnamefont {Longhi}},\ and\ \bibinfo {author} {\bibfnamefont {L.}~\bibnamefont {Feng}},\ }\bibfield  {title} {\bibinfo {title} {Photonic zero mode in a non-hermitian photonic lattice},\ }\href@noop {} {\bibfield  {journal} {\bibinfo  {journal} {Nature communications}\ }\textbf {\bibinfo {volume} {9}},\ \bibinfo {pages} {1308} (\bibinfo {year} {2018})}\BibitemShut {NoStop}%
\bibitem [{\citenamefont {Wang}\ \emph {et~al.}(2021)\citenamefont {Wang}, \citenamefont {Li}, \citenamefont {Xiao}, \citenamefont {Han}, \citenamefont {Yi},\ and\ \citenamefont {Xue}}]{wang2021detecting}%
  \BibitemOpen
  \bibfield  {author} {\bibinfo {author} {\bibfnamefont {K.}~\bibnamefont {Wang}}, \bibinfo {author} {\bibfnamefont {T.}~\bibnamefont {Li}}, \bibinfo {author} {\bibfnamefont {L.}~\bibnamefont {Xiao}}, \bibinfo {author} {\bibfnamefont {Y.}~\bibnamefont {Han}}, \bibinfo {author} {\bibfnamefont {W.}~\bibnamefont {Yi}},\ and\ \bibinfo {author} {\bibfnamefont {P.}~\bibnamefont {Xue}},\ }\bibfield  {title} {\bibinfo {title} {Detecting non-bloch topological invariants in quantum dynamics},\ }\href@noop {} {\bibfield  {journal} {\bibinfo  {journal} {Physical Review Letters}\ }\textbf {\bibinfo {volume} {127}},\ \bibinfo {pages} {270602} (\bibinfo {year} {2021})}\BibitemShut {NoStop}%
\bibitem [{\citenamefont {Lin}\ \emph {et~al.}(2022)\citenamefont {Lin}, \citenamefont {Li}, \citenamefont {Xiao}, \citenamefont {Wang}, \citenamefont {Yi},\ and\ \citenamefont {Xue}}]{lin2022topological}%
  \BibitemOpen
  \bibfield  {author} {\bibinfo {author} {\bibfnamefont {Q.}~\bibnamefont {Lin}}, \bibinfo {author} {\bibfnamefont {T.}~\bibnamefont {Li}}, \bibinfo {author} {\bibfnamefont {L.}~\bibnamefont {Xiao}}, \bibinfo {author} {\bibfnamefont {K.}~\bibnamefont {Wang}}, \bibinfo {author} {\bibfnamefont {W.}~\bibnamefont {Yi}},\ and\ \bibinfo {author} {\bibfnamefont {P.}~\bibnamefont {Xue}},\ }\bibfield  {title} {\bibinfo {title} {Topological phase transitions and mobility edges in non-hermitian quasicrystals},\ }\href@noop {} {\bibfield  {journal} {\bibinfo  {journal} {Physical Review Letters}\ }\textbf {\bibinfo {volume} {129}},\ \bibinfo {pages} {113601} (\bibinfo {year} {2022})}\BibitemShut {NoStop}%
\bibitem [{\citenamefont {Xiao}\ \emph {et~al.}(2024)\citenamefont {Xiao}, \citenamefont {Xue}, \citenamefont {Song}, \citenamefont {Hu}, \citenamefont {Yi}, \citenamefont {Wang},\ and\ \citenamefont {Xue}}]{xiao2024observation}%
  \BibitemOpen
  \bibfield  {author} {\bibinfo {author} {\bibfnamefont {L.}~\bibnamefont {Xiao}}, \bibinfo {author} {\bibfnamefont {W.-T.}\ \bibnamefont {Xue}}, \bibinfo {author} {\bibfnamefont {F.}~\bibnamefont {Song}}, \bibinfo {author} {\bibfnamefont {Y.-M.}\ \bibnamefont {Hu}}, \bibinfo {author} {\bibfnamefont {W.}~\bibnamefont {Yi}}, \bibinfo {author} {\bibfnamefont {Z.}~\bibnamefont {Wang}},\ and\ \bibinfo {author} {\bibfnamefont {P.}~\bibnamefont {Xue}},\ }\bibfield  {title} {\bibinfo {title} {Observation of non-hermitian edge burst in quantum dynamics},\ }\href@noop {} {\bibfield  {journal} {\bibinfo  {journal} {Physical Review Letters}\ }\textbf {\bibinfo {volume} {133}},\ \bibinfo {pages} {070801} (\bibinfo {year} {2024})}\BibitemShut {NoStop}%
\bibitem [{\citenamefont {Lei}\ \emph {et~al.}(2024)\citenamefont {Lei}, \citenamefont {Lee},\ and\ \citenamefont {Li}}]{lei2024activating}%
  \BibitemOpen
  \bibfield  {author} {\bibinfo {author} {\bibfnamefont {Z.}~\bibnamefont {Lei}}, \bibinfo {author} {\bibfnamefont {C.~H.}\ \bibnamefont {Lee}},\ and\ \bibinfo {author} {\bibfnamefont {L.}~\bibnamefont {Li}},\ }\bibfield  {title} {\bibinfo {title} {Activating non-hermitian skin modes by parity-time symmetry breaking},\ }\href@noop {} {\bibfield  {journal} {\bibinfo  {journal} {Communications Physics}\ }\textbf {\bibinfo {volume} {7}},\ \bibinfo {pages} {100} (\bibinfo {year} {2024})}\BibitemShut {NoStop}%
\bibitem [{\citenamefont {Rafi-Ul-Islam}\ \emph {et~al.}(2022)\citenamefont {Rafi-Ul-Islam}, \citenamefont {Siu}, \citenamefont {Sahin}, \citenamefont {Lee},\ and\ \citenamefont {Jalil}}]{rafi2022system}%
  \BibitemOpen
  \bibfield  {author} {\bibinfo {author} {\bibfnamefont {S.}~\bibnamefont {Rafi-Ul-Islam}}, \bibinfo {author} {\bibfnamefont {Z.~B.}\ \bibnamefont {Siu}}, \bibinfo {author} {\bibfnamefont {H.}~\bibnamefont {Sahin}}, \bibinfo {author} {\bibfnamefont {C.~H.}\ \bibnamefont {Lee}},\ and\ \bibinfo {author} {\bibfnamefont {M.~B.}\ \bibnamefont {Jalil}},\ }\bibfield  {title} {\bibinfo {title} {System size dependent topological zero modes in coupled topolectrical chains},\ }\href@noop {} {\bibfield  {journal} {\bibinfo  {journal} {Physical Review B}\ }\textbf {\bibinfo {volume} {106}},\ \bibinfo {pages} {075158} (\bibinfo {year} {2022})}\BibitemShut {NoStop}%
\bibitem [{\citenamefont {Liu}\ \emph {et~al.}(2024{\natexlab{b}})\citenamefont {Liu}, \citenamefont {Jiang}, \citenamefont {Xue}, \citenamefont {Li}, \citenamefont {Gong}, \citenamefont {Liu},\ and\ \citenamefont {Lee}}]{liu2024non}%
  \BibitemOpen
  \bibfield  {author} {\bibinfo {author} {\bibfnamefont {S.}~\bibnamefont {Liu}}, \bibinfo {author} {\bibfnamefont {H.}~\bibnamefont {Jiang}}, \bibinfo {author} {\bibfnamefont {W.-T.}\ \bibnamefont {Xue}}, \bibinfo {author} {\bibfnamefont {Q.}~\bibnamefont {Li}}, \bibinfo {author} {\bibfnamefont {J.}~\bibnamefont {Gong}}, \bibinfo {author} {\bibfnamefont {X.}~\bibnamefont {Liu}},\ and\ \bibinfo {author} {\bibfnamefont {C.~H.}\ \bibnamefont {Lee}},\ }\bibfield  {title} {\bibinfo {title} {Non-hermitian entanglement dip from scaling-induced exceptional criticality},\ }\href@noop {} {\bibfield  {journal} {\bibinfo  {journal} {arXiv preprint arXiv:2408.02736}\ } (\bibinfo {year} {2024}{\natexlab{b}})}\BibitemShut {NoStop}%
\bibitem [{\citenamefont {Hatano}\ and\ \citenamefont {Nelson}(1996)}]{hatano1996localization}%
  \BibitemOpen
  \bibfield  {author} {\bibinfo {author} {\bibfnamefont {N.}~\bibnamefont {Hatano}}\ and\ \bibinfo {author} {\bibfnamefont {D.~R.}\ \bibnamefont {Nelson}},\ }\bibfield  {title} {\bibinfo {title} {Localization transitions in non-hermitian quantum mechanics},\ }\href@noop {} {\bibfield  {journal} {\bibinfo  {journal} {Physical review letters}\ }\textbf {\bibinfo {volume} {77}},\ \bibinfo {pages} {570} (\bibinfo {year} {1996})}\BibitemShut {NoStop}%
\bibitem [{\citenamefont {Ou}\ \emph {et~al.}(2023)\citenamefont {Ou}, \citenamefont {Wang},\ and\ \citenamefont {Li}}]{ou2023non}%
  \BibitemOpen
  \bibfield  {author} {\bibinfo {author} {\bibfnamefont {Z.}~\bibnamefont {Ou}}, \bibinfo {author} {\bibfnamefont {Y.}~\bibnamefont {Wang}},\ and\ \bibinfo {author} {\bibfnamefont {L.}~\bibnamefont {Li}},\ }\bibfield  {title} {\bibinfo {title} {Non-hermitian boundary spectral winding},\ }\href@noop {} {\bibfield  {journal} {\bibinfo  {journal} {Physical Review B}\ }\textbf {\bibinfo {volume} {107}},\ \bibinfo {pages} {L161404} (\bibinfo {year} {2023})}\BibitemShut {NoStop}%
\bibitem [{\citenamefont {Wang}\ \emph {et~al.}(2024)\citenamefont {Wang}, \citenamefont {Zhong},\ and\ \citenamefont {Fan}}]{wang2024non}%
  \BibitemOpen
  \bibfield  {author} {\bibinfo {author} {\bibfnamefont {H.}~\bibnamefont {Wang}}, \bibinfo {author} {\bibfnamefont {J.}~\bibnamefont {Zhong}},\ and\ \bibinfo {author} {\bibfnamefont {S.}~\bibnamefont {Fan}},\ }\bibfield  {title} {\bibinfo {title} {Non-hermitian photonic band winding and skin effects: a tutorial},\ }\href@noop {} {\bibfield  {journal} {\bibinfo  {journal} {Advances in Optics and Photonics}\ }\textbf {\bibinfo {volume} {16}},\ \bibinfo {pages} {659} (\bibinfo {year} {2024})}\BibitemShut {NoStop}%
\bibitem [{\citenamefont {Yang}\ \emph {et~al.}(2022{\natexlab{c}})\citenamefont {Yang}, \citenamefont {Tan}, \citenamefont {Tai}, \citenamefont {Koh}, \citenamefont {Li}, \citenamefont {Longhi},\ and\ \citenamefont {Lee}}]{yang2022designing}%
  \BibitemOpen
  \bibfield  {author} {\bibinfo {author} {\bibfnamefont {R.}~\bibnamefont {Yang}}, \bibinfo {author} {\bibfnamefont {J.~W.}\ \bibnamefont {Tan}}, \bibinfo {author} {\bibfnamefont {T.}~\bibnamefont {Tai}}, \bibinfo {author} {\bibfnamefont {J.~M.}\ \bibnamefont {Koh}}, \bibinfo {author} {\bibfnamefont {L.}~\bibnamefont {Li}}, \bibinfo {author} {\bibfnamefont {S.}~\bibnamefont {Longhi}},\ and\ \bibinfo {author} {\bibfnamefont {C.~H.}\ \bibnamefont {Lee}},\ }\bibfield  {title} {\bibinfo {title} {Designing non-hermitian real spectra through electrostatics},\ }\href@noop {} {\bibfield  {journal} {\bibinfo  {journal} {Science Bulletin}\ }\textbf {\bibinfo {volume} {67}},\ \bibinfo {pages} {1865} (\bibinfo {year} {2022}{\natexlab{c}})}\BibitemShut {NoStop}%
\bibitem [{\citenamefont {Xiong}\ and\ \citenamefont {Hu}(2024)}]{xiong2024graph}%
  \BibitemOpen
  \bibfield  {author} {\bibinfo {author} {\bibfnamefont {Y.}~\bibnamefont {Xiong}}\ and\ \bibinfo {author} {\bibfnamefont {H.}~\bibnamefont {Hu}},\ }\bibfield  {title} {\bibinfo {title} {Graph morphology of non-hermitian bands},\ }\href@noop {} {\bibfield  {journal} {\bibinfo  {journal} {Physical Review B}\ }\textbf {\bibinfo {volume} {109}},\ \bibinfo {pages} {L100301} (\bibinfo {year} {2024})}\BibitemShut {NoStop}%
\bibitem [{Note1()}]{Note1}%
  \BibitemOpen
  \bibinfo {note} {We choose not to use $\beta _x=1$ because the $x$-PBC loops are too large for the $x$-OBC spectral details to be visible.}\BibitemShut {Stop}%
\bibitem [{\citenamefont {Fang}\ \emph {et~al.}(2022)\citenamefont {Fang}, \citenamefont {Hu}, \citenamefont {Zhou},\ and\ \citenamefont {Ding}}]{fang2022geometry}%
  \BibitemOpen
  \bibfield  {author} {\bibinfo {author} {\bibfnamefont {Z.}~\bibnamefont {Fang}}, \bibinfo {author} {\bibfnamefont {M.}~\bibnamefont {Hu}}, \bibinfo {author} {\bibfnamefont {L.}~\bibnamefont {Zhou}},\ and\ \bibinfo {author} {\bibfnamefont {K.}~\bibnamefont {Ding}},\ }\bibfield  {title} {\bibinfo {title} {Geometry-dependent skin effects in reciprocal photonic crystals},\ }\href@noop {} {\bibfield  {journal} {\bibinfo  {journal} {Nanophotonics}\ }\textbf {\bibinfo {volume} {11}},\ \bibinfo {pages} {3447} (\bibinfo {year} {2022})}\BibitemShut {NoStop}%
\bibitem [{\citenamefont {Wang}\ \emph {et~al.}(2023{\natexlab{b}})\citenamefont {Wang}, \citenamefont {Hu}, \citenamefont {Wang}, \citenamefont {Ma},\ and\ \citenamefont {Ding}}]{wang2023experimental}%
  \BibitemOpen
  \bibfield  {author} {\bibinfo {author} {\bibfnamefont {W.}~\bibnamefont {Wang}}, \bibinfo {author} {\bibfnamefont {M.}~\bibnamefont {Hu}}, \bibinfo {author} {\bibfnamefont {X.}~\bibnamefont {Wang}}, \bibinfo {author} {\bibfnamefont {G.}~\bibnamefont {Ma}},\ and\ \bibinfo {author} {\bibfnamefont {K.}~\bibnamefont {Ding}},\ }\bibfield  {title} {\bibinfo {title} {Experimental realization of geometry-dependent skin effect in a reciprocal two-dimensional lattice},\ }\href@noop {} {\bibfield  {journal} {\bibinfo  {journal} {Physical Review Letters}\ }\textbf {\bibinfo {volume} {131}},\ \bibinfo {pages} {207201} (\bibinfo {year} {2023}{\natexlab{b}})}\BibitemShut {NoStop}%
\bibitem [{\citenamefont {Wan}\ \emph {et~al.}(2023)\citenamefont {Wan}, \citenamefont {Zhang}, \citenamefont {Li}, \citenamefont {Yang},\ and\ \citenamefont {Yang}}]{wan2023observation}%
  \BibitemOpen
  \bibfield  {author} {\bibinfo {author} {\bibfnamefont {T.}~\bibnamefont {Wan}}, \bibinfo {author} {\bibfnamefont {K.}~\bibnamefont {Zhang}}, \bibinfo {author} {\bibfnamefont {J.}~\bibnamefont {Li}}, \bibinfo {author} {\bibfnamefont {Z.}~\bibnamefont {Yang}},\ and\ \bibinfo {author} {\bibfnamefont {Z.}~\bibnamefont {Yang}},\ }\bibfield  {title} {\bibinfo {title} {Observation of the geometry-dependent skin effect and dynamical degeneracy splitting},\ }\href@noop {} {\bibfield  {journal} {\bibinfo  {journal} {Science Bulletin}\ }\textbf {\bibinfo {volume} {68}},\ \bibinfo {pages} {2330} (\bibinfo {year} {2023})}\BibitemShut {NoStop}%
\bibitem [{\citenamefont {Qin}\ \emph {et~al.}(2024)\citenamefont {Qin}, \citenamefont {Zhang},\ and\ \citenamefont {Li}}]{qin2024geometry}%
  \BibitemOpen
  \bibfield  {author} {\bibinfo {author} {\bibfnamefont {Y.}~\bibnamefont {Qin}}, \bibinfo {author} {\bibfnamefont {K.}~\bibnamefont {Zhang}},\ and\ \bibinfo {author} {\bibfnamefont {L.}~\bibnamefont {Li}},\ }\bibfield  {title} {\bibinfo {title} {Geometry-dependent skin effect and anisotropic bloch oscillations in a non-hermitian optical lattice},\ }\href@noop {} {\bibfield  {journal} {\bibinfo  {journal} {Physical Review A}\ }\textbf {\bibinfo {volume} {109}},\ \bibinfo {pages} {023317} (\bibinfo {year} {2024})}\BibitemShut {NoStop}%
\bibitem [{\citenamefont {Yang}\ \emph {et~al.}(2025)\citenamefont {Yang}, \citenamefont {Fang}, \citenamefont {Zhang},\ and\ \citenamefont {Fang}}]{yang2025tailoring}%
  \BibitemOpen
  \bibfield  {author} {\bibinfo {author} {\bibfnamefont {A.}~\bibnamefont {Yang}}, \bibinfo {author} {\bibfnamefont {Z.}~\bibnamefont {Fang}}, \bibinfo {author} {\bibfnamefont {K.}~\bibnamefont {Zhang}},\ and\ \bibinfo {author} {\bibfnamefont {C.}~\bibnamefont {Fang}},\ }\bibfield  {title} {\bibinfo {title} {Tailoring bound state geometry in high-dimensional non-hermitian systems},\ }\href@noop {} {\bibfield  {journal} {\bibinfo  {journal} {Communications Physics}\ }\textbf {\bibinfo {volume} {8}},\ \bibinfo {pages} {124} (\bibinfo {year} {2025})}\BibitemShut {NoStop}%
\bibitem [{\citenamefont {Hofmann}\ \emph {et~al.}(2019)\citenamefont {Hofmann}, \citenamefont {Helbig}, \citenamefont {Lee}, \citenamefont {Greiter},\ and\ \citenamefont {Thomale}}]{hofmann2019chiral}%
  \BibitemOpen
  \bibfield  {author} {\bibinfo {author} {\bibfnamefont {T.}~\bibnamefont {Hofmann}}, \bibinfo {author} {\bibfnamefont {T.}~\bibnamefont {Helbig}}, \bibinfo {author} {\bibfnamefont {C.~H.}\ \bibnamefont {Lee}}, \bibinfo {author} {\bibfnamefont {M.}~\bibnamefont {Greiter}},\ and\ \bibinfo {author} {\bibfnamefont {R.}~\bibnamefont {Thomale}},\ }\bibfield  {title} {\bibinfo {title} {Chiral voltage propagation and calibration in a topolectrical chern circuit},\ }\href@noop {} {\bibfield  {journal} {\bibinfo  {journal} {Physical review letters}\ }\textbf {\bibinfo {volume} {122}},\ \bibinfo {pages} {247702} (\bibinfo {year} {2019})}\BibitemShut {NoStop}%
\bibitem [{\citenamefont {Ezawa}(2019)}]{ezawa2019electric}%
  \BibitemOpen
  \bibfield  {author} {\bibinfo {author} {\bibfnamefont {M.}~\bibnamefont {Ezawa}},\ }\bibfield  {title} {\bibinfo {title} {Electric circuits for non-hermitian chern insulators},\ }\href@noop {} {\bibfield  {journal} {\bibinfo  {journal} {Physical Review B}\ }\textbf {\bibinfo {volume} {100}},\ \bibinfo {pages} {081401} (\bibinfo {year} {2019})}\BibitemShut {NoStop}%
\bibitem [{\citenamefont {Liu}\ \emph {et~al.}(2020{\natexlab{b}})\citenamefont {Liu}, \citenamefont {Ma}, \citenamefont {Yang}, \citenamefont {Zhang}, \citenamefont {Gao}, \citenamefont {Xiang}, \citenamefont {Cui},\ and\ \citenamefont {Zhang}}]{liu2020gain}%
  \BibitemOpen
  \bibfield  {author} {\bibinfo {author} {\bibfnamefont {S.}~\bibnamefont {Liu}}, \bibinfo {author} {\bibfnamefont {S.}~\bibnamefont {Ma}}, \bibinfo {author} {\bibfnamefont {C.}~\bibnamefont {Yang}}, \bibinfo {author} {\bibfnamefont {L.}~\bibnamefont {Zhang}}, \bibinfo {author} {\bibfnamefont {W.}~\bibnamefont {Gao}}, \bibinfo {author} {\bibfnamefont {Y.~J.}\ \bibnamefont {Xiang}}, \bibinfo {author} {\bibfnamefont {T.~J.}\ \bibnamefont {Cui}},\ and\ \bibinfo {author} {\bibfnamefont {S.}~\bibnamefont {Zhang}},\ }\bibfield  {title} {\bibinfo {title} {Gain-and loss-induced topological insulating phase in a non-hermitian electrical circuit},\ }\href@noop {} {\bibfield  {journal} {\bibinfo  {journal} {Physical Review Applied}\ }\textbf {\bibinfo {volume} {13}},\ \bibinfo {pages} {014047} (\bibinfo {year} {2020}{\natexlab{b}})}\BibitemShut {NoStop}%
\bibitem [{\citenamefont {Zhang}\ and\ \citenamefont {Franz}(2020)}]{zhang2020non}%
  \BibitemOpen
  \bibfield  {author} {\bibinfo {author} {\bibfnamefont {X.-X.}\ \bibnamefont {Zhang}}\ and\ \bibinfo {author} {\bibfnamefont {M.}~\bibnamefont {Franz}},\ }\bibfield  {title} {\bibinfo {title} {Non-hermitian exceptional landau quantization in electric circuits},\ }\href@noop {} {\bibfield  {journal} {\bibinfo  {journal} {Physical Review Letters}\ }\textbf {\bibinfo {volume} {124}},\ \bibinfo {pages} {046401} (\bibinfo {year} {2020})}\BibitemShut {NoStop}%
\bibitem [{\citenamefont {Zhang}\ \emph {et~al.}(2022{\natexlab{a}})\citenamefont {Zhang}, \citenamefont {Zhang}, \citenamefont {Zhao},\ and\ \citenamefont {Lee}}]{zhang2022observation}%
  \BibitemOpen
  \bibfield  {author} {\bibinfo {author} {\bibfnamefont {X.}~\bibnamefont {Zhang}}, \bibinfo {author} {\bibfnamefont {B.}~\bibnamefont {Zhang}}, \bibinfo {author} {\bibfnamefont {W.}~\bibnamefont {Zhao}},\ and\ \bibinfo {author} {\bibfnamefont {C.~H.}\ \bibnamefont {Lee}},\ }\bibfield  {title} {\bibinfo {title} {Observation of non-local impedance response in a passive electrical circuit},\ }\href@noop {} {\bibfield  {journal} {\bibinfo  {journal} {arXiv preprint arXiv:2211.09152}\ } (\bibinfo {year} {2022}{\natexlab{a}})}\BibitemShut {NoStop}%
\bibitem [{\citenamefont {Yuan}\ \emph {et~al.}(2023)\citenamefont {Yuan}, \citenamefont {Zhang}, \citenamefont {Zhou}, \citenamefont {Wang}, \citenamefont {Pan}, \citenamefont {Feng}, \citenamefont {Sun},\ and\ \citenamefont {Zhang}}]{yuan2023non}%
  \BibitemOpen
  \bibfield  {author} {\bibinfo {author} {\bibfnamefont {H.}~\bibnamefont {Yuan}}, \bibinfo {author} {\bibfnamefont {W.}~\bibnamefont {Zhang}}, \bibinfo {author} {\bibfnamefont {Z.}~\bibnamefont {Zhou}}, \bibinfo {author} {\bibfnamefont {W.}~\bibnamefont {Wang}}, \bibinfo {author} {\bibfnamefont {N.}~\bibnamefont {Pan}}, \bibinfo {author} {\bibfnamefont {Y.}~\bibnamefont {Feng}}, \bibinfo {author} {\bibfnamefont {H.}~\bibnamefont {Sun}},\ and\ \bibinfo {author} {\bibfnamefont {X.}~\bibnamefont {Zhang}},\ }\bibfield  {title} {\bibinfo {title} {Non-hermitian topolectrical circuit sensor with high sensitivity},\ }\href@noop {} {\bibfield  {journal} {\bibinfo  {journal} {Advanced Science}\ ,\ \bibinfo {pages} {2301128}} (\bibinfo {year} {2023})}\BibitemShut {NoStop}%
\bibitem [{\citenamefont {Stegmaier}\ \emph {et~al.}(2024)\citenamefont {Stegmaier}, \citenamefont {Fritzsche}, \citenamefont {Sorbello}, \citenamefont {Greiter}, \citenamefont {Brand}, \citenamefont {Barko}, \citenamefont {Hofer}, \citenamefont {Schwingenschl{\"o}gl}, \citenamefont {Moessner}, \citenamefont {Lee} \emph {et~al.}}]{stegmaier2024topological}%
  \BibitemOpen
  \bibfield  {author} {\bibinfo {author} {\bibfnamefont {A.}~\bibnamefont {Stegmaier}}, \bibinfo {author} {\bibfnamefont {A.}~\bibnamefont {Fritzsche}}, \bibinfo {author} {\bibfnamefont {R.}~\bibnamefont {Sorbello}}, \bibinfo {author} {\bibfnamefont {M.}~\bibnamefont {Greiter}}, \bibinfo {author} {\bibfnamefont {H.}~\bibnamefont {Brand}}, \bibinfo {author} {\bibfnamefont {C.}~\bibnamefont {Barko}}, \bibinfo {author} {\bibfnamefont {M.}~\bibnamefont {Hofer}}, \bibinfo {author} {\bibfnamefont {U.}~\bibnamefont {Schwingenschl{\"o}gl}}, \bibinfo {author} {\bibfnamefont {R.}~\bibnamefont {Moessner}}, \bibinfo {author} {\bibfnamefont {C.~H.}\ \bibnamefont {Lee}}, \emph {et~al.},\ }\bibfield  {title} {\bibinfo {title} {Topological edge state nucleation in frequency space and its realization with floquet electrical circuits},\ }\href@noop {} {\bibfield  {journal} {\bibinfo  {journal} {arXiv preprint arXiv:2407.10191}\ } (\bibinfo {year} {2024})}\BibitemShut {NoStop}%
\bibitem [{\citenamefont {Zou}\ \emph {et~al.}(2024)\citenamefont {Zou}, \citenamefont {Chen}, \citenamefont {Meng}, \citenamefont {Ang}, \citenamefont {Zhang},\ and\ \citenamefont {Lee}}]{zou2024experimental}%
  \BibitemOpen
  \bibfield  {author} {\bibinfo {author} {\bibfnamefont {D.}~\bibnamefont {Zou}}, \bibinfo {author} {\bibfnamefont {T.}~\bibnamefont {Chen}}, \bibinfo {author} {\bibfnamefont {H.}~\bibnamefont {Meng}}, \bibinfo {author} {\bibfnamefont {Y.~S.}\ \bibnamefont {Ang}}, \bibinfo {author} {\bibfnamefont {X.}~\bibnamefont {Zhang}},\ and\ \bibinfo {author} {\bibfnamefont {C.~H.}\ \bibnamefont {Lee}},\ }\bibfield  {title} {\bibinfo {title} {Experimental observation of exceptional bound states in a classical circuit network},\ }\href@noop {} {\bibfield  {journal} {\bibinfo  {journal} {Science Bulletin}\ } (\bibinfo {year} {2024})}\BibitemShut {NoStop}%
\bibitem [{\citenamefont {Zhu}\ \emph {et~al.}(2023)\citenamefont {Zhu}, \citenamefont {Sun}, \citenamefont {Hughes},\ and\ \citenamefont {Bahl}}]{zhu2023higher}%
  \BibitemOpen
  \bibfield  {author} {\bibinfo {author} {\bibfnamefont {P.}~\bibnamefont {Zhu}}, \bibinfo {author} {\bibfnamefont {X.-Q.}\ \bibnamefont {Sun}}, \bibinfo {author} {\bibfnamefont {T.~L.}\ \bibnamefont {Hughes}},\ and\ \bibinfo {author} {\bibfnamefont {G.}~\bibnamefont {Bahl}},\ }\bibfield  {title} {\bibinfo {title} {Higher rank chirality and non-hermitian skin effect in a topolectrical circuit},\ }\href@noop {} {\bibfield  {journal} {\bibinfo  {journal} {Nature communications}\ }\textbf {\bibinfo {volume} {14}},\ \bibinfo {pages} {720} (\bibinfo {year} {2023})}\BibitemShut {NoStop}%
\bibitem [{\citenamefont {Sahin}\ \emph {et~al.}(2025)\citenamefont {Sahin}, \citenamefont {Jalil},\ and\ \citenamefont {Lee}}]{sahin2025topolectrical}%
  \BibitemOpen
  \bibfield  {author} {\bibinfo {author} {\bibfnamefont {H.}~\bibnamefont {Sahin}}, \bibinfo {author} {\bibfnamefont {M.}~\bibnamefont {Jalil}},\ and\ \bibinfo {author} {\bibfnamefont {C.~H.}\ \bibnamefont {Lee}},\ }\bibfield  {title} {\bibinfo {title} {Topolectrical circuits—recent experimental advances and developments},\ }\href@noop {} {\bibfield  {journal} {\bibinfo  {journal} {APL Electronic Devices}\ }\textbf {\bibinfo {volume} {1}} (\bibinfo {year} {2025})}\BibitemShut {NoStop}%
\bibitem [{\citenamefont {Ao}\ \emph {et~al.}(2020)\citenamefont {Ao}, \citenamefont {Hu}, \citenamefont {You}, \citenamefont {Lu}, \citenamefont {Fu}, \citenamefont {Wang},\ and\ \citenamefont {Gong}}]{ao2020topological}%
  \BibitemOpen
  \bibfield  {author} {\bibinfo {author} {\bibfnamefont {Y.}~\bibnamefont {Ao}}, \bibinfo {author} {\bibfnamefont {X.}~\bibnamefont {Hu}}, \bibinfo {author} {\bibfnamefont {Y.}~\bibnamefont {You}}, \bibinfo {author} {\bibfnamefont {C.}~\bibnamefont {Lu}}, \bibinfo {author} {\bibfnamefont {Y.}~\bibnamefont {Fu}}, \bibinfo {author} {\bibfnamefont {X.}~\bibnamefont {Wang}},\ and\ \bibinfo {author} {\bibfnamefont {Q.}~\bibnamefont {Gong}},\ }\bibfield  {title} {\bibinfo {title} {Topological phase transition in the non-hermitian coupled resonator array},\ }\href@noop {} {\bibfield  {journal} {\bibinfo  {journal} {Physical Review Letters}\ }\textbf {\bibinfo {volume} {125}},\ \bibinfo {pages} {013902} (\bibinfo {year} {2020})}\BibitemShut {NoStop}%
\bibitem [{\citenamefont {Smith}\ \emph {et~al.}(2019)\citenamefont {Smith}, \citenamefont {Kim}, \citenamefont {Pollmann},\ and\ \citenamefont {Knolle}}]{smith2019simulating}%
  \BibitemOpen
  \bibfield  {author} {\bibinfo {author} {\bibfnamefont {A.}~\bibnamefont {Smith}}, \bibinfo {author} {\bibfnamefont {M.}~\bibnamefont {Kim}}, \bibinfo {author} {\bibfnamefont {F.}~\bibnamefont {Pollmann}},\ and\ \bibinfo {author} {\bibfnamefont {J.}~\bibnamefont {Knolle}},\ }\bibfield  {title} {\bibinfo {title} {Simulating quantum many-body dynamics on a current digital quantum computer},\ }\href@noop {} {\bibfield  {journal} {\bibinfo  {journal} {npj Quantum Information}\ }\textbf {\bibinfo {volume} {5}},\ \bibinfo {pages} {106} (\bibinfo {year} {2019})}\BibitemShut {NoStop}%
\bibitem [{\citenamefont {Gou}\ \emph {et~al.}(2020)\citenamefont {Gou}, \citenamefont {Chen}, \citenamefont {Xie}, \citenamefont {Xiao}, \citenamefont {Deng}, \citenamefont {Gadway}, \citenamefont {Yi},\ and\ \citenamefont {Yan}}]{gou2020tunable}%
  \BibitemOpen
  \bibfield  {author} {\bibinfo {author} {\bibfnamefont {W.}~\bibnamefont {Gou}}, \bibinfo {author} {\bibfnamefont {T.}~\bibnamefont {Chen}}, \bibinfo {author} {\bibfnamefont {D.}~\bibnamefont {Xie}}, \bibinfo {author} {\bibfnamefont {T.}~\bibnamefont {Xiao}}, \bibinfo {author} {\bibfnamefont {T.-S.}\ \bibnamefont {Deng}}, \bibinfo {author} {\bibfnamefont {B.}~\bibnamefont {Gadway}}, \bibinfo {author} {\bibfnamefont {W.}~\bibnamefont {Yi}},\ and\ \bibinfo {author} {\bibfnamefont {B.}~\bibnamefont {Yan}},\ }\bibfield  {title} {\bibinfo {title} {Tunable nonreciprocal quantum transport through a dissipative aharonov-bohm ring in ultracold atoms},\ }\href@noop {} {\bibfield  {journal} {\bibinfo  {journal} {Physical review letters}\ }\textbf {\bibinfo {volume} {124}},\ \bibinfo {pages} {070402} (\bibinfo {year} {2020})}\BibitemShut {NoStop}%
\bibitem [{\citenamefont {Koh}\ \emph {et~al.}(2022{\natexlab{a}})\citenamefont {Koh}, \citenamefont {Tai},\ and\ \citenamefont {Lee}}]{koh2022simulation}%
  \BibitemOpen
  \bibfield  {author} {\bibinfo {author} {\bibfnamefont {J.~M.}\ \bibnamefont {Koh}}, \bibinfo {author} {\bibfnamefont {T.}~\bibnamefont {Tai}},\ and\ \bibinfo {author} {\bibfnamefont {C.~H.}\ \bibnamefont {Lee}},\ }\bibfield  {title} {\bibinfo {title} {Simulation of interaction-induced chiral topological dynamics on a digital quantum computer},\ }\href@noop {} {\bibfield  {journal} {\bibinfo  {journal} {Physical Review Letters}\ }\textbf {\bibinfo {volume} {129}},\ \bibinfo {pages} {140502} (\bibinfo {year} {2022}{\natexlab{a}})}\BibitemShut {NoStop}%
\bibitem [{\citenamefont {Kirmani}\ \emph {et~al.}(2022)\citenamefont {Kirmani}, \citenamefont {Bull}, \citenamefont {Hou}, \citenamefont {Saravanan}, \citenamefont {Saeed}, \citenamefont {Papi{\'c}}, \citenamefont {Rahmani},\ and\ \citenamefont {Ghaemi}}]{kirmani2022probing}%
  \BibitemOpen
  \bibfield  {author} {\bibinfo {author} {\bibfnamefont {A.}~\bibnamefont {Kirmani}}, \bibinfo {author} {\bibfnamefont {K.}~\bibnamefont {Bull}}, \bibinfo {author} {\bibfnamefont {C.-Y.}\ \bibnamefont {Hou}}, \bibinfo {author} {\bibfnamefont {V.}~\bibnamefont {Saravanan}}, \bibinfo {author} {\bibfnamefont {S.~M.}\ \bibnamefont {Saeed}}, \bibinfo {author} {\bibfnamefont {Z.}~\bibnamefont {Papi{\'c}}}, \bibinfo {author} {\bibfnamefont {A.}~\bibnamefont {Rahmani}},\ and\ \bibinfo {author} {\bibfnamefont {P.}~\bibnamefont {Ghaemi}},\ }\bibfield  {title} {\bibinfo {title} {Probing geometric excitations of fractional quantum hall states on quantum computers},\ }\href@noop {} {\bibfield  {journal} {\bibinfo  {journal} {Physical Review Letters}\ }\textbf {\bibinfo {volume} {129}},\ \bibinfo {pages} {056801} (\bibinfo {year} {2022})}\BibitemShut {NoStop}%
\bibitem [{\citenamefont {Frey}\ and\ \citenamefont {Rachel}(2022)}]{frey2022realization}%
  \BibitemOpen
  \bibfield  {author} {\bibinfo {author} {\bibfnamefont {P.}~\bibnamefont {Frey}}\ and\ \bibinfo {author} {\bibfnamefont {S.}~\bibnamefont {Rachel}},\ }\bibfield  {title} {\bibinfo {title} {Realization of a discrete time crystal on 57 qubits of a quantum computer},\ }\href@noop {} {\bibfield  {journal} {\bibinfo  {journal} {Science advances}\ }\textbf {\bibinfo {volume} {8}},\ \bibinfo {pages} {eabm7652} (\bibinfo {year} {2022})}\BibitemShut {NoStop}%
\bibitem [{\citenamefont {Chen}\ \emph {et~al.}(2023)\citenamefont {Chen}, \citenamefont {Shen}, \citenamefont {Lee}, \citenamefont {Yang},\ and\ \citenamefont {Bomantara}}]{chen2023robust}%
  \BibitemOpen
  \bibfield  {author} {\bibinfo {author} {\bibfnamefont {T.}~\bibnamefont {Chen}}, \bibinfo {author} {\bibfnamefont {R.}~\bibnamefont {Shen}}, \bibinfo {author} {\bibfnamefont {C.~H.}\ \bibnamefont {Lee}}, \bibinfo {author} {\bibfnamefont {B.}~\bibnamefont {Yang}},\ and\ \bibinfo {author} {\bibfnamefont {R.~W.}\ \bibnamefont {Bomantara}},\ }\bibfield  {title} {\bibinfo {title} {A robust large-period discrete time crystal and its signature in a digital quantum computer},\ }\href@noop {} {\bibfield  {journal} {\bibinfo  {journal} {arXiv preprint arXiv:2309.11560}\ } (\bibinfo {year} {2023})}\BibitemShut {NoStop}%
\bibitem [{\citenamefont {Yang}\ \emph {et~al.}(2023)\citenamefont {Yang}, \citenamefont {Christianen}, \citenamefont {Coll-Vinent}, \citenamefont {Smelyanskiy}, \citenamefont {Ba{\~n}uls}, \citenamefont {O'Brien}, \citenamefont {Wild},\ and\ \citenamefont {Cirac}}]{yang2023simulating}%
  \BibitemOpen
  \bibfield  {author} {\bibinfo {author} {\bibfnamefont {Y.}~\bibnamefont {Yang}}, \bibinfo {author} {\bibfnamefont {A.}~\bibnamefont {Christianen}}, \bibinfo {author} {\bibfnamefont {S.}~\bibnamefont {Coll-Vinent}}, \bibinfo {author} {\bibfnamefont {V.}~\bibnamefont {Smelyanskiy}}, \bibinfo {author} {\bibfnamefont {M.~C.}\ \bibnamefont {Ba{\~n}uls}}, \bibinfo {author} {\bibfnamefont {T.~E.}\ \bibnamefont {O'Brien}}, \bibinfo {author} {\bibfnamefont {D.~S.}\ \bibnamefont {Wild}},\ and\ \bibinfo {author} {\bibfnamefont {J.~I.}\ \bibnamefont {Cirac}},\ }\bibfield  {title} {\bibinfo {title} {Simulating prethermalization using near-term quantum computers},\ }\href@noop {} {\bibfield  {journal} {\bibinfo  {journal} {PRX Quantum}\ }\textbf {\bibinfo {volume} {4}},\ \bibinfo {pages} {030320} (\bibinfo {year} {2023})}\BibitemShut {NoStop}%
\bibitem [{\citenamefont {Iqbal}\ \emph {et~al.}(2023)\citenamefont {Iqbal}, \citenamefont {Tantivasadakarn}, \citenamefont {Verresen}, \citenamefont {Campbell}, \citenamefont {Dreiling}, \citenamefont {Figgatt}, \citenamefont {Gaebler}, \citenamefont {Johansen}, \citenamefont {Mills}, \citenamefont {Moses} \emph {et~al.}}]{iqbal2023creation}%
  \BibitemOpen
  \bibfield  {author} {\bibinfo {author} {\bibfnamefont {M.}~\bibnamefont {Iqbal}}, \bibinfo {author} {\bibfnamefont {N.}~\bibnamefont {Tantivasadakarn}}, \bibinfo {author} {\bibfnamefont {R.}~\bibnamefont {Verresen}}, \bibinfo {author} {\bibfnamefont {S.~L.}\ \bibnamefont {Campbell}}, \bibinfo {author} {\bibfnamefont {J.~M.}\ \bibnamefont {Dreiling}}, \bibinfo {author} {\bibfnamefont {C.}~\bibnamefont {Figgatt}}, \bibinfo {author} {\bibfnamefont {J.~P.}\ \bibnamefont {Gaebler}}, \bibinfo {author} {\bibfnamefont {J.}~\bibnamefont {Johansen}}, \bibinfo {author} {\bibfnamefont {M.}~\bibnamefont {Mills}}, \bibinfo {author} {\bibfnamefont {S.~A.}\ \bibnamefont {Moses}}, \emph {et~al.},\ }\bibfield  {title} {\bibinfo {title} {Creation of non-abelian topological order and anyons on a trapped-ion processor},\ }\href@noop {} {\bibfield  {journal} {\bibinfo  {journal} {arXiv preprint arXiv:2305.03766}\ } (\bibinfo {year} {2023})}\BibitemShut {NoStop}%
\bibitem [{\citenamefont {Shen}\ \emph {et~al.}(2025{\natexlab{c}})\citenamefont {Shen}, \citenamefont {Chen}, \citenamefont {Yang},\ and\ \citenamefont {Lee}}]{shen2023observation}%
  \BibitemOpen
  \bibfield  {author} {\bibinfo {author} {\bibfnamefont {R.}~\bibnamefont {Shen}}, \bibinfo {author} {\bibfnamefont {T.}~\bibnamefont {Chen}}, \bibinfo {author} {\bibfnamefont {B.}~\bibnamefont {Yang}},\ and\ \bibinfo {author} {\bibfnamefont {C.~H.}\ \bibnamefont {Lee}},\ }\bibfield  {title} {\bibinfo {title} {Observation of the non-hermitian skin effect and fermi skin on a digital quantum computer},\ }\href@noop {} {\bibfield  {journal} {\bibinfo  {journal} {Nature Communications}\ }\textbf {\bibinfo {volume} {16}},\ \bibinfo {pages} {1340} (\bibinfo {year} {2025}{\natexlab{c}})}\BibitemShut {NoStop}%
\bibitem [{\citenamefont {Chertkov}\ \emph {et~al.}(2023)\citenamefont {Chertkov}, \citenamefont {Cheng}, \citenamefont {Potter}, \citenamefont {Gopalakrishnan}, \citenamefont {Gatterman}, \citenamefont {Gerber}, \citenamefont {Gilmore}, \citenamefont {Gresh}, \citenamefont {Hall}, \citenamefont {Hankin} \emph {et~al.}}]{chertkov2023characterizing}%
  \BibitemOpen
  \bibfield  {author} {\bibinfo {author} {\bibfnamefont {E.}~\bibnamefont {Chertkov}}, \bibinfo {author} {\bibfnamefont {Z.}~\bibnamefont {Cheng}}, \bibinfo {author} {\bibfnamefont {A.~C.}\ \bibnamefont {Potter}}, \bibinfo {author} {\bibfnamefont {S.}~\bibnamefont {Gopalakrishnan}}, \bibinfo {author} {\bibfnamefont {T.~M.}\ \bibnamefont {Gatterman}}, \bibinfo {author} {\bibfnamefont {J.~A.}\ \bibnamefont {Gerber}}, \bibinfo {author} {\bibfnamefont {K.}~\bibnamefont {Gilmore}}, \bibinfo {author} {\bibfnamefont {D.}~\bibnamefont {Gresh}}, \bibinfo {author} {\bibfnamefont {A.}~\bibnamefont {Hall}}, \bibinfo {author} {\bibfnamefont {A.}~\bibnamefont {Hankin}}, \emph {et~al.},\ }\bibfield  {title} {\bibinfo {title} {Characterizing a non-equilibrium phase transition on a quantum computer},\ }\href@noop {} {\bibfield  {journal} {\bibinfo  {journal} {Nature Physics}\ ,\ \bibinfo {pages} {1}} (\bibinfo {year} {2023})}\BibitemShut {NoStop}%
\bibitem [{\citenamefont {Koh}\ \emph {et~al.}(2023{\natexlab{a}})\citenamefont {Koh}, \citenamefont {Tai},\ and\ \citenamefont {Lee}}]{koh2023observation}%
  \BibitemOpen
  \bibfield  {author} {\bibinfo {author} {\bibfnamefont {J.~M.}\ \bibnamefont {Koh}}, \bibinfo {author} {\bibfnamefont {T.}~\bibnamefont {Tai}},\ and\ \bibinfo {author} {\bibfnamefont {C.~H.}\ \bibnamefont {Lee}},\ }\bibfield  {title} {\bibinfo {title} {Observation of higher-order topological states on a quantum computer},\ }\href@noop {} {\bibfield  {journal} {\bibinfo  {journal} {arXiv preprint arXiv:2303.02179}\ } (\bibinfo {year} {2023}{\natexlab{a}})}\BibitemShut {NoStop}%
\bibitem [{\citenamefont {Koh}\ \emph {et~al.}(2022{\natexlab{b}})\citenamefont {Koh}, \citenamefont {Tai}, \citenamefont {Phee}, \citenamefont {Ng},\ and\ \citenamefont {Lee}}]{koh2022stabilizing}%
  \BibitemOpen
  \bibfield  {author} {\bibinfo {author} {\bibfnamefont {J.~M.}\ \bibnamefont {Koh}}, \bibinfo {author} {\bibfnamefont {T.}~\bibnamefont {Tai}}, \bibinfo {author} {\bibfnamefont {Y.~H.}\ \bibnamefont {Phee}}, \bibinfo {author} {\bibfnamefont {W.~E.}\ \bibnamefont {Ng}},\ and\ \bibinfo {author} {\bibfnamefont {C.~H.}\ \bibnamefont {Lee}},\ }\bibfield  {title} {\bibinfo {title} {Stabilizing multiple topological fermions on a quantum computer},\ }\href@noop {} {\bibfield  {journal} {\bibinfo  {journal} {npj Quantum Information}\ }\textbf {\bibinfo {volume} {8}},\ \bibinfo {pages} {16} (\bibinfo {year} {2022}{\natexlab{b}})}\BibitemShut {NoStop}%
\bibitem [{\citenamefont {Koukoutsis}\ \emph {et~al.}(2024)\citenamefont {Koukoutsis}, \citenamefont {Papagiannis}, \citenamefont {Hizanidis}, \citenamefont {Ram}, \citenamefont {Vahala}, \citenamefont {Amaro}, \citenamefont {Gamiz},\ and\ \citenamefont {Vallis}}]{koukoutsis2024quantum}%
  \BibitemOpen
  \bibfield  {author} {\bibinfo {author} {\bibfnamefont {E.}~\bibnamefont {Koukoutsis}}, \bibinfo {author} {\bibfnamefont {P.}~\bibnamefont {Papagiannis}}, \bibinfo {author} {\bibfnamefont {K.}~\bibnamefont {Hizanidis}}, \bibinfo {author} {\bibfnamefont {A.~K.}\ \bibnamefont {Ram}}, \bibinfo {author} {\bibfnamefont {G.}~\bibnamefont {Vahala}}, \bibinfo {author} {\bibfnamefont {O.}~\bibnamefont {Amaro}}, \bibinfo {author} {\bibfnamefont {L.~I.~I.}\ \bibnamefont {Gamiz}},\ and\ \bibinfo {author} {\bibfnamefont {D.}~\bibnamefont {Vallis}},\ }\bibfield  {title} {\bibinfo {title} {Quantum implementation of non-unitary operations with biorthogonal representations},\ }\href@noop {} {\bibfield  {journal} {\bibinfo  {journal} {arXiv preprint arXiv:2410.22505}\ } (\bibinfo {year} {2024})}\BibitemShut {NoStop}%
\bibitem [{\citenamefont {Koh}\ \emph {et~al.}(2025)\citenamefont {Koh}, \citenamefont {Xue}, \citenamefont {Tai}, \citenamefont {Koh},\ and\ \citenamefont {Lee}}]{koh2025interacting}%
  \BibitemOpen
  \bibfield  {author} {\bibinfo {author} {\bibfnamefont {J.~M.}\ \bibnamefont {Koh}}, \bibinfo {author} {\bibfnamefont {W.-T.}\ \bibnamefont {Xue}}, \bibinfo {author} {\bibfnamefont {T.}~\bibnamefont {Tai}}, \bibinfo {author} {\bibfnamefont {D.~E.}\ \bibnamefont {Koh}},\ and\ \bibinfo {author} {\bibfnamefont {C.~H.}\ \bibnamefont {Lee}},\ }\bibfield  {title} {\bibinfo {title} {Interacting non-hermitian edge and cluster bursts on a digital quantum processor},\ }\href@noop {} {\bibfield  {journal} {\bibinfo  {journal} {arXiv preprint arXiv:2503.14595}\ } (\bibinfo {year} {2025})}\BibitemShut {NoStop}%
\bibitem [{\citenamefont {Abdo}\ \emph {et~al.}(2013)\citenamefont {Abdo}, \citenamefont {Sliwa}, \citenamefont {Frunzio},\ and\ \citenamefont {Devoret}}]{abdo2013directional}%
  \BibitemOpen
  \bibfield  {author} {\bibinfo {author} {\bibfnamefont {B.}~\bibnamefont {Abdo}}, \bibinfo {author} {\bibfnamefont {K.}~\bibnamefont {Sliwa}}, \bibinfo {author} {\bibfnamefont {L.}~\bibnamefont {Frunzio}},\ and\ \bibinfo {author} {\bibfnamefont {M.}~\bibnamefont {Devoret}},\ }\bibfield  {title} {\bibinfo {title} {Directional amplification with a josephson circuit},\ }\href@noop {} {\bibfield  {journal} {\bibinfo  {journal} {Physical Review X}\ }\textbf {\bibinfo {volume} {3}},\ \bibinfo {pages} {031001} (\bibinfo {year} {2013})}\BibitemShut {NoStop}%
\bibitem [{\citenamefont {Li}\ \emph {et~al.}(2017)\citenamefont {Li}, \citenamefont {Huang}, \citenamefont {Zhang},\ and\ \citenamefont {Tian}}]{li2017optical}%
  \BibitemOpen
  \bibfield  {author} {\bibinfo {author} {\bibfnamefont {Y.}~\bibnamefont {Li}}, \bibinfo {author} {\bibfnamefont {Y.}~\bibnamefont {Huang}}, \bibinfo {author} {\bibfnamefont {X.}~\bibnamefont {Zhang}},\ and\ \bibinfo {author} {\bibfnamefont {L.}~\bibnamefont {Tian}},\ }\bibfield  {title} {\bibinfo {title} {Optical directional amplification in a three-mode optomechanical system},\ }\href@noop {} {\bibfield  {journal} {\bibinfo  {journal} {Optics express}\ }\textbf {\bibinfo {volume} {25}},\ \bibinfo {pages} {18907} (\bibinfo {year} {2017})}\BibitemShut {NoStop}%
\bibitem [{\citenamefont {Mercier~de L{\'e}pinay}\ \emph {et~al.}(2019)\citenamefont {Mercier~de L{\'e}pinay}, \citenamefont {Damsk{\"a}gg}, \citenamefont {Ockeloen-Korppi},\ and\ \citenamefont {Sillanp{\"a}{\"a}}}]{mercier2019realization}%
  \BibitemOpen
  \bibfield  {author} {\bibinfo {author} {\bibfnamefont {L.}~\bibnamefont {Mercier~de L{\'e}pinay}}, \bibinfo {author} {\bibfnamefont {E.}~\bibnamefont {Damsk{\"a}gg}}, \bibinfo {author} {\bibfnamefont {C.~F.}\ \bibnamefont {Ockeloen-Korppi}},\ and\ \bibinfo {author} {\bibfnamefont {M.~A.}\ \bibnamefont {Sillanp{\"a}{\"a}}},\ }\bibfield  {title} {\bibinfo {title} {Realization of directional amplification in a microwave optomechanical device},\ }\href@noop {} {\bibfield  {journal} {\bibinfo  {journal} {Physical Review Applied}\ }\textbf {\bibinfo {volume} {11}},\ \bibinfo {pages} {034027} (\bibinfo {year} {2019})}\BibitemShut {NoStop}%
\bibitem [{\citenamefont {Wanjura}\ \emph {et~al.}(2020)\citenamefont {Wanjura}, \citenamefont {Brunelli},\ and\ \citenamefont {Nunnenkamp}}]{wanjura2020topological}%
  \BibitemOpen
  \bibfield  {author} {\bibinfo {author} {\bibfnamefont {C.~C.}\ \bibnamefont {Wanjura}}, \bibinfo {author} {\bibfnamefont {M.}~\bibnamefont {Brunelli}},\ and\ \bibinfo {author} {\bibfnamefont {A.}~\bibnamefont {Nunnenkamp}},\ }\bibfield  {title} {\bibinfo {title} {Topological framework for directional amplification in driven-dissipative cavity arrays},\ }\href@noop {} {\bibfield  {journal} {\bibinfo  {journal} {Nature communications}\ }\textbf {\bibinfo {volume} {11}},\ \bibinfo {pages} {3149} (\bibinfo {year} {2020})}\BibitemShut {NoStop}%
\bibitem [{\citenamefont {Ramos}\ \emph {et~al.}(2021)\citenamefont {Ramos}, \citenamefont {Garcia-Ripoll},\ and\ \citenamefont {Porras}}]{ramos2021topological}%
  \BibitemOpen
  \bibfield  {author} {\bibinfo {author} {\bibfnamefont {T.}~\bibnamefont {Ramos}}, \bibinfo {author} {\bibfnamefont {J.~J.}\ \bibnamefont {Garcia-Ripoll}},\ and\ \bibinfo {author} {\bibfnamefont {D.}~\bibnamefont {Porras}},\ }\bibfield  {title} {\bibinfo {title} {Topological input-output theory for directional amplification},\ }\href {https://doi.org/10.1103/PhysRevA.103.033513} {\bibfield  {journal} {\bibinfo  {journal} {Phys. Rev. A}\ }\textbf {\bibinfo {volume} {103}},\ \bibinfo {pages} {033513} (\bibinfo {year} {2021})}\BibitemShut {NoStop}%
\bibitem [{\citenamefont {Wanjura}\ \emph {et~al.}(2021)\citenamefont {Wanjura}, \citenamefont {Brunelli},\ and\ \citenamefont {Nunnenkamp}}]{wanjura2021correspondence}%
  \BibitemOpen
  \bibfield  {author} {\bibinfo {author} {\bibfnamefont {C.~C.}\ \bibnamefont {Wanjura}}, \bibinfo {author} {\bibfnamefont {M.}~\bibnamefont {Brunelli}},\ and\ \bibinfo {author} {\bibfnamefont {A.}~\bibnamefont {Nunnenkamp}},\ }\bibfield  {title} {\bibinfo {title} {Correspondence between non-hermitian topology and directional amplification in the presence of disorder},\ }\href@noop {} {\bibfield  {journal} {\bibinfo  {journal} {Physical Review Letters}\ }\textbf {\bibinfo {volume} {127}},\ \bibinfo {pages} {213601} (\bibinfo {year} {2021})}\BibitemShut {NoStop}%
\bibitem [{\citenamefont {Bandres}\ \emph {et~al.}(2018)\citenamefont {Bandres}, \citenamefont {Wittek}, \citenamefont {Harari}, \citenamefont {Parto}, \citenamefont {Ren}, \citenamefont {Segev}, \citenamefont {Christodoulides},\ and\ \citenamefont {Khajavikhan}}]{bandres2018topological}%
  \BibitemOpen
  \bibfield  {author} {\bibinfo {author} {\bibfnamefont {M.~A.}\ \bibnamefont {Bandres}}, \bibinfo {author} {\bibfnamefont {S.}~\bibnamefont {Wittek}}, \bibinfo {author} {\bibfnamefont {G.}~\bibnamefont {Harari}}, \bibinfo {author} {\bibfnamefont {M.}~\bibnamefont {Parto}}, \bibinfo {author} {\bibfnamefont {J.}~\bibnamefont {Ren}}, \bibinfo {author} {\bibfnamefont {M.}~\bibnamefont {Segev}}, \bibinfo {author} {\bibfnamefont {D.~N.}\ \bibnamefont {Christodoulides}},\ and\ \bibinfo {author} {\bibfnamefont {M.}~\bibnamefont {Khajavikhan}},\ }\bibfield  {title} {\bibinfo {title} {Topological insulator laser: Experiments},\ }\href@noop {} {\bibfield  {journal} {\bibinfo  {journal} {Science}\ }\textbf {\bibinfo {volume} {359}},\ \bibinfo {pages} {eaar4005} (\bibinfo {year} {2018})}\BibitemShut {NoStop}%
\bibitem [{\citenamefont {Harari}\ \emph {et~al.}(2018)\citenamefont {Harari}, \citenamefont {Bandres}, \citenamefont {Lumer}, \citenamefont {Rechtsman}, \citenamefont {Chong}, \citenamefont {Khajavikhan}, \citenamefont {Christodoulides},\ and\ \citenamefont {Segev}}]{harari2018topological}%
  \BibitemOpen
  \bibfield  {author} {\bibinfo {author} {\bibfnamefont {G.}~\bibnamefont {Harari}}, \bibinfo {author} {\bibfnamefont {M.~A.}\ \bibnamefont {Bandres}}, \bibinfo {author} {\bibfnamefont {Y.}~\bibnamefont {Lumer}}, \bibinfo {author} {\bibfnamefont {M.~C.}\ \bibnamefont {Rechtsman}}, \bibinfo {author} {\bibfnamefont {Y.~D.}\ \bibnamefont {Chong}}, \bibinfo {author} {\bibfnamefont {M.}~\bibnamefont {Khajavikhan}}, \bibinfo {author} {\bibfnamefont {D.~N.}\ \bibnamefont {Christodoulides}},\ and\ \bibinfo {author} {\bibfnamefont {M.}~\bibnamefont {Segev}},\ }\bibfield  {title} {\bibinfo {title} {Topological insulator laser: Theory},\ }\href@noop {} {\bibfield  {journal} {\bibinfo  {journal} {Science}\ }\textbf {\bibinfo {volume} {359}},\ \bibinfo {pages} {eaar4003} (\bibinfo {year} {2018})}\BibitemShut {NoStop}%
\bibitem [{\citenamefont {Zeng}\ \emph {et~al.}(2020)\citenamefont {Zeng}, \citenamefont {Chattopadhyay}, \citenamefont {Zhu}, \citenamefont {Qiang}, \citenamefont {Li}, \citenamefont {Jin}, \citenamefont {Li}, \citenamefont {Davies}, \citenamefont {Linfield}, \citenamefont {Zhang} \emph {et~al.}}]{zeng2020electrically}%
  \BibitemOpen
  \bibfield  {author} {\bibinfo {author} {\bibfnamefont {Y.}~\bibnamefont {Zeng}}, \bibinfo {author} {\bibfnamefont {U.}~\bibnamefont {Chattopadhyay}}, \bibinfo {author} {\bibfnamefont {B.}~\bibnamefont {Zhu}}, \bibinfo {author} {\bibfnamefont {B.}~\bibnamefont {Qiang}}, \bibinfo {author} {\bibfnamefont {J.}~\bibnamefont {Li}}, \bibinfo {author} {\bibfnamefont {Y.}~\bibnamefont {Jin}}, \bibinfo {author} {\bibfnamefont {L.}~\bibnamefont {Li}}, \bibinfo {author} {\bibfnamefont {A.~G.}\ \bibnamefont {Davies}}, \bibinfo {author} {\bibfnamefont {E.~H.}\ \bibnamefont {Linfield}}, \bibinfo {author} {\bibfnamefont {B.}~\bibnamefont {Zhang}}, \emph {et~al.},\ }\bibfield  {title} {\bibinfo {title} {Electrically pumped topological laser with valley edge modes},\ }\href@noop {} {\bibfield  {journal} {\bibinfo  {journal} {Nature}\ }\textbf {\bibinfo {volume} {578}},\ \bibinfo {pages} {246} (\bibinfo {year} {2020})}\BibitemShut {NoStop}%
\bibitem [{\citenamefont {Yang}\ \emph {et~al.}(2022{\natexlab{d}})\citenamefont {Yang}, \citenamefont {Li}, \citenamefont {Gao},\ and\ \citenamefont {Lu}}]{yang2022topological}%
  \BibitemOpen
  \bibfield  {author} {\bibinfo {author} {\bibfnamefont {L.}~\bibnamefont {Yang}}, \bibinfo {author} {\bibfnamefont {G.}~\bibnamefont {Li}}, \bibinfo {author} {\bibfnamefont {X.}~\bibnamefont {Gao}},\ and\ \bibinfo {author} {\bibfnamefont {L.}~\bibnamefont {Lu}},\ }\bibfield  {title} {\bibinfo {title} {Topological-cavity surface-emitting laser},\ }\href@noop {} {\bibfield  {journal} {\bibinfo  {journal} {Nature Photonics}\ }\textbf {\bibinfo {volume} {16}},\ \bibinfo {pages} {279} (\bibinfo {year} {2022}{\natexlab{d}})}\BibitemShut {NoStop}%
\bibitem [{\citenamefont {Ota}\ \emph {et~al.}(2020)\citenamefont {Ota}, \citenamefont {Takata}, \citenamefont {Ozawa}, \citenamefont {Amo}, \citenamefont {Jia}, \citenamefont {Kante}, \citenamefont {Notomi}, \citenamefont {Arakawa},\ and\ \citenamefont {Iwamoto}}]{ota2020active}%
  \BibitemOpen
  \bibfield  {author} {\bibinfo {author} {\bibfnamefont {Y.}~\bibnamefont {Ota}}, \bibinfo {author} {\bibfnamefont {K.}~\bibnamefont {Takata}}, \bibinfo {author} {\bibfnamefont {T.}~\bibnamefont {Ozawa}}, \bibinfo {author} {\bibfnamefont {A.}~\bibnamefont {Amo}}, \bibinfo {author} {\bibfnamefont {Z.}~\bibnamefont {Jia}}, \bibinfo {author} {\bibfnamefont {B.}~\bibnamefont {Kante}}, \bibinfo {author} {\bibfnamefont {M.}~\bibnamefont {Notomi}}, \bibinfo {author} {\bibfnamefont {Y.}~\bibnamefont {Arakawa}},\ and\ \bibinfo {author} {\bibfnamefont {S.}~\bibnamefont {Iwamoto}},\ }\bibfield  {title} {\bibinfo {title} {Active topological photonics},\ }\href@noop {} {\bibfield  {journal} {\bibinfo  {journal} {Nanophotonics}\ }\textbf {\bibinfo {volume} {9}},\ \bibinfo {pages} {547} (\bibinfo {year} {2020})}\BibitemShut {NoStop}%
\bibitem [{\citenamefont {Kulishov}\ \emph {et~al.}(2005)\citenamefont {Kulishov}, \citenamefont {Laniel}, \citenamefont {B{\'e}langer}, \citenamefont {Aza{\~n}a},\ and\ \citenamefont {Plant}}]{kulishov2005nonreciprocal}%
  \BibitemOpen
  \bibfield  {author} {\bibinfo {author} {\bibfnamefont {M.}~\bibnamefont {Kulishov}}, \bibinfo {author} {\bibfnamefont {J.~M.}\ \bibnamefont {Laniel}}, \bibinfo {author} {\bibfnamefont {N.}~\bibnamefont {B{\'e}langer}}, \bibinfo {author} {\bibfnamefont {J.}~\bibnamefont {Aza{\~n}a}},\ and\ \bibinfo {author} {\bibfnamefont {D.~V.}\ \bibnamefont {Plant}},\ }\bibfield  {title} {\bibinfo {title} {Nonreciprocal waveguide bragg gratings},\ }\href@noop {} {\bibfield  {journal} {\bibinfo  {journal} {Optics express}\ }\textbf {\bibinfo {volume} {13}},\ \bibinfo {pages} {3068} (\bibinfo {year} {2005})}\BibitemShut {NoStop}%
\bibitem [{\citenamefont {Shemuly}\ \emph {et~al.}(2013)\citenamefont {Shemuly}, \citenamefont {Ruff}, \citenamefont {Stolyarov}, \citenamefont {Spektor}, \citenamefont {Johnson}, \citenamefont {Fink},\ and\ \citenamefont {Shapira}}]{shemuly2013asymmetric}%
  \BibitemOpen
  \bibfield  {author} {\bibinfo {author} {\bibfnamefont {D.}~\bibnamefont {Shemuly}}, \bibinfo {author} {\bibfnamefont {Z.~M.}\ \bibnamefont {Ruff}}, \bibinfo {author} {\bibfnamefont {A.~M.}\ \bibnamefont {Stolyarov}}, \bibinfo {author} {\bibfnamefont {G.}~\bibnamefont {Spektor}}, \bibinfo {author} {\bibfnamefont {S.~G.}\ \bibnamefont {Johnson}}, \bibinfo {author} {\bibfnamefont {Y.}~\bibnamefont {Fink}},\ and\ \bibinfo {author} {\bibfnamefont {O.}~\bibnamefont {Shapira}},\ }\bibfield  {title} {\bibinfo {title} {Asymmetric wave propagation in planar chiral fibers},\ }\href@noop {} {\bibfield  {journal} {\bibinfo  {journal} {Optics Express}\ }\textbf {\bibinfo {volume} {21}},\ \bibinfo {pages} {1465} (\bibinfo {year} {2013})}\BibitemShut {NoStop}%
\bibitem [{\citenamefont {Sounas}\ and\ \citenamefont {Al{\`u}}(2017)}]{sounas2017non}%
  \BibitemOpen
  \bibfield  {author} {\bibinfo {author} {\bibfnamefont {D.~L.}\ \bibnamefont {Sounas}}\ and\ \bibinfo {author} {\bibfnamefont {A.}~\bibnamefont {Al{\`u}}},\ }\bibfield  {title} {\bibinfo {title} {Non-reciprocal photonics based on time modulation},\ }\href@noop {} {\bibfield  {journal} {\bibinfo  {journal} {Nature Photonics}\ }\textbf {\bibinfo {volume} {11}},\ \bibinfo {pages} {774} (\bibinfo {year} {2017})}\BibitemShut {NoStop}%
\bibitem [{\citenamefont {Hu}\ \emph {et~al.}(2019)\citenamefont {Hu}, \citenamefont {Liu}, \citenamefont {Hu}, \citenamefont {Liu},\ and\ \citenamefont {Gao}}]{hu2019routing}%
  \BibitemOpen
  \bibfield  {author} {\bibinfo {author} {\bibfnamefont {H.}~\bibnamefont {Hu}}, \bibinfo {author} {\bibfnamefont {L.}~\bibnamefont {Liu}}, \bibinfo {author} {\bibfnamefont {X.}~\bibnamefont {Hu}}, \bibinfo {author} {\bibfnamefont {D.}~\bibnamefont {Liu}},\ and\ \bibinfo {author} {\bibfnamefont {D.}~\bibnamefont {Gao}},\ }\bibfield  {title} {\bibinfo {title} {Routing emission with a multi-channel nonreciprocal waveguide},\ }\href@noop {} {\bibfield  {journal} {\bibinfo  {journal} {Photonics Research}\ }\textbf {\bibinfo {volume} {7}},\ \bibinfo {pages} {642} (\bibinfo {year} {2019})}\BibitemShut {NoStop}%
\bibitem [{\citenamefont {Meng}\ \emph {et~al.}(2025)\citenamefont {Meng}, \citenamefont {Ang},\ and\ \citenamefont {Lee}}]{meng2025generalized}%
  \BibitemOpen
  \bibfield  {author} {\bibinfo {author} {\bibfnamefont {H.}~\bibnamefont {Meng}}, \bibinfo {author} {\bibfnamefont {Y.~S.}\ \bibnamefont {Ang}},\ and\ \bibinfo {author} {\bibfnamefont {C.~H.}\ \bibnamefont {Lee}},\ }\bibfield  {title} {\bibinfo {title} {Generalized brillouin zone fragmentation},\ }\href@noop {} {\bibfield  {journal} {\bibinfo  {journal} {arXiv preprint arXiv:2508.13275}\ } (\bibinfo {year} {2025})}\BibitemShut {NoStop}%
\bibitem [{\citenamefont {Li}\ \emph {et~al.}(2021{\natexlab{b}})\citenamefont {Li}, \citenamefont {Li}, \citenamefont {Naik}, \citenamefont {Xie}, \citenamefont {Li}, \citenamefont {Wang}, \citenamefont {Regan}, \citenamefont {Wang}, \citenamefont {Zhao}, \citenamefont {Zhao} \emph {et~al.}}]{li2021imaging}%
  \BibitemOpen
  \bibfield  {author} {\bibinfo {author} {\bibfnamefont {H.}~\bibnamefont {Li}}, \bibinfo {author} {\bibfnamefont {S.}~\bibnamefont {Li}}, \bibinfo {author} {\bibfnamefont {M.~H.}\ \bibnamefont {Naik}}, \bibinfo {author} {\bibfnamefont {J.}~\bibnamefont {Xie}}, \bibinfo {author} {\bibfnamefont {X.}~\bibnamefont {Li}}, \bibinfo {author} {\bibfnamefont {J.}~\bibnamefont {Wang}}, \bibinfo {author} {\bibfnamefont {E.}~\bibnamefont {Regan}}, \bibinfo {author} {\bibfnamefont {D.}~\bibnamefont {Wang}}, \bibinfo {author} {\bibfnamefont {W.}~\bibnamefont {Zhao}}, \bibinfo {author} {\bibfnamefont {S.}~\bibnamefont {Zhao}}, \emph {et~al.},\ }\bibfield  {title} {\bibinfo {title} {Imaging moir{\'e} flat bands in three-dimensional reconstructed wse2/ws2 superlattices},\ }\href@noop {} {\bibfield  {journal} {\bibinfo  {journal} {Nature materials}\ }\textbf {\bibinfo {volume} {20}},\ \bibinfo {pages} {945} (\bibinfo {year} {2021}{\natexlab{b}})}\BibitemShut {NoStop}%
\bibitem [{\citenamefont {Gu}\ \emph {et~al.}(2016)\citenamefont {Gu}, \citenamefont {Lee}, \citenamefont {Wen}, \citenamefont {Cho}, \citenamefont {Ryu},\ and\ \citenamefont {Qi}}]{gu2016holographic}%
  \BibitemOpen
  \bibfield  {author} {\bibinfo {author} {\bibfnamefont {Y.}~\bibnamefont {Gu}}, \bibinfo {author} {\bibfnamefont {C.~H.}\ \bibnamefont {Lee}}, \bibinfo {author} {\bibfnamefont {X.}~\bibnamefont {Wen}}, \bibinfo {author} {\bibfnamefont {G.~Y.}\ \bibnamefont {Cho}}, \bibinfo {author} {\bibfnamefont {S.}~\bibnamefont {Ryu}},\ and\ \bibinfo {author} {\bibfnamefont {X.-L.}\ \bibnamefont {Qi}},\ }\bibfield  {title} {\bibinfo {title} {Holographic duality between (2+ 1)-dimensional quantum anomalous hall state and (3+ 1)-dimensional topological insulators},\ }\href@noop {} {\bibfield  {journal} {\bibinfo  {journal} {Physical Review B}\ }\textbf {\bibinfo {volume} {94}},\ \bibinfo {pages} {125107} (\bibinfo {year} {2016})}\BibitemShut {NoStop}%
\bibitem [{\citenamefont {Liu}\ \emph {et~al.}(2010)\citenamefont {Liu}, \citenamefont {Zhang}, \citenamefont {Yan}, \citenamefont {Qi}, \citenamefont {Frauenheim}, \citenamefont {Dai}, \citenamefont {Fang},\ and\ \citenamefont {Zhang}}]{liu2010oscillatory}%
  \BibitemOpen
  \bibfield  {author} {\bibinfo {author} {\bibfnamefont {C.-X.}\ \bibnamefont {Liu}}, \bibinfo {author} {\bibfnamefont {H.}~\bibnamefont {Zhang}}, \bibinfo {author} {\bibfnamefont {B.}~\bibnamefont {Yan}}, \bibinfo {author} {\bibfnamefont {X.-L.}\ \bibnamefont {Qi}}, \bibinfo {author} {\bibfnamefont {T.}~\bibnamefont {Frauenheim}}, \bibinfo {author} {\bibfnamefont {X.}~\bibnamefont {Dai}}, \bibinfo {author} {\bibfnamefont {Z.}~\bibnamefont {Fang}},\ and\ \bibinfo {author} {\bibfnamefont {S.-C.}\ \bibnamefont {Zhang}},\ }\bibfield  {title} {\bibinfo {title} {Oscillatory crossover from two-dimensional to three-dimensional topological insulators},\ }\href@noop {} {\bibfield  {journal} {\bibinfo  {journal} {Physical Review B—Condensed Matter and Materials Physics}\ }\textbf {\bibinfo {volume} {81}},\ \bibinfo {pages} {041307} (\bibinfo {year} {2010})}\BibitemShut {NoStop}%
\bibitem [{\citenamefont {Liu}\ and\ \citenamefont {Wen}(2013)}]{liu2013symmetry}%
  \BibitemOpen
  \bibfield  {author} {\bibinfo {author} {\bibfnamefont {Z.-X.}\ \bibnamefont {Liu}}\ and\ \bibinfo {author} {\bibfnamefont {X.-G.}\ \bibnamefont {Wen}},\ }\bibfield  {title} {\bibinfo {title} {Symmetry-protected quantum spin hall phases in two dimensions},\ }\href@noop {} {\bibfield  {journal} {\bibinfo  {journal} {Physical Review Letters}\ }\textbf {\bibinfo {volume} {110}},\ \bibinfo {pages} {067205} (\bibinfo {year} {2013})}\BibitemShut {NoStop}%
\bibitem [{\citenamefont {Liu}\ \emph {et~al.}(2024{\natexlab{c}})\citenamefont {Liu}, \citenamefont {Shtengel},\ and\ \citenamefont {Pollmann}}]{liu2024simulating}%
  \BibitemOpen
  \bibfield  {author} {\bibinfo {author} {\bibfnamefont {Y.-J.}\ \bibnamefont {Liu}}, \bibinfo {author} {\bibfnamefont {K.}~\bibnamefont {Shtengel}},\ and\ \bibinfo {author} {\bibfnamefont {F.}~\bibnamefont {Pollmann}},\ }\bibfield  {title} {\bibinfo {title} {Simulating two-dimensional topological quantum phase transitions on a digital quantum computer},\ }\href@noop {} {\bibfield  {journal} {\bibinfo  {journal} {Physical Review Research}\ }\textbf {\bibinfo {volume} {6}},\ \bibinfo {pages} {043256} (\bibinfo {year} {2024}{\natexlab{c}})}\BibitemShut {NoStop}%
\bibitem [{\citenamefont {Zhao}\ \emph {et~al.}(2025)\citenamefont {Zhao}, \citenamefont {Wang}, \citenamefont {He}, \citenamefont {Poon}, \citenamefont {Pak}, \citenamefont {Liu}, \citenamefont {Ren}, \citenamefont {Liu},\ and\ \citenamefont {Jo}}]{zhao2025two}%
  \BibitemOpen
  \bibfield  {author} {\bibinfo {author} {\bibfnamefont {E.}~\bibnamefont {Zhao}}, \bibinfo {author} {\bibfnamefont {Z.}~\bibnamefont {Wang}}, \bibinfo {author} {\bibfnamefont {C.}~\bibnamefont {He}}, \bibinfo {author} {\bibfnamefont {T.~F.~J.}\ \bibnamefont {Poon}}, \bibinfo {author} {\bibfnamefont {K.~K.}\ \bibnamefont {Pak}}, \bibinfo {author} {\bibfnamefont {Y.-J.}\ \bibnamefont {Liu}}, \bibinfo {author} {\bibfnamefont {P.}~\bibnamefont {Ren}}, \bibinfo {author} {\bibfnamefont {X.-J.}\ \bibnamefont {Liu}},\ and\ \bibinfo {author} {\bibfnamefont {G.-B.}\ \bibnamefont {Jo}},\ }\bibfield  {title} {\bibinfo {title} {Two-dimensional non-hermitian skin effect in an ultracold fermi gas},\ }\href@noop {} {\bibfield  {journal} {\bibinfo  {journal} {Nature}\ }\textbf {\bibinfo {volume} {637}},\ \bibinfo {pages} {565} (\bibinfo {year} {2025})}\BibitemShut {NoStop}%
\bibitem [{\citenamefont {Kawabata}\ \emph {et~al.}(2020)\citenamefont {Kawabata}, \citenamefont {Sato},\ and\ \citenamefont {Shiozaki}}]{kawabata2020higher}%
  \BibitemOpen
  \bibfield  {author} {\bibinfo {author} {\bibfnamefont {K.}~\bibnamefont {Kawabata}}, \bibinfo {author} {\bibfnamefont {M.}~\bibnamefont {Sato}},\ and\ \bibinfo {author} {\bibfnamefont {K.}~\bibnamefont {Shiozaki}},\ }\bibfield  {title} {\bibinfo {title} {Higher-order non-hermitian skin effect},\ }\href@noop {} {\bibfield  {journal} {\bibinfo  {journal} {Physical Review B}\ }\textbf {\bibinfo {volume} {102}},\ \bibinfo {pages} {205118} (\bibinfo {year} {2020})}\BibitemShut {NoStop}%
\bibitem [{\citenamefont {Zhang}\ \emph {et~al.}(2022{\natexlab{b}})\citenamefont {Zhang}, \citenamefont {Yang},\ and\ \citenamefont {Fang}}]{zhang2022universal}%
  \BibitemOpen
  \bibfield  {author} {\bibinfo {author} {\bibfnamefont {K.}~\bibnamefont {Zhang}}, \bibinfo {author} {\bibfnamefont {Z.}~\bibnamefont {Yang}},\ and\ \bibinfo {author} {\bibfnamefont {C.}~\bibnamefont {Fang}},\ }\bibfield  {title} {\bibinfo {title} {Universal non-hermitian skin effect in two and higher dimensions},\ }\href@noop {} {\bibfield  {journal} {\bibinfo  {journal} {Nature communications}\ }\textbf {\bibinfo {volume} {13}},\ \bibinfo {pages} {2496} (\bibinfo {year} {2022}{\natexlab{b}})}\BibitemShut {NoStop}%
\bibitem [{\citenamefont {Schindler}\ and\ \citenamefont {Prem}(2021)}]{schindler2021dislocation}%
  \BibitemOpen
  \bibfield  {author} {\bibinfo {author} {\bibfnamefont {F.}~\bibnamefont {Schindler}}\ and\ \bibinfo {author} {\bibfnamefont {A.}~\bibnamefont {Prem}},\ }\bibfield  {title} {\bibinfo {title} {Dislocation non-hermitian skin effect},\ }\href@noop {} {\bibfield  {journal} {\bibinfo  {journal} {Physical Review B}\ }\textbf {\bibinfo {volume} {104}},\ \bibinfo {pages} {L161106} (\bibinfo {year} {2021})}\BibitemShut {NoStop}%
\bibitem [{\citenamefont {Kunst}\ \emph {et~al.}(2018{\natexlab{b}})\citenamefont {Kunst}, \citenamefont {Edvardsson}, \citenamefont {Budich},\ and\ \citenamefont {Bergholtz}}]{kunst2018a}%
  \BibitemOpen
  \bibfield  {author} {\bibinfo {author} {\bibfnamefont {F.~K.}\ \bibnamefont {Kunst}}, \bibinfo {author} {\bibfnamefont {E.}~\bibnamefont {Edvardsson}}, \bibinfo {author} {\bibfnamefont {J.~C.}\ \bibnamefont {Budich}},\ and\ \bibinfo {author} {\bibfnamefont {E.~J.}\ \bibnamefont {Bergholtz}},\ }\bibfield  {title} {\bibinfo {title} {Biorthogonal bulk-boundary correspondence in non-hermitian systems},\ }\href {https://doi.org/10.1103/PhysRevLett.121.026808} {\bibfield  {journal} {\bibinfo  {journal} {Phys. Rev. Lett.}\ }\textbf {\bibinfo {volume} {121}},\ \bibinfo {pages} {026808} (\bibinfo {year} {2018}{\natexlab{b}})}\BibitemShut {NoStop}%
\bibitem [{\citenamefont {Kunst}\ and\ \citenamefont {Dwivedi}(2019)}]{kunst2019non}%
  \BibitemOpen
  \bibfield  {author} {\bibinfo {author} {\bibfnamefont {F.~K.}\ \bibnamefont {Kunst}}\ and\ \bibinfo {author} {\bibfnamefont {V.}~\bibnamefont {Dwivedi}},\ }\bibfield  {title} {\bibinfo {title} {Non-hermitian systems and topology: A transfer-matrix perspective},\ }\href@noop {} {\bibfield  {journal} {\bibinfo  {journal} {Physical Review B}\ }\textbf {\bibinfo {volume} {99}},\ \bibinfo {pages} {245116} (\bibinfo {year} {2019})}\BibitemShut {NoStop}%
\bibitem [{\citenamefont {Ashida}\ \emph {et~al.}(2020)\citenamefont {Ashida}, \citenamefont {Gong},\ and\ \citenamefont {Ueda}}]{ashida2020non}%
  \BibitemOpen
  \bibfield  {author} {\bibinfo {author} {\bibfnamefont {Y.}~\bibnamefont {Ashida}}, \bibinfo {author} {\bibfnamefont {Z.}~\bibnamefont {Gong}},\ and\ \bibinfo {author} {\bibfnamefont {M.}~\bibnamefont {Ueda}},\ }\bibfield  {title} {\bibinfo {title} {Non-hermitian physics},\ }\href@noop {} {\bibfield  {journal} {\bibinfo  {journal} {Advances in Physics}\ }\textbf {\bibinfo {volume} {69}},\ \bibinfo {pages} {249} (\bibinfo {year} {2020})}\BibitemShut {NoStop}%
\bibitem [{\citenamefont {Bergholtz}\ \emph {et~al.}(2021)\citenamefont {Bergholtz}, \citenamefont {Budich},\ and\ \citenamefont {Kunst}}]{bergholtz2021exceptional}%
  \BibitemOpen
  \bibfield  {author} {\bibinfo {author} {\bibfnamefont {E.~J.}\ \bibnamefont {Bergholtz}}, \bibinfo {author} {\bibfnamefont {J.~C.}\ \bibnamefont {Budich}},\ and\ \bibinfo {author} {\bibfnamefont {F.~K.}\ \bibnamefont {Kunst}},\ }\bibfield  {title} {\bibinfo {title} {Exceptional topology of non-hermitian systems},\ }\href@noop {} {\bibfield  {journal} {\bibinfo  {journal} {Reviews of Modern Physics}\ }\textbf {\bibinfo {volume} {93}},\ \bibinfo {pages} {015005} (\bibinfo {year} {2021})}\BibitemShut {NoStop}%
\bibitem [{\citenamefont {Longhi}(2019{\natexlab{a}})}]{longhi2019metal}%
  \BibitemOpen
  \bibfield  {author} {\bibinfo {author} {\bibfnamefont {S.}~\bibnamefont {Longhi}},\ }\bibfield  {title} {\bibinfo {title} {Metal-insulator phase transition in a non-hermitian aubry-andr{\'e}-harper model},\ }\href@noop {} {\bibfield  {journal} {\bibinfo  {journal} {Physical Review B}\ }\textbf {\bibinfo {volume} {100}},\ \bibinfo {pages} {125157} (\bibinfo {year} {2019}{\natexlab{a}})}\BibitemShut {NoStop}%
\bibitem [{\citenamefont {Mandal}\ \emph {et~al.}(2020)\citenamefont {Mandal}, \citenamefont {Banerjee}, \citenamefont {Ostrovskaya},\ and\ \citenamefont {Liew}}]{mandal2020nonreciprocal}%
  \BibitemOpen
  \bibfield  {author} {\bibinfo {author} {\bibfnamefont {S.}~\bibnamefont {Mandal}}, \bibinfo {author} {\bibfnamefont {R.}~\bibnamefont {Banerjee}}, \bibinfo {author} {\bibfnamefont {E.~A.}\ \bibnamefont {Ostrovskaya}},\ and\ \bibinfo {author} {\bibfnamefont {T.~C.~H.}\ \bibnamefont {Liew}},\ }\bibfield  {title} {\bibinfo {title} {Nonreciprocal transport of exciton polaritons in a non-hermitian chain},\ }\href@noop {} {\bibfield  {journal} {\bibinfo  {journal} {Physical Review Letters}\ }\textbf {\bibinfo {volume} {125}},\ \bibinfo {pages} {123902} (\bibinfo {year} {2020})}\BibitemShut {NoStop}%
\bibitem [{\citenamefont {Longhi}(2019{\natexlab{b}})}]{longhi2019topological}%
  \BibitemOpen
  \bibfield  {author} {\bibinfo {author} {\bibfnamefont {S.}~\bibnamefont {Longhi}},\ }\bibfield  {title} {\bibinfo {title} {Topological phase transition in non-hermitian quasicrystals},\ }\href@noop {} {\bibfield  {journal} {\bibinfo  {journal} {Physical review letters}\ }\textbf {\bibinfo {volume} {122}},\ \bibinfo {pages} {237601} (\bibinfo {year} {2019}{\natexlab{b}})}\BibitemShut {NoStop}%
\bibitem [{\citenamefont {Weidemann}\ \emph {et~al.}(2022)\citenamefont {Weidemann}, \citenamefont {Kremer}, \citenamefont {Longhi},\ and\ \citenamefont {Szameit}}]{weidemann2022topological}%
  \BibitemOpen
  \bibfield  {author} {\bibinfo {author} {\bibfnamefont {S.}~\bibnamefont {Weidemann}}, \bibinfo {author} {\bibfnamefont {M.}~\bibnamefont {Kremer}}, \bibinfo {author} {\bibfnamefont {S.}~\bibnamefont {Longhi}},\ and\ \bibinfo {author} {\bibfnamefont {A.}~\bibnamefont {Szameit}},\ }\bibfield  {title} {\bibinfo {title} {Topological triple phase transition in non-hermitian floquet quasicrystals},\ }\href@noop {} {\bibfield  {journal} {\bibinfo  {journal} {Nature}\ }\textbf {\bibinfo {volume} {601}},\ \bibinfo {pages} {354} (\bibinfo {year} {2022})}\BibitemShut {NoStop}%
\bibitem [{\citenamefont {Weidemann}\ \emph {et~al.}(2020)\citenamefont {Weidemann}, \citenamefont {Kremer}, \citenamefont {Helbig}, \citenamefont {Hofmann}, \citenamefont {Stegmaier}, \citenamefont {Greiter}, \citenamefont {Thomale},\ and\ \citenamefont {Szameit}}]{weidemann2020topological}%
  \BibitemOpen
  \bibfield  {author} {\bibinfo {author} {\bibfnamefont {S.}~\bibnamefont {Weidemann}}, \bibinfo {author} {\bibfnamefont {M.}~\bibnamefont {Kremer}}, \bibinfo {author} {\bibfnamefont {T.}~\bibnamefont {Helbig}}, \bibinfo {author} {\bibfnamefont {T.}~\bibnamefont {Hofmann}}, \bibinfo {author} {\bibfnamefont {A.}~\bibnamefont {Stegmaier}}, \bibinfo {author} {\bibfnamefont {M.}~\bibnamefont {Greiter}}, \bibinfo {author} {\bibfnamefont {R.}~\bibnamefont {Thomale}},\ and\ \bibinfo {author} {\bibfnamefont {A.}~\bibnamefont {Szameit}},\ }\bibfield  {title} {\bibinfo {title} {Topological funneling of light},\ }\href@noop {} {\bibfield  {journal} {\bibinfo  {journal} {Science}\ }\textbf {\bibinfo {volume} {368}},\ \bibinfo {pages} {311} (\bibinfo {year} {2020})}\BibitemShut {NoStop}%
\bibitem [{\citenamefont {Ghatak}\ \emph {et~al.}(2020)\citenamefont {Ghatak}, \citenamefont {Brandenbourger}, \citenamefont {Van~Wezel},\ and\ \citenamefont {Coulais}}]{ghatak2020observation}%
  \BibitemOpen
  \bibfield  {author} {\bibinfo {author} {\bibfnamefont {A.}~\bibnamefont {Ghatak}}, \bibinfo {author} {\bibfnamefont {M.}~\bibnamefont {Brandenbourger}}, \bibinfo {author} {\bibfnamefont {J.}~\bibnamefont {Van~Wezel}},\ and\ \bibinfo {author} {\bibfnamefont {C.}~\bibnamefont {Coulais}},\ }\bibfield  {title} {\bibinfo {title} {Observation of non-hermitian topology and its bulk--edge correspondence in an active mechanical metamaterial},\ }\href@noop {} {\bibfield  {journal} {\bibinfo  {journal} {Proceedings of the National Academy of Sciences}\ }\textbf {\bibinfo {volume} {117}},\ \bibinfo {pages} {29561} (\bibinfo {year} {2020})}\BibitemShut {NoStop}%
\bibitem [{\citenamefont {Wang}\ \emph {et~al.}(2022)\citenamefont {Wang}, \citenamefont {Wang},\ and\ \citenamefont {Ma}}]{wang2022non}%
  \BibitemOpen
  \bibfield  {author} {\bibinfo {author} {\bibfnamefont {W.}~\bibnamefont {Wang}}, \bibinfo {author} {\bibfnamefont {X.}~\bibnamefont {Wang}},\ and\ \bibinfo {author} {\bibfnamefont {G.}~\bibnamefont {Ma}},\ }\bibfield  {title} {\bibinfo {title} {Non-hermitian morphing of topological modes},\ }\href@noop {} {\bibfield  {journal} {\bibinfo  {journal} {Nature}\ }\textbf {\bibinfo {volume} {608}},\ \bibinfo {pages} {50} (\bibinfo {year} {2022})}\BibitemShut {NoStop}%
\bibitem [{\citenamefont {Liang}\ \emph {et~al.}(2022)\citenamefont {Liang}, \citenamefont {Xie}, \citenamefont {Dong}, \citenamefont {Li}, \citenamefont {Li}, \citenamefont {Gadway}, \citenamefont {Yi},\ and\ \citenamefont {Yan}}]{liang2022dynamic}%
  \BibitemOpen
  \bibfield  {author} {\bibinfo {author} {\bibfnamefont {Q.}~\bibnamefont {Liang}}, \bibinfo {author} {\bibfnamefont {D.}~\bibnamefont {Xie}}, \bibinfo {author} {\bibfnamefont {Z.}~\bibnamefont {Dong}}, \bibinfo {author} {\bibfnamefont {H.}~\bibnamefont {Li}}, \bibinfo {author} {\bibfnamefont {H.}~\bibnamefont {Li}}, \bibinfo {author} {\bibfnamefont {B.}~\bibnamefont {Gadway}}, \bibinfo {author} {\bibfnamefont {W.}~\bibnamefont {Yi}},\ and\ \bibinfo {author} {\bibfnamefont {B.}~\bibnamefont {Yan}},\ }\bibfield  {title} {\bibinfo {title} {Dynamic signatures of non-hermitian skin effect and topology in ultracold atoms},\ }\href@noop {} {\bibfield  {journal} {\bibinfo  {journal} {Physical review letters}\ }\textbf {\bibinfo {volume} {129}},\ \bibinfo {pages} {070401} (\bibinfo {year} {2022})}\BibitemShut {NoStop}%
\bibitem [{\citenamefont {Xiao}\ \emph {et~al.}(2022)\citenamefont {Xiao}, \citenamefont {Wang}, \citenamefont {Li}, \citenamefont {Chen},\ and\ \citenamefont {Yuan}}]{xiao2022bound}%
  \BibitemOpen
  \bibfield  {author} {\bibinfo {author} {\bibfnamefont {H.}~\bibnamefont {Xiao}}, \bibinfo {author} {\bibfnamefont {L.}~\bibnamefont {Wang}}, \bibinfo {author} {\bibfnamefont {Z.-H.}\ \bibnamefont {Li}}, \bibinfo {author} {\bibfnamefont {X.}~\bibnamefont {Chen}},\ and\ \bibinfo {author} {\bibfnamefont {L.}~\bibnamefont {Yuan}},\ }\bibfield  {title} {\bibinfo {title} {Bound state in a giant atom-modulated resonators system},\ }\href@noop {} {\bibfield  {journal} {\bibinfo  {journal} {npj Quantum Information}\ }\textbf {\bibinfo {volume} {8}},\ \bibinfo {pages} {80} (\bibinfo {year} {2022})}\BibitemShut {NoStop}%
\bibitem [{\citenamefont {Zhang}\ \emph {et~al.}(2023{\natexlab{b}})\citenamefont {Zhang}, \citenamefont {Dong}, \citenamefont {Gao}, \citenamefont {Zhao}, \citenamefont {Hao}, \citenamefont {Desaules}, \citenamefont {Guo}, \citenamefont {Chen}, \citenamefont {Deng}, \citenamefont {Liu} \emph {et~al.}}]{zhang2023many}%
  \BibitemOpen
  \bibfield  {author} {\bibinfo {author} {\bibfnamefont {P.}~\bibnamefont {Zhang}}, \bibinfo {author} {\bibfnamefont {H.}~\bibnamefont {Dong}}, \bibinfo {author} {\bibfnamefont {Y.}~\bibnamefont {Gao}}, \bibinfo {author} {\bibfnamefont {L.}~\bibnamefont {Zhao}}, \bibinfo {author} {\bibfnamefont {J.}~\bibnamefont {Hao}}, \bibinfo {author} {\bibfnamefont {J.-Y.}\ \bibnamefont {Desaules}}, \bibinfo {author} {\bibfnamefont {Q.}~\bibnamefont {Guo}}, \bibinfo {author} {\bibfnamefont {J.}~\bibnamefont {Chen}}, \bibinfo {author} {\bibfnamefont {J.}~\bibnamefont {Deng}}, \bibinfo {author} {\bibfnamefont {B.}~\bibnamefont {Liu}}, \emph {et~al.},\ }\bibfield  {title} {\bibinfo {title} {Many-body hilbert space scarring on a superconducting processor},\ }\href@noop {} {\bibfield  {journal} {\bibinfo  {journal} {Nature Physics}\ }\textbf {\bibinfo {volume} {19}},\ \bibinfo {pages} {120} (\bibinfo {year} {2023}{\natexlab{b}})}\BibitemShut {NoStop}%
\bibitem [{\citenamefont {Shen}\ \emph {et~al.}(2023)\citenamefont {Shen}, \citenamefont {Chen}, \citenamefont {Aliyu}, \citenamefont {Qin}, \citenamefont {Zhong}, \citenamefont {Loh},\ and\ \citenamefont {Lee}}]{shen2023proposal}%
  \BibitemOpen
  \bibfield  {author} {\bibinfo {author} {\bibfnamefont {R.}~\bibnamefont {Shen}}, \bibinfo {author} {\bibfnamefont {T.}~\bibnamefont {Chen}}, \bibinfo {author} {\bibfnamefont {M.~M.}\ \bibnamefont {Aliyu}}, \bibinfo {author} {\bibfnamefont {F.}~\bibnamefont {Qin}}, \bibinfo {author} {\bibfnamefont {Y.}~\bibnamefont {Zhong}}, \bibinfo {author} {\bibfnamefont {H.}~\bibnamefont {Loh}},\ and\ \bibinfo {author} {\bibfnamefont {C.~H.}\ \bibnamefont {Lee}},\ }\bibfield  {title} {\bibinfo {title} {Proposal for observing yang-lee criticality in rydberg atomic arrays},\ }\href@noop {} {\bibfield  {journal} {\bibinfo  {journal} {Physical Review Letters}\ }\textbf {\bibinfo {volume} {131}},\ \bibinfo {pages} {080403} (\bibinfo {year} {2023})}\BibitemShut {NoStop}%
\bibitem [{\citenamefont {Koh}\ \emph {et~al.}(2023{\natexlab{b}})\citenamefont {Koh}, \citenamefont {Sun}, \citenamefont {Motta},\ and\ \citenamefont {Minnich}}]{koh2023measurement}%
  \BibitemOpen
  \bibfield  {author} {\bibinfo {author} {\bibfnamefont {J.~M.}\ \bibnamefont {Koh}}, \bibinfo {author} {\bibfnamefont {S.-N.}\ \bibnamefont {Sun}}, \bibinfo {author} {\bibfnamefont {M.}~\bibnamefont {Motta}},\ and\ \bibinfo {author} {\bibfnamefont {A.~J.}\ \bibnamefont {Minnich}},\ }\bibfield  {title} {\bibinfo {title} {Measurement-induced entanglement phase transition on a superconducting quantum processor with mid-circuit readout},\ }\href@noop {} {\bibfield  {journal} {\bibinfo  {journal} {Nature Physics}\ }\textbf {\bibinfo {volume} {19}},\ \bibinfo {pages} {1314} (\bibinfo {year} {2023}{\natexlab{b}})}\BibitemShut {NoStop}%
\end{thebibliography}
\end{document}